\documentclass[11pt]{article}
\usepackage{epsf}
\usepackage{color}
\usepackage{amsmath,amsfonts,amsbsy,amssymb}

\begin{document}


\begin{titlepage}

\def\slash#1{{\rlap{$#1$} \thinspace/}}

\begin{flushright} 
\end{flushright} 

\vspace{0.1cm}

\begin{Large}
\begin{center}

{\bf Graded Hopf Maps and Fuzzy Superspheres}
\end{center}
\end{Large}
\vspace{1cm}

\begin{center}
{\bf Kazuki Hasebe}   \\ 

\vspace{0.5cm} 
\it{Kagawa National College of Technology, Takuma, 
Mitoyo, Kagawa 769-1192, Japan} \\

\vspace{0.5cm} 
{\sf
hasebe@dg.kagawa-nct.ac.jp} 

\vspace{0.8cm} 


\end{center}

\vspace{1.5cm}

\begin{abstract}
\noindent

\baselineskip=18pt

We argue supersymmetric generalizations of fuzzy two- and four-spheres based on the unitary-orthosymplectic algebras, $uosp(N|2)$ and $uosp(N|4)$, respectively. Supersymmetric version of Schwinger construction is applied to derive graded fully symmetric representation for fuzzy superspheres.  
As a classical counterpart of fuzzy superspheres, graded versions of 1st and 2nd Hopf maps are introduced, and their basic geometrical structures are studied.   
It is shown that  fuzzy superspheres are represented as a ``superposition'' of fuzzy superspheres with lower supersymmetries.  
We also investigate algebraic structures of fuzzy two- and four-superspheres to identify $su(2|N)$ and $su(4|N)$ as their enhanced algebraic structures, respectively. Evaluation of correlation functions manifests such enhanced structure as  quantum fluctuations of fuzzy supersphere. 

\end{abstract}

\end{titlepage}

\newpage

\tableofcontents

\section{Introduction}


Concrete idea and technique of quantization of two-sphere may be traced back to the work of Berezin \cite{berezin1975} in 70s.  In the beginning of 80s, 
the algebraic structure of fuzzy two-sphere and field theory on it were first argued by Hoppe \cite{Hoppe1982} and in the early 90s subsequently explored by Madore \cite{madore1992}.  The fuzzy two-sphere is one of the simplest curved fuzzy manifolds whose coordinates satisfy the $SU(2)$ algebra. 
Field theory defined on fuzzy manifolds naturally contain a ``cut-off'', and such non-commutative field theory was expected to have weaker infinity than that of the conventional field theory.  
 Few years after the work of Madore, Grosse et al. introduced four-dimensional fuzzy spheres \cite{hep-th/9602115} and supersymmetric (SUSY) generalizations of fuzzy spheres in sequel works \cite{hep-th/9507074, math-ph/9804013}. In the developments of string theory in late 90s, researchers recognized that the geometry of D-branes is described by fuzzy geometry \cite{hep-th/0007170,hep-th/0101126,hep-th/0512054} (as reviews) and fuzzy manifolds arise as classical solutions of Matrix theory, e.g.  \cite{hep-th/9711078,hep-th/9910053}.  It is also known that fuzzy superspheres provide a set-up for field theory on SUSY lattice regularization \cite{hep-th/9507074,hep-th/9903112,hep-th/9903202}, and realize as a classical solution of supermatrix model \cite{hep-th/0311005,hep-th/0312307}.   For such important properties, fuzzy spheres and their variants have attracted a great deal of attentions \cite
{hep-th/0106048,Azumathesis,hep-th/0511114,Abethesis} (as 
reviews).  Non-commutative geometry and fuzzy physics also found their applications to gravity \cite{arXiv:hep-th/0504183,arXiv:hep-th/0506157,hep-th/0606197} and even to condensed matter physics \cite{Girvin1984,hep-th/0209198}.  Recently, the mathematics of fuzzy geometry is applied to construction of topologically non-trivial many-body states on bosonic manifolds \cite{ZhangHu2001,hep-th/0203264,hep-th/0310274} and on supermanifolds \cite{arXiv:hep-th/0411137,arXiv:0901.1498} as well. 

 In this paper, we apply close relations between fuzzy spheres and Hopf maps \cite{arXiv:1009.1192} to generalize fuzzy superspheres in higher dimensions. 
A useful mathematical tool for that construction is the Schwinger operator formalism \cite{hep-th/0204170,hep-th/0511114}. Specifically,  
the two-dimensional fuzzy sphere coordinates are simply obtained by sandwiching the Pauli matrices with two-component Schwinger operators: 
\begin{equation}
X_i=\Phi^{\dagger}\sigma_i\Phi. 
\label{1stquantumHopfmap}
\end{equation}
With the Schwinger operator, it is quite straightforward to derive fully symmetric representation, which corresponds to a finite number of states on fuzzy sphere.   
In general,  a finite number of states on $2k$-dimensional fuzzy spheres are given by fully symmetric representation of $SO(2k+1)$ \cite{AzumaBagnoud2003}. 
  The Schwinger operator is regarded as  the ``square root'' of the fuzzy sphere coordinates, and play fundamental roles rather than the fuzzy sphere coordinates themselves. 
Meanwhile, with $\phi$ denoting a normalized two-component complex spinor, 
 the (1st) Hopf map is represented as 
\begin{equation}
x_i=\phi^{\dagger}\sigma_i\phi. 
\label{1stclassicalHopfmap}
\end{equation}
Comparison between (\ref{1stquantumHopfmap}) and (\ref{1stclassicalHopfmap}) finds that the (1st) Hopf map can be regarded as the ``classical'' counterpart of the (Schwinger) operator construction of fuzzy two-sphere.    

 In the construction of fuzzy superspheres, nice algebraic structures and  relations between the Hopf map and fuzzy sphere are inherited \cite{hep-th/9510083,hep-th/0409230}.   
The fuzzy two-superspheres\footnote{In this paper, two-supersphere is referred to as the supersphere whose body is two-dimensional sphere. Two-supersphere with $N$ supersymmetry is denoted as $S^{2|2N}$ whose bosonic dimension is two and the fermionic dimension is $2N$, and hence the total dimension is $2+2N$. Similarly, fuzzy four-supersphere consists of four-sphere body and extra fermionic coordinates.} constructed by Grosse et al. \cite{hep-th/9507074, math-ph/9804013} are based on the $UOSp(1|2)$ algebra that includes $su(2)\simeq usp(2)$: 
\begin{equation}
su(2) \subset uosp(1|2). 
\end{equation}
(The classical counter part of the fuzzy two-supersphere, the graded 1st Hopf map, was first given in Refs.\cite{LandiMarmo1987,Bartocci1990}. See also Refs.\cite{hep-th/0409230,Landi2001}.)  
The coordinates of the fuzzy two-supersphere are introduced by replacing the $SU(2)$ Pauli matrices with the $UOSp(1|2)$ matrices of fundamental representation. 
 As  $uosp(1|2)$ contains $su(2)$ as its maximal bosonic subalgebra, the fuzzy two-supersphere ``contains'' the fuzzy two-sphere as its fuzzy body. 
Such construction is based on the graded Lie algebra, and hence the structure of fuzzy super-geometry is transparent.  We want to maintain such nice features.  To this end, 
we utilize a graded Lie algebra whose maximal bosonic subalgebra is $so(5)$. The minimal graded Lie algebra that suffices for this requirement is $uosp(1|4)$, since $so(5)\simeq usp(4)$: 
\begin{equation}
so(5) \subset uosp(1|4). 
\end{equation}
We adopt $UOSp(1|4)$ version of Schwinger operator in the construction of fuzzy four-supersphere and also introduce the graded 2nd Hopf map as its classical counterpart.    
We further extend such formulation to include more supersymmetries with use of $UOSp(N|2)$ and $UOSp(N|4)$.     
Representation theory of the graded Lie algebra is rather complicated,  however if restricted to graded fully symmetric representation\footnote{ 
 We adopt the terminology, ``graded fully symmetric representation'' to indicate a representation constructed by a supersymmetric version of Schwinger operator. 
 The graded fully symmetric representation is totally symmetric for the bosonic part and totally $\it{antisymmetric}$ for the fermionic part.  It is also referred to as harmonic oscillator representation in several literatures. For general representation theory of graded Lie groups, one may for instance consult Ref.\cite{BookFrappat} and references therein.}, investigations are greatly simplified.  
 By dealing with the Schwinger operator as fundamental quantity, we observe ``enhancement'' of symmetry of fuzzy superspheres. This mechanism is similar to the symmetry enhancement reported in  higher-dimensional fuzzy spheres \cite{Ho2002,Kimura2002,Kimura2003}. We also reconsider such enhancement in view of quantum fluctuations of fuzzy superspheres. 
   
Some comments are added to clarify difference to related works.  
  In Ref.\cite{arXiv:0912.3279}, supersymmetric Hopf maps were introduced in the context of SUSY non-linear sigma models. 
  In the construction, the fermionic parts are introduced to incorporate   ${N}=4$ supersymmetry. 
Though the bosonic parts are related to Hopf maps, the fermionic parts themselves are not directly related.   
In the present construction, 
together with bosonic components, the fermionic components themselves constitute graded Hopf maps.   Supersymmetric quantum mechanics in monopole background related to the Hopf map is well investigated recently \cite{hep-th/0606152,arXiv:0902.2682,arXiv:0905.3461,arXiv:0905.4951,arXiv:0911.3257,arXiv:1004.4597,arXiv:1001.2659}.  
Works about higher-dimensional fuzzy super-manifolds of which the author is aware are 
 Ref.\cite{arXiv:hep-th/0311159,arXiv:hep-th/0611328,arXiv:0811.4743,DeBellisetal2010}. 
The fuzzy complex projective space was constructed in Ref.\cite{arXiv:hep-th/0311159} based on the super unitary algebra. Such construction is similar to the spirit of the present work, and is indeed closed related as we shall discuss. In \cite{DeBellisetal2010},
 fuzzy superspheres  are formulated in any dimensions. However, the fuzzy two-supersphere provided by the formulation is not same as of Grosse et al.   In the present, though the construction is restricted to two and four-dimensions,  the algebraic structure underlying fuzzy geometry is transparent and the fuzzy two-supersphere of Grosse et al. is naturally reproduced.

  The paper is organized as follows. 
  In Sec.\ref{SecUOSp(MN)}, we briefly introduce the unitary-orthosymplectic algebra,  $uosp(N|M)$. 
In Sec.\ref{Sec1stgradedHopffuzzysuperspheres}, we review the construction of fuzzy two-supersphere as well as 1st graded Hopf map. ${N}=2$ fuzzy two-supersphere and the corresponding 1st graded Hopf map are also discussed. 
In Sec.\ref{SectGraded2nd},  we argue construction of  ${N}=1$ and ${N}=2$ fuzzy superspheres and the graded 2nd Hopf maps.  
  More supersymmetric extensions are explored in Sec.\ref{SectMoreSUSY}.  
 In Sec.\ref{Sectsupercoherent}, we give 
   supercoherent states on fuzzy two- and four-superspheres and investigate quantum fluctuations of fuzzy superspheres.    
  Sec.\ref{SectSummary} is devoted to summary and discussions.

\section{$UOSp(N|M)$}\label{SecUOSp(MN)}

Generators of the orthosymplectic group $OSp(N|M)$ are defined so as to satisfy  
\begin{equation}
\Sigma_{AB}^{st}
\begin{pmatrix}
J & 0 \\
0 & 1_N
\end{pmatrix} + \begin{pmatrix}
J & 0 \\
0 & 1_N
\end{pmatrix}\Sigma_{AB}=0, 
\label{defospmn}
\end{equation}
where $1_N$ denotes $N\times N$ unit matrix and  $J$ represents the invariant matrix of the symplectic group   
\begin{equation}
J=
\begin{pmatrix}
0 & 1_{M/2} \\
-1_{M/2} & 0 
\end{pmatrix}, 
\label{sympecticinvmat} 
\end{equation}
and the supertranspose, $st$, is defined as  
\begin{equation}
\begin{pmatrix}
B & F \\
F' & B'
\end{pmatrix}^{st}
\equiv 
\begin{pmatrix}
B^t & {F'}^t \\
-{F}^t & {B'}^t
\end{pmatrix}.  
\end{equation}
Here, $t$ stands for the ordinary transpose, and $B$ and $B'$ signify bosonic components while $F$ and $F'$ fermionic components.    
$\Sigma_{AB}$ can be expressed by a linear combination of 
\begin{equation}
\Sigma_{\alpha\beta}=\begin{pmatrix}
\sigma_{\alpha\beta} & 0 \\
0 & 0 
\end{pmatrix},~~\Sigma_{lm}=\begin{pmatrix}
0 & 0 \\
0 & \sigma_{lm}
\end{pmatrix}
,~~\Sigma_{l\alpha}=
\begin{pmatrix}
0 & \sigma_{l\alpha} \\
-(J\sigma_{l\alpha})^t & 0 
\end{pmatrix}, 
\end{equation}
where   $\alpha,\beta$ are the indices of  $Sp(M)$ $(\alpha,\beta=1,2,\cdots,M)$ and $l,m$ those of $O(N)$ ($l,m=1,2,\cdots,N$).  
$\sigma_{l\alpha}$ denote arbitrary $M\times N$ matrices, while  
$\sigma_{\alpha\beta}$ and $\sigma_{lm}$ signify $M\times M$ and $N\times N$ matrices that  respectively satisfy 
\begin{subequations}
\begin{align}
&{\sigma_{lm}}^t+\sigma_{lm}=0,\label{defortho}\\
&{\sigma_{\alpha\beta}}^t J + J\sigma_{\alpha\beta}=0\label{defsymp}.   
\end{align} 
\end{subequations}
The $OSp(N|M)$ algebra contains the maximal bosonic subalgebra, $sp(M) \oplus o(N)$, whose generators are $\Sigma_{\alpha\beta}$ and $\Sigma_{lm}$.  The off-diagonal block matrices  $\Sigma_{l\alpha}$ are called fermionic generators that  transform as fundamental representation under each of $Sp(M)$ and $O(N)$.   
Then, the $so(N)$ matrix $\sigma_{lm}$ is an antisymmetric real matrix (\ref{defortho}) with real degrees of freedom  $N(N-1)/2$. The indices of $\sigma_{lm}$ can be taken to be antisymmetric,    
$\sigma_{lm}=-\sigma_{ml}$.    
Meanwhile, from the relation (\ref{defsymp}) $\sigma_{\alpha\beta}$ takes the form of 
\begin{equation}
\sigma_{\alpha\beta}=
\begin{pmatrix}
k & s \\
s' & -k^t
\end{pmatrix},  
\end{equation}
where $k$ stands for a $M/2 \times M/2$ complex matrix, and $s$ and $s'$ 
 are $M/2 \times M/2$ symmetric complex matrices.  
If the hermiticity condition is further imposed, $\sigma_{\alpha\beta}$ are reduced to 
the generators of $USp(M)$ and take the form of 
\begin{equation}
\sigma_{\alpha\beta}=
\begin{pmatrix}
h & s \\
s^{\dagger} & -h^* 
\end{pmatrix},  
\label{sp4generators}
\end{equation}
where $h$ represents hermitian matrix and $s$ symmetric complex  matrix. The real independent degrees of freedom of   $\sigma_{\alpha\beta}$ is $M(M+1)/2$.  Then, for $usp(M)$, the indices can be taken to be  symmetric,   
$\sigma_{\alpha\beta}=\sigma_{\beta\alpha}. $ 
Meanwhile, the real degrees of freedom of the fermionic generators 
$\Sigma_{l\alpha}$ is $MN$. 
Consequently, th real degrees of freedom of $uosp(N|M)$ are given by  
\begin{equation}
\dim[uosp(N|M)]=\frac{1}{2}(M^2+N^2+M-N)|MN=\frac{1}{2}((M+N)^2+M-N).
\end{equation}

 There are isometries between the unitary-symplectic and orthogonal algebras only for 
 \begin{equation}
 usp(2)\simeq so(3),~~~~~~~~usp(4)\simeq so(5).
 \end{equation}
 Taking advantage of such isomorphism, we construct fuzzy two- and four-superspheres based on  $uosp(N|2)$ and $uosp(N|4)$. 

\section{Graded 1st Hopf maps and fuzzy two-superspheres}\label{Sec1stgradedHopffuzzysuperspheres}

Here, we review relations between fuzzy two-sphere and 1st Hopf map, and their supersymmetric version. We also explore a construction of ${N}=2$ fuzzy supersphere with use of typical representation of $UOSp(2|2)$ algebra.    

\subsection{The 1st Hopf map and fuzzy two-sphere}

To begin with, we introduce relations between fuzzy two-sphere and 1st Hopf map  
\begin{equation}
S^3\overset{S^1}\longrightarrow S^2.   
\end{equation}
With a normalized complex two-component spinor $\phi=(\phi_1,\phi_2)^t$ subject to $\phi^{\dagger}\phi=1$,  
the 1st Hopf map is realized as  
\begin{equation}
\phi \rightarrow x_i =\phi^{\dagger}\sigma_i\phi,
\label{theexplicitmapof1stHopf}
\end{equation}
where $\sigma_i$ $(i=1,2,3)$ are the Pauli matrices,
\begin{equation}
\sigma_1=\begin{pmatrix}
0 & 1 \\
1 & 0 
\end{pmatrix},~~\sigma_2=
\begin{pmatrix}
0 & -i \\
i & 0 
\end{pmatrix},~~
\sigma_3=
\begin{pmatrix}
1 & 0 \\
0 & -1 
\end{pmatrix}. 
\end{equation}
$\phi$ is regarded as coordinates on $S^3$ from the normalization condition, and $x_i$ denote coordinates of $S^2$: 
\begin{equation}
x_ix_i=(\phi^{\dagger}\phi)^2=1. 
\label{twossphereradius}
\end{equation}
Coordinates of fuzzy two-sphere $S_F^2$ are constructed as 
\begin{equation}
X_i=\Phi^{\dagger}\sigma_i \Phi, 
\label{Schwingersf2}
\end{equation}
where $\Phi=(\Phi_1,\Phi_2)^t$ stands for two-component Schwinger operator 
that satisfies $[\Phi_{\alpha},\Phi^{\dagger}_{\beta}]=\delta_{\alpha\beta}$ and 
$[\Phi_{\alpha},\Phi_{\beta}]=0$ $(\alpha,\beta=1,2)$.
 Usually, in front of the right-hand side of (\ref{Schwingersf2}), 
the non-commutative parameter of dimension of length is added, however 
for notational brevity, we omit it throughout the paper. $X_i$ satisfy 
\begin{equation}
[X_i,X_j]=2i\epsilon_{ijk}X_k, 
\end{equation}
and square of the radius of fuzzy two-sphere is given by 
\begin{equation}
X_iX_i=(\Phi^{\dagger}\Phi)(\Phi^{\dagger}\Phi+2)=\hat{n}(\hat{n}+2).  
\label{fuzzytwosphereradius}
\end{equation}
Here, $\hat{n}$ is the number operator $\hat{n}=\Phi^{\dagger}\Phi$ and its eigenvalues are non-negative integers that specify fully symmetric representation.   
The fully symmetric representation is simply obtained by acting the components of the Schwinger operator to the vacuum:  
\begin{equation}
|l_1,l_2\rangle=
\frac{1}{\sqrt{l_1!~l_2!}}
{\Phi_1^{\dagger}}^{l_1}{\Phi_2^{\dagger}}^{l_2}|0\rangle, 
\label{su2irredu}
\end{equation}
where $l_1$ and $l_2$ are non-negative integers satisfying $l_1+l_2=n$.  Physically, $|l_1,l_2\rangle$ represent a finite number of states on fuzzy two-sphere, and their 3rd-components are   
\begin{equation}
X_3=l_1-l_2=n-2k,  
\label{X3fuzzysphere}
\end{equation}
where $k=l_2=0,1,2,\cdots,n$.
 The dimension of (\ref{su2irredu}) is 
\begin{equation}
d(n)=n+1.  
\label{su2repre}
\end{equation}

The Hopf map (\ref{theexplicitmapof1stHopf}) is regarded as a classical counterpart of the  Schwinger construction of fuzzy sphere (\ref{Schwingersf2}) with the replacement   
\begin{align}
\Phi_{\alpha}\rightarrow \phi_{\alpha},~~~~~\Phi^{\dagger}_{\alpha}\rightarrow \phi^{*}_{\alpha},  
\end{align}
and (\ref{fuzzytwosphereradius}) is reduced to 
 (\ref{twossphereradius}) except for the ``zero-point energy'',   stemming from the non-commutativity of two bosonic components of the Schwinger operator.

\subsection{${N}=1$ fuzzy two-supersphere }\label{Subsecn1twosphere}

Here, we extend the above discussions to the graded 1st Hopf map \cite{LandiMarmo1987,Bartocci1990} and  ${N}=1$ fuzzy two-supersphere \cite{hep-th/9507074, math-ph/9804013} along Refs.\cite{hep-th/0409230,Landi2001}. 

\subsubsection{$UOSp(1|2)$ algebra}

The $UOSp(1|2)$ algebra contains the $SU(2)$ algebra as its maximal bosonic subalgebra, and consists of five generators three of which are bosonic $L_i$ $(i=1,2,3)$ and two of which are 
fermionic $L_{\alpha}$ $(\alpha=\theta_1,\theta_2)$. They satisfy 
\begin{equation}
[L_i,L_j]=i\epsilon_{ijk}L_k,~~~
[L_i,L_{\alpha}]=\frac{1}{2}(\sigma_{i})_{\beta\alpha}L_{\beta},~~~\{L_{\alpha},L_{\beta}\}=\frac{1}{2}(\epsilon\sigma_i)_{\alpha\beta}L_i,\label{osp3}
\end{equation}
 where $\epsilon=i\sigma_2$ is the $SU(2)$ charge conjugation matrix. One may find that $L_i$ transform as an $SU(2)$ vector, while $L_{\alpha}$ an $SU(2)$ spinor.  
The $UOSp(1|2)$ Casimir is constructed as  
\begin{equation}
\mathcal{C}=L_iL_i+\epsilon_{\alpha\beta}L_{\alpha}L_{\beta}, 
\end{equation}
and its eigenvalues are given by $j(j+1/2)$ with 
$j$ referred to as superspin that takes non-negative integers and half-integers, $j=0,1,2,1,3/2,\cdots$. 
The $UOSp(1|2)$ irreducible representation specified by the superspin index $j$ consists of $SU(2)$ $j$ and $j-1/2$ spin representations and hence the dimension of the $UOSp(1|2)$  representation with superspin $j=n/2$ is   
\begin{equation}
d(n)+d(n-1)=2n+1, 
\end{equation}
where $d(n)$ is the dimension of the $SU(2)$ spin $n/2$  (\ref{su2repre}). 
For $UOSp(1|M)$, there exists a  ``square root'' of the Casimir, the Scasimir \cite{q-alg/9605021,BookFrappat}. In the present, Scasimir is given by 
\begin{equation}
\mathcal{S}=\frac{1}{4}(1-8\epsilon_{\alpha\beta}L_{\alpha}L_{\beta}),
\end{equation}
which satisfies  
\begin{equation}
\mathcal{S}^2=\mathcal{C}+\frac{1}{16}. 
\label{relationScasimirandcasimir}
\end{equation}
Then, the eigenvalues of Scasimir are $\pm j(j+{1}/{4})$.  
Interestingly, the Scasimir is commutative with the bosonic generators and {\it{anticommutative}} with the fermionic ones,  
\begin{equation}
[L_i,\mathcal{S}]=\{L_{\alpha},\mathcal{S}\}=0. 
\end{equation}

\subsubsection{${N}=1$ graded 1st Hopf map}\label{subsec:n=11stHopffuzzytwo}
 
The graded 1st Hopf map is given by   
\begin{equation}
S^{3|2}\overset{S^1}\longrightarrow S^{2|2},  
\label{absgraded1sthopf}
\end{equation}
where left index to the slash indicates the number of bosonic coordinates, while the right index fermionic coordinates.  The bosonic part of (\ref{absgraded1sthopf}) is exactly equivalent to the 1st Hopf map. 
The coordinates on the total manifold $S^{3|2}$ is represented by a normalized three-component superspinor  $\psi=(
\psi_1,
\psi_2, 
\eta)^t$ whose first two components are Grassmann even and the third component is Grassmann odd.  A normalization condition is imposed as   
\begin{equation}
\psi^{\ddagger}\psi=\psi_1^*\psi_1+\psi_2^*\psi_2-\eta^*\eta=1,   
\end{equation}
where $\psi^{\ddagger}=(\psi_1^*,\psi_2^*,-\eta^*)$ and $*$ represents the pseudo-conjugation\footnote{The pseudo-conjugation is imposed as  $(\eta^*)^*=-\eta$ and $(\eta_1\eta_2)^* =\eta_1^*\eta_2^*$ for Grassmann odd quantities. See Ref.\cite{BookFrappat} for instance. }. 
The graded 1st Hopf map is realized as \cite{LandiMarmo1987,Bartocci1990} 
\begin{equation}
\psi ~~\rightarrow~~ x_i =2\psi^{\ddagger}L_i\psi,~~\theta_{\alpha}=2\psi^{\ddagger}L_{\alpha}\psi, 
\label{1stgradedHopf}
\end{equation}
where $L_i$ and $L_{\alpha}$ are the fundamental representation matrices of $uosp(1|2)$  
\begin{align}
&L_i=\frac{1}{2}\begin{pmatrix}
\sigma_i & 0 \\
0 & 0 
\end{pmatrix},~~L_{\alpha}=\frac{1}{2}\begin{pmatrix}
0_2 & \tau_{\alpha} \\
-(\epsilon\tau_{\alpha})^t & 0 
\end{pmatrix}, 
\label{superPaulimat}
\end{align}
with $\epsilon=i\sigma_2$,  $\tau_1=(1,0)^t$ and $\tau_2=(0,1)^t$. 
One may regard (\ref{superPaulimat}) as a supersymmetric extension of the Pauli matrices. 
They are ``hermitian'' in the sense 
\begin{equation}
L_{i}^{\ddagger}=L_i,~~~L_{\alpha}^{\ddagger}=\epsilon_{\alpha\beta}L_{\beta}, \label{adjointofosp12generators}
\end{equation}
where $\ddagger$ is the super-adjoint defined by 
\begin{align}
\begin{pmatrix}
A & B \\
C & D
\end{pmatrix}^{\ddagger}= \begin{pmatrix}
A^{\dagger} & C^{\dagger} \\
-B^{\dagger} & D^{\dagger}
\end{pmatrix}. 
\end{align}
From (\ref{1stgradedHopf}), we see that $x_i$ and $\theta_{\alpha}$ are coordinates on $S^{2|2}$:  
\begin{equation}
x_ix_i +\epsilon_{\alpha\beta}\theta_{\alpha}\theta_{\beta}=(\psi^{\ddagger}\psi)^2=1, 
\label{normalization2supersphere}
\end{equation}
and from (\ref{adjointofosp12generators}), 
\begin{equation}
x_i^*=x_i,~~~\theta_{\alpha}^*=\epsilon_{\alpha\beta}\theta_{\beta}. 
\end{equation}
Notice that $x_i$ are Grassmann even but not usual c-number, since the square of $x_i$ is not c-number as observed in (\ref{normalization2supersphere}). 
Instead, we can introduce c-number $y_i$ as 
\begin{equation}
y_i=\frac{1}{\sqrt{1-\epsilon_{\alpha\beta}\theta_{\alpha} \theta_{\beta}}} ~ x_i, 
\end{equation}
which satisfy $y_iy_i=1$ and denote coordinates on $S^2$, the body of $S^{2|2}$.   
The original normalized $SU(2)$ spinor is ``embedded'' in $\psi$ as 
\begin{equation}
\begin{pmatrix}
\phi_1 \\
\phi_2 
\end{pmatrix}
=
\frac{2}{2+\eta^*\eta}
\begin{pmatrix}
\psi_1 \\
\psi_2
\end{pmatrix}. 
\end{equation}
With  $y_i$, $\phi$ can be written as 
\begin{equation}
\begin{pmatrix}
\phi_1 \\
\phi_2
\end{pmatrix}=
\frac{1}{\sqrt{2(1+y_3)}}
\begin{pmatrix}
1+y_3 \\
y_1+iy_2
\end{pmatrix}e^{i\chi}, 
\label{explitphi2com}
\end{equation}
where $e^{i\chi}$ denotes arbitrary $U(1)$ phase. Represent the Grassmann odd component $\eta$ as 
\begin{equation}
\eta=\phi_1\mu+\phi_2\nu,
\end{equation}
where $\mu$ and $\nu$ are real and imaginary components of $\eta$, which satisfy 
\begin{equation}
\mu^*=\nu,~~~~~~~~~~\nu^*=-\mu.
\end{equation}
Therefore, 
\begin{equation}
\eta^*\eta=-\mu\nu.
\end{equation}
The map (\ref{1stgradedHopf}) immediately determines the relations between $\theta_1$, $\theta_2$ and $\mu$, $\nu$: 
\begin{equation}
\mu=\theta_1,~~~~~~~\nu=\theta_2.
\end{equation}
Consequently, $\psi$ can be expressed as 
\begin{align}
\psi&=\frac{1}{\sqrt{1-\eta^*\eta}}\begin{pmatrix}
\phi_1\\
\phi_2\\
\eta
\end{pmatrix}=\frac{1}{\sqrt{1+\theta_1\theta_2}}\begin{pmatrix}
\phi_1\\
\phi_2\\
\phi_1\theta_1+\phi_2\theta_2
\end{pmatrix}\nonumber\\
&=\frac{1}{\sqrt{2(1+y_3)(1+\theta_1\theta_2)}}\begin{pmatrix}
1+y_3\\
y_1+iy_2\\
(1+y_3)\theta_1+(y_1+iy_2)\theta_2
\end{pmatrix}e^{i\chi}.
\end{align}
The last expression on the right-hand side manifests the $N=1$ graded Hopf fibration, $S^{3|2}\sim S^{2|2}\otimes S^1$: the $S^1(\simeq U(1))$-fibre, $e^{i\chi}$, is canceled in the graded 
Hopf map (\ref{1stgradedHopf}), and the remaining quantities, $y_i$ and $\theta_{\alpha}$, correspond to the coordinates on $S^{2|2}$.

\subsubsection{${N}=1$ fuzzy two-supersphere}\label{subsec:n=11stfuzzytwo}
 
Coordinates on fuzzy supersphere are constructed by the graded version of the Schwinger construction\footnote{In (\ref{SchRepF2sphere}) we adopted the ordinary definition of the Hermitian conjugate $\dagger$, so $\Theta_{\alpha}^{\dagger}\neq \epsilon_{\alpha\beta}\Theta_{\beta}$ unlike $\theta_{\alpha}^*=\epsilon_{\alpha\beta}\theta_{\beta}$.} \cite{hep-th/9510083}: 
\begin{equation}
X_i= 2\Psi^{\dagger}L_i\Psi,~~~~\Theta_{\alpha}=2\Psi^{\dagger}L_{\alpha}\Psi,  
\label{SchRepF2sphere}
\end{equation}
where $\Psi$ stands for a graded Schwinger operator 
\begin{equation}
\Psi=
(\Psi_1, 
\Psi_2, 
\tilde{\Psi})^t, 
\end{equation}
with  bosonic operators $\Psi_1$ and $\Psi_2$ and fermionic one $\tilde{\Psi}$ satisfying    
\begin{align}
&[\Psi_{\alpha},\Psi_{\beta}^{\dagger}]=\delta_{\alpha\beta},~~\{\tilde{\Psi},\tilde{\Psi}^{\dagger}\}=1,~~[\Psi_{\alpha},\tilde{\Psi}^{\dagger}]=0,\nonumber\\
&[\Psi_{\alpha},\Psi_{\beta}]=\{\tilde{\Psi},\tilde{\Psi}\}=[\Psi_{\alpha},\tilde{\Psi}]=0. 
\end{align}
It is straightforward to see that 
(\ref{SchRepF2sphere}) satisfy the algebra 
\begin{align}
[X_i,X_j]=2i\epsilon_{ijk}X_k,~~~~~[X_i,\Theta_{\alpha}]=(\sigma_i)_{\beta\alpha}\Theta_{\beta},~~~~~\{\Theta_{\alpha},\Theta_{\beta}\}=(\epsilon\sigma_i)_{\alpha\beta}X_i. 
\end{align}
Square of the radius of fuzzy supersphere is given by the $UOSp(1|2)$ Casimir  
\begin{equation}
X_i X_i +\epsilon_{\alpha\beta}\Theta_{\alpha}\Theta_{\beta}=(\Psi^{\dagger}\Psi)(\Psi^{\dagger}\Psi+1), 
\label{casimirofF2sphere}
\end{equation}
where we used 
\begin{align}
&X_iX_i=\hat{n}_B(\hat{n}_B+2),\nonumber\\
&\epsilon_{\alpha\beta}\Theta_{\alpha}\Theta_{\beta}=-\hat{n}_B+2\hat{n}_B\hat{n}_F+2\hat{n}_F, 
\end{align}
with $\hat{n}_B=\Psi_1^{\dagger}\Psi_1+\Psi_2^{\dagger}\Psi_2$, $\hat{n}_F=\tilde{\Psi}^{\dagger}\tilde{\Psi}$, and $\hat{n}_F^2=\hat{n}_F$. 
$\Psi^{\dagger}\Psi$ denotes the total number-operator   
$\hat{n}=\Psi^{\dagger}\Psi=\hat{n}_B+\hat{n}_F.$ 
Notice the zero-point energy in (\ref{casimirofF2sphere}) reflects the difference between the bosonic and fermionic degrees of freedom of the Schwinger operator.  
The Scasimir is expressed as 
\begin{equation}
\mathcal{S}=(\frac{1}{2}-\hat{n}_F)(\hat{n}+\frac{1}{2}). 
\label{ScasimirSchwinger}
\end{equation}
From (\ref{casimirofF2sphere}) and (\ref{ScasimirSchwinger}), one may readily show (\ref{relationScasimirandcasimir}). 

Graded fully symmetric representation specified by the superspin $j=n/2$ is given by 
\begin{subequations}
\begin{align}
&|l_1,l_2\rangle= \frac{1}{\sqrt{l_1!~l_2!}}
 {\Psi_1^{\dagger}}^{l_1}{\Psi_2^{\dagger}}^{l_2}|0\rangle,\label{bosonicosp12}\\
&|m_1,m_2) = \frac{1}{\sqrt{m_1!~m_2!}} 
{\Psi_1^{\dagger}}^{m_1} {\Psi_2^{\dagger}}^{m_2}\tilde{\Psi}^{\dagger}|0\rangle, \label{fermionicosp12}  
\end{align}\label{UOSp12irredu}
\end{subequations}
where $l_1+l_2=m_1+m_2+1=n$ with non-negative integers,  $l_1, l_2 , m_1$ and $m_2$. 
$|m_1,m_2)$ are the fermionic counterpart of $|l_1,l_2\rangle$, and thus they exhibit ${N}=1$ SUSY. 
The bosonic and fermionic states\footnote{ In this paper, the bosonic and fermionic states   refer to states with even and odd number of fermion operators, respectively. 
They are eigenstates of the fermion parity $(-1)^{\hat{n}_F}$ with the eigenvalues $+1$ and $-1$.}  are classified by the sign of Scasimir (\ref{ScasimirSchwinger}).  Scasimir takes the values  
\begin{equation} 
\mathcal{S}=\pm \frac{1}{4}(2{n}+1), 
\end{equation}
with $+$ and $-$ for the bosonic (\ref{bosonicosp12}) and fermionic (\ref{fermionicosp12}) states, respectively.  
The degrees of freedom of bosonic and fermionic states are respectively 
\begin{equation}
d_B=d(n)=n+1, ~~~d_F=d(n-1)=n,
\label{usp21bosferdof}
\end{equation}
and then the total degrees of freedom is 
\begin{equation}
d_T=d_B+d_F=2n+1. 
\label{usp21totdof}
\end{equation}
 $X_3$-coordinates of these states are   
\begin{equation}
X_3=n-{k}, 
\label{X3fuzzysupersphere}
\end{equation}
where $k=0,1,2,\cdots,2n$.  For even $k$, the eigenvalues of $X_3$ correspond to the bosonic states (\ref{bosonicosp12}), while for odd $k$, the fermionic states (\ref{fermionicosp12}). Compare the $X_3$ eigenvalues of fuzzy supersphere (\ref{X3fuzzysupersphere}) and those of the fuzzy (bosonic) sphere (\ref{X3fuzzysphere}): 
the degrees of freedom of fuzzy supersphere for even $k$ are accounted for by those of fuzzy sphere with radius $n$, while those for odd $k$ are by fuzzy sphere with radius $n-1$.  Thus, the bosonic and fermionic degrees of freedom are same as of the fuzzy spheres with radius $n$ and radius $n-1$, respectively.  
Consequently, the fuzzy two-supersphere of radius ${n}$ is intuitively understood as a ``superposition'' of  two fuzzy  spheres whose radii are $n$ and $n-1$. Schematically, 
\begin{equation}
S_F^{2|2}(n)\simeq S_F^{2}(n)\oplus S_F^{2}(n-1). 
\end{equation}

It is noted that though we only utilized the $UOSp(1|2)$ algebra, fuzzy two-supersphere itself is invariant under the larger $SU(2|1)$ symmetry: indeed, the right-hand side of (\ref{casimirofF2sphere}) is invariant under the $SU(2|1)$ rotation of the Schwinger operator $\Psi$. 
In this sense, the symmetry of fuzzy two-supersphere is $SU(2|1)$ rather than $UOSp(1|2)$.   
Also notice that the graded fully symmetric representation (\ref{UOSp12irredu}) is regarded as a (atypical)  representation of $SU(2|1)$.

\subsection{${N}=2$ fuzzy two-supersphere}

 We utilized the $UOSp(1|2)$ algebra to construct ${N}=1$ fuzzy supersphere $S_F^{2|2}$. Here, we apply  $UOSp(2|2)$ algebra to construct ${N}=2$ fuzzy supersphere $S_F^{2|4}$.  

\subsubsection{$UOSp(2|2)$ algebra}

$UOSp(2|2)$ algebra contains $usp(2)\simeq su(2)$ and $o(2)\simeq u(1)$ as its bosonic algebras, and the fermionic generators transform as a $SU(2)$ spinor and carry  $U(1)$ charge as well. Thus, $uosp(2|2)$ is isomorphic to $su(2|1)$, and its dimension is  
\begin{equation}
\dim[uosp(2|2)]=\dim[su(2|1)]=4|4=8. 
\end{equation}
We denote the four bosonic generators as $L_i$ $(i=1,2,3)$ and $\Gamma$, and the four fermionic generators as $L_{\alpha}$ and $L_{\alpha}'$ $(\alpha=\theta_1,\theta_2)$.   
The $UOSp(2|2)$ algebra is given by 
\begin{align}
&[L_i,L_j]=i\epsilon_{ijk}L_k,~~~~
[L_i,L_{\alpha\sigma}]=\frac{1}{2}(\sigma_i)_{\beta\alpha}L_{\beta\sigma},~~~~\{L_{\alpha\sigma},L_{\beta\tau}\}=\frac{1}{2}\delta_{\sigma\tau}(\epsilon\sigma_i)_{\alpha\beta}L_i+\frac{1}{2}\epsilon_{\sigma\tau}\epsilon_{\alpha\beta}\Gamma,\nonumber\\
&[\Gamma,L_i]=0,~~~~~~~~~~~~~[\Gamma,L_{\alpha\sigma}]=\frac{1}{2}\epsilon_{\tau\sigma}L_{\alpha\tau}, 
\label{osp22algebranew}
\end{align}
where $L_{\alpha\sigma}=(L_{\alpha},L'_{\alpha})$ 
\footnote{The algebra (\ref{osp22algebranew}) coincides with the $UOSp(2|2)$ algebra usually found in literature by the following redefinitions,   
\begin{equation}
L_i \rightarrow L_i,~~L_{\alpha}\rightarrow L_{\alpha},~~L'_{\alpha}\rightarrow iD_{\alpha},~~\Gamma\rightarrow -i\Gamma. 
\end{equation}
}. 
 $L_i$ and $L_{\alpha}$ form the $UOSp(1|2)$ subalgebra.  
There are two sets of fermionic generators, $L_{\alpha}$ and  $L_{\alpha}'$, which bring ${N}=2$ SUSY. 
The fundamental representation is 3 dimensional representation, as expected from $uosp(2|2)\simeq su(2|1)$.   
The $UOSp(2|2)$ algebra has two Casimirs, quadratic and cubic \cite{Scheunert1977}. 
The quadratic Casimir  is given by 
\begin{equation}
\mathcal{C}=L_iL_i+\epsilon_{\alpha\beta}L_{\alpha}L_{\beta}+\epsilon_{\alpha\beta}L'_{\alpha}L'_{\beta}+\Gamma^2. 
\label{osp22casimir}
\end{equation}
The irreducible representation is classified into two categories;  typical representation and atypical representation (see Appendix \ref{appendatyposp(2|2)} for details). 
Since the Casimir eigenvalues of (\ref{osp22casimir}) are identically zero for atypical representation,  we utilize  typical representation to construct ${N}=2$ fuzzy two-superspheres.  
The minimal dimension matrices of typical representation are the following $4\times 4$ matrices: 
\begin{align}
L_i=\frac{1}{2}
\begin{pmatrix}
\sigma_i & 0 _2 \\
0_2 & 0_2
\end{pmatrix},~~~L_{\alpha}=\frac{1}{2}
\begin{pmatrix}
0_2 & \tau_{\alpha} & 0 \\
-(\epsilon\tau_{\alpha})^t & 0 & 0  \\
0 & 0 & 0 
\end{pmatrix},~~~L'_{\alpha}=\frac{1}{2}
\begin{pmatrix}
0_2& 0  & \tau_{\alpha}  \\
0 & 0 & 0  \\
-(\epsilon\tau_{\alpha})^t & 0 & 0 
\end{pmatrix},~~~\Gamma=\frac{1}{2}\begin{pmatrix}
0_2 & 0_2 \\
0_2 & \epsilon
\end{pmatrix}.
\label{osp22typicalmatrices}
\end{align}
(These are equivalent to those given in Ref.\cite{Scheunert1977}.)

\subsubsection{${N}=2$ fuzzy two-supersphere}\label{subsec:n=21fuzzysph}

Applying the Schwinger construction to (\ref{osp22typicalmatrices}), we introduce ${N}=2$ fuzzy supersphere coordinates as  
\begin{equation}
X_i=2\Psi^{\dagger}L_i\Psi,~~~~\Theta_{\alpha}=2\Psi^{\dagger}L_{\alpha}\Psi,~~~~
\Theta'_{\alpha}=2\Psi^{\dagger}L'_{\alpha}\Psi,~~~~G=2\Psi^{\dagger}\Gamma\Psi, 
\label{Schwingerconstrfuzzytwosphere}
\end{equation}
where $\Psi$ denotes the four-component Schwinger operator  
\begin{equation}
\Psi=
(\Psi_1,
\Psi_2, 
\tilde{\Psi}_1, 
\tilde{\Psi}_2)^t.   
\end{equation}
 $\Psi_{\alpha}$ 
$(\alpha=1,2)$ are bosonic operators while $\tilde{\Psi}_{\sigma}$ $(\sigma=1,2)$ are fermionic ones   
satisfying  
\begin{align}
&[\Psi_{\alpha},\Psi_{\beta}^{\dagger}]=\delta_{\alpha\beta},~~\{\tilde{\Psi}_{\sigma},\tilde{\Psi}_{\tau}^{\dagger}\}=\delta_{\sigma\tau},\nonumber\\
&[\Psi_{\alpha},\Psi_{\beta}]=\{\tilde{\Psi}_{\sigma},\tilde{\Psi}_{\tau}\}=[\Psi_{\alpha},\tilde{\Psi}_{\sigma}]=0. 
\end{align}
Square of the radius of ${N}=2$ fuzzy two-supersphere is evaluated as  
\begin{equation}
X_iX_i+\epsilon_{\alpha\beta}\Theta_{\alpha}\Theta_{\beta}+\epsilon_{\alpha\beta}\Theta'_{\alpha}\Theta'_{\beta}+G^2=(\Psi^{\dagger}\Psi)^2. 
\label{radiusoftyposp22ori}
\end{equation}
 Here, we used 
\begin{align}
&X_iX_i=\hat{n}_B(\hat{n}_B+2),\nonumber\\
&\epsilon_{\alpha\beta}\Theta_{\alpha}\Theta_{\beta}+\epsilon_{\alpha\beta}\Theta'_{\alpha}\Theta'_{\beta} = -\hat{n}_B+2 \hat{n}_{B}\hat{n}_{F}+2\hat{n}_{F},\nonumber\\
&G^2=4\hat{n}_F(\hat{n}_F-2),  
\end{align}
where $\hat{n}_B=\sum_{\alpha=1}^2\Psi_{\alpha}^{\dagger}\Psi_{\alpha}$, $\hat{n}_{F}=\sum_{\sigma=1}^2\tilde{\Psi}_{\sigma}^{\dagger}\tilde{\Psi}_{\sigma}$. 
For $\Psi^{\dagger}\Psi=n$, the graded fully symmetric  representation is derived as  
\begin{subequations}
\begin{align}
&|l_1,l_2\rangle= \frac{1}{\sqrt{l_1!~l_2!}}
{\Psi_1^{\dagger}}^{l_1}{\Psi_2^{\dagger}}^{l_2}|0\rangle,\label{grosp22b1}\\
&|m_1,m_2)= \frac{1}{\sqrt{m_1!~m_2!}} 
 {\Psi_1^{\dagger}}^{m_1}{\Psi_2^{\dagger}}^{m_2}\tilde{\Psi}_1^{\dagger}|0\rangle,\label{grosp22f1}\\
&|m'_1,m'_2)=\frac{1}{\sqrt{m'_1!~m'_2!}}
{\Psi_1^{\dagger}}^{m'_1}{\Psi_2^{\dagger}}^{m'_2}\tilde{\Psi}_2^{\dagger}|0\rangle,\label{grosp22f2}\\
&|n_1,n_2\rangle= \frac{1}{\sqrt{n_1!~n_2!}}
{\Psi_1^{\dagger}}^{n_1}{\Psi_2^{\dagger}}^{n_2}\tilde{\Psi}_1^{\dagger}\tilde{\Psi}_2^{\dagger}|0\rangle, \label{grosp22b2}
\end{align}\label{fullysymmosp22rep}
\end{subequations}
where $l_1+l_2=m_1+m_2+1=m_1'+m_2'+1=n_1+n_2+2=n$ with non-negative integers, $l_1$, $l_2$, $m_1$, $m_2$, $m'_1$, $m'_2$, $n_1$, $n_2$. We have two sets of bosonic states, $|l_1,l_2\rangle$ and $|n_1,n_2\rangle$, and two sets of fermionic states, $|m_1,m_2)$ and $|m_1',m_2')$ as well.  The degrees of freedom of bosonic and fermionic states are equally given by  
\begin{align}
&d_B=d(n)+d(n-2)=2n,\nonumber\\
&d_F=2\times d(n-1)=2n, 
\end{align}
with $d(n)=n+1$, and the total is 
\begin{equation}
d_T=d_B+d_F=4n.
\end{equation}
Square of the radius of ${N}=2$ fuzzy two-supersphere (\ref{radiusoftyposp22ori}) does not have the zero-point energy since the bosonic and fermionic degrees of freedom are equal. 
The first two sets, (\ref{grosp22b1}) and (\ref{grosp22f1}), are $UOSp(1|2)$ $j=n/2$ irreducible representation, and the other two, (\ref{grosp22f2}) and (\ref{grosp22b2}), are $UOSp(1|2)$ $j={n}/2-1/2$ irreducible representation. In this sense, the ${N}=2$ fuzzy two-supersphere with radius $n$ is regarded as a ``superposition'' of two ${N}=1$ fuzzy superspheres whose radii are $n$ and $n-1$.  
Remember that $N=1$ fuzzy two-supersphere can also be regarded as a superposition of two bosonic fuzzy spheres. Consequently, $N=2$ fuzzy sphere is realized as a superposition of four fuzzy spheres whose radii are 
$n$, $n-1$, $n-1$ and $n-2$. Schematically, 
\begin{align}
S_F^{2|4}(n)&\simeq S_F^{2|2}(n)\oplus S_F^{2|2}(n-1)\nonumber\\
&\simeq 
S_F^{2}(n)\oplus S_F^2(n-1)\oplus S_F^2(n-1)\oplus S_F^2(n-2). 
\end{align}
Notice that such particular feature is a consequence of the adoption of graded fully symmetric representation. 
The corresponding latitudes of the states (\ref{fullysymmosp22rep}) are given by 
\begin{equation}
X_3=n-k
\label{x3eigenfuzzyfour}
\end{equation}
with $k=0,1,2,\cdots,2n$. The even $k$ correspond to the bosonic states, (\ref{grosp22b1}) and (\ref{grosp22b2}), while odd $k$ the fermionic states, (\ref{grosp22f1}) and (\ref{grosp22f2}).  
Except for non-degenerate states at the north and south poles $X_3=\pm n$,  the eigenvalues of $X_3$ (\ref{x3eigenfuzzyfour}) are doubly-degenerate. 

Since the right-hand side of (\ref{radiusoftyposp22ori}) is invariant under the $SU(2|2)$ rotation of $\Psi$, the symmetry of  ${N}=2$ fuzzy two-supersphere is considered as $SU(2|2)$ rather than $UOSp(2|2)$.  

\subsubsection{${N}=2$ graded 1st Hopf map}\label{subsec:n=21sthopf}

Based on the Schwinger construction of ${N}=2$ fuzzy two-supersphere, we introduce ${N}=2$ version of the graded 1st Hopf map.  
With (\ref{osp22typicalmatrices}), we define
\begin{align}
x_i =2\psi^{\ddagger}L_i\psi,~~~~\theta_{\alpha}=2\psi^{\ddagger}L_{\alpha}\psi,~~~~\theta'_{\alpha}=2\psi^{\ddagger}L'_{\alpha}\psi,~~~~{g}=\psi^{\ddagger}\Gamma\psi. 
\label{1stgradedmapn2ori}
\end{align}
Here, $\psi$ denotes a four-component spinor $\psi=(\psi_1,\psi_2,\eta_1,\eta_2)^t$ normalized as  
\begin{equation}
\psi^{\ddagger}\psi=\psi_1^*\psi_1+\psi_2^*\psi_2-\eta^*_1\eta_1-\eta_2^*\eta_2=1,
\end{equation}
and then is regarded as coordinates on $S^{3|4}$.  The coordinates (\ref{1stgradedmapn2ori}) satisfy the relation 
\begin{equation}
x_ix_i+\epsilon_{\alpha\beta}\theta_{\alpha}\theta_{\beta}+\epsilon_{\alpha\beta}\theta'_{\alpha}\theta'_{\beta}+g^2=(\psi^{\ddagger}\psi)^2=1. 
\label{classicaldeffuzzyn2twosphere}
\end{equation}
Notice all of the 
quantities (\ref{1stgradedmapn2ori}) are not independent\footnote{ This situation is similar to  Schwinger construction of fuzzy complex projective space. The coordinates on fuzzy $CP^{N-1}$ are represented by the $SU(N)$ generators sandwiched by Schwinger operators. Though the real dimension of $CP^{N-1}$ is $2N-2$, the dimension of $SU(N)$ generator is  $N^2-1$. This ``discrepancy'' is resolved by noticing   
all of the $SU(N)$ generators in the Schwinger construction are not independent and satisfy a set of constraints.  
See \cite{arXiv:hep-th/0103023} for more details.}. 
This can typically be seen from  $\theta_1\theta_2\theta_1'\theta'_2 g=0$. (If $\theta_{\alpha}$, $\theta'_{\alpha}$ and $g$ were independent, their product would not be zero.)  Rewrite  $\psi$ as 
\begin{align}
\psi=
\begin{pmatrix}
\psi_1 \\
\psi_2\\
\eta_1 \\
\eta_2
\end{pmatrix}
={\sqrt{1+\eta_1^*\eta_1+\eta_2^*\eta_2}}
\begin{pmatrix}
\phi_1 \\
\phi_2\\
\sqrt{1-\eta_2^*\eta_2}~\eta_1 \\
\sqrt{1-\eta_1^*\eta_1}~\eta_2
\end{pmatrix},
\end{align}
where $(\phi_1,\phi_2)^t$ denotes the normalized $SU(2)$ spinor (\ref{explitphi2com}). Also, we express $\eta_1$ and $\eta_2$ as 
\begin{align}
&\eta_1=\phi_1\mu_1+\phi_2\nu_1,\nonumber\\
&\eta_2=\phi_1\mu_2+\phi_2\nu_2,
\end{align}
where $\mu_1$ and $\mu_2$ represent the real parts of the Grassmann odd quantities, and $\nu_1$ and $\nu_2$ represent the imaginary parts. 
The map (\ref{1stgradedmapn2ori}) determines the relations between $\mu_{1,2}$, $\nu_{1,2}$ and $\theta_{1,2}$, $\theta'_{1,2}$ as   
\begin{align}
&\theta_1={\sqrt{1+\eta_1^*\eta_1+\eta_2^*\eta_2}}~\mu_1,~~~~~~\theta_2={\sqrt{1+\eta_1^*\eta_1+\eta_2^*\eta_2}}~\nu_1,\nonumber\\
&\theta'_1={\sqrt{1+\eta_1^*\eta_1+\eta_2^*\eta_2}}~\mu_2,~~~~~~\theta'_2={\sqrt{1+\eta_1^*\eta_1+\eta_2^*\eta_2}}~\nu_2. 
\label{relthetaetan=1}
\end{align}
Then, 
\begin{equation}
\theta_1\theta_2+\theta_1'\theta'_2=-(1+\eta_1^*\eta_1+\eta_2^*\eta_2)(\eta_1^*\eta_1+\eta_2^*\eta_2)=-\frac{\eta_1^*\eta_1+\eta_2^*\eta_2}{1-\eta_1^*\eta_1-\eta_2^*\eta_2},
\end{equation}
or inversely, 
\begin{equation}
\eta_1^*\eta_1+\eta_2^*\eta_2=-\frac{\theta_1\theta_2+\theta_1'\theta_2'}{1-\theta_1\theta_2-\theta_1'\theta_2'}=-\theta_1\theta_2-\theta_1'\theta_2'-2\theta_1\theta_2\theta_1'\theta_2'. 
\label{thetasqareetan=1}
\end{equation}
Therefore, from (\ref{relthetaetan=1}) and (\ref{thetasqareetan=1}),  $\mu_1$, $\nu_1$, $\mu_2$ and $\nu_2$ are represented as 
\begin{align}
&\mu_1=\frac{1}{\sqrt{1-\theta'_1\theta'_2}}\theta_1,~~~~~~\nu_1=\frac{1}{\sqrt{1-\theta'_1\theta'_2}}\theta_2,\nonumber\\
&\mu_2=\frac{1}{\sqrt{1-\theta_1\theta_2}}\theta'_1,~~~~~~\nu_2=\frac{1}{\sqrt{1-\theta_1\theta_2}}\theta'_2. 
\end{align}
Consequently, $\psi$ is given by 
\begin{align}
\psi&
   =  {\frac{1}{\sqrt{2(1+y_3)(1+\theta_1\theta_2+\theta_1'\theta_2'+4\theta_1\theta_2\theta_1'\theta_2')}}}
\begin{pmatrix}
1+y_3 \\
y_1+iy_2 \\
(1+\theta_1'\theta_2')(\theta_1(1+y_3)+\theta_2(y_1+iy_2))\\
(1+\theta_1\theta_2)(\theta'_1(1+y_3)+\theta'_2(y_1+iy_2))
\end{pmatrix}e^{i\chi}, 
\label{explicitn=2hopfspinor}
\end{align}
where $e^{i\chi}$ denotes arbitrary $U(1)$ phase factor. 
$x_i$ and $y_i$ are related as 
\begin{equation}
y_i=\frac{1}{1+\eta_1^*\eta_1+\eta_2^*\eta_2}x_i=\frac{1}{{1-\theta_{1}\theta_{2}-\theta'_{1}\theta'_{2}-2\theta_{1}\theta_{2}\theta_{1}'\theta_{2}' }}x_i.
\end{equation}
Thus, $\psi$ can be expressed by $x_i$, $\theta_{\alpha}$, $\theta'_{\alpha}$, the coordinates on $S^{2|4}$, and arbitrary $U(1)$ phase factor.  Obviously, the $U(1)(\simeq S^1)$ phase is canceled in (\ref{1stgradedmapn2ori}).  
Then, the bilinear map (\ref{1stgradedmapn2ori}) represents  
\begin{equation}
S^{3|4}~~\overset{S^1}\longrightarrow~~ S^{2|4},  
\label{n=21sthopfabst}
\end{equation}
which we call the ${N}=2$ graded 1st Hopf map. 
We have four bosonic and four fermionic coordinates in (\ref{1stgradedmapn2ori}), but 
 $g=-\eta_1^*\eta_2+\eta_2^*\eta_1$ is a redundant coordinate. 
Indeed, with (\ref{explicitn=2hopfspinor}), $g$ is expressed by $y_i$, $\theta_{\alpha}$ and $\theta'_{\alpha}$ as 
\begin{equation}
g=y_1(\theta_1\theta_1'-\theta_2\theta_2')-iy_2(\theta_1\theta_1'+\theta_2\theta_2')-y_3(\theta_1\theta_2'+\theta_2\theta_1').
\end{equation}
It can also be shown that the following ``renormalization'', 
\begin{align}
&x_i \rightarrow {\sqrt{1-g^2}}~x_i =(1-\frac{1}{2}g^2) x_i,~~~~\theta_{\alpha}\rightarrow {\sqrt{1-g^2}}~\theta_{\alpha}=\theta_{\alpha},\nonumber\\
&\theta'_{\alpha}\rightarrow {\sqrt{1-g^2}}~\theta'_{\alpha}=\theta_{\alpha}', 
\end{align}
 eliminates $g$: 
the renormalized coordinates satisfy the ordinary condition of $S^{2|4}$,  
\begin{equation}
x_ix_i+\epsilon_{\alpha\beta}\theta_{\alpha}\theta_{\beta}+\epsilon_{\alpha\beta}\theta'_{\alpha}\theta'_{\beta}=1. 
\label{classicaln2twosphere}
\end{equation}

One might attempt to introduce more supersymmetry. 
In principle, it is probable to do so by utilizing $UOSp(N|2)$ algebras for $N\ge 3$.  
 However, the radius of the ${N}=2$ fuzzy two-supersphere (\ref{radiusoftyposp22ori}) already saturates the ``classical bound'' (\ref{classicaldeffuzzyn2twosphere}). In general, square of the radius of fuzzy supersphere with $N$-SUSY is proportional to $n(n+2-N)$ and becomes negative for ``sufficiently small'' $n$ that satisfies $n < N-2$. Hence we stop at $N=2$. 

\section{Graded 2nd Hopf maps and fuzzy four-superspheres}\label{SectGraded2nd}

In this section, we extend the previous formulation to fuzzy four-supersphere. 

\subsection{The 2nd Hopf map and fuzzy four-sphere}\label{subsec2ndHopfandfuzzy}

The 2nd Hopf map 
\begin{equation}
S^7~\overset{S^3}\longrightarrow~ S^4  
\label{2ndHopfmapabstract}
\end{equation}
is represented as 
\begin{equation}
\phi ~~\rightarrow ~~x_a =\phi^{\dagger}\gamma_a\phi,
\label{2ndHopfmapexplicit}
\end{equation}
where $\phi=(\phi_1,\phi_2,\phi_3,\phi_4)^t$ is a normalized four-component complex spinor $\phi^{\dagger}\phi=1$, representing  coordinates on $S^7$. $\gamma_a$ $(a=1,2,3,4,5)$ are $SO(5)$ gamma matrices that satisfy 
$\{\gamma_a,\gamma_b\}=2\delta_{ab}$ with Kronecker delta $\delta_{ab}$.  $\gamma_{a}$ can be taken as     
\begin{align}
&\gamma_1=
\begin{pmatrix}
&0 & i\sigma_1\\
&-i\sigma_1 & 0
\end{pmatrix}
,~~\gamma_2=
\begin{pmatrix}
&0 & i\sigma_2\\
&-i\sigma_2 & 0
\end{pmatrix}
,~~\gamma_3=
\begin{pmatrix}
&0 & i\sigma_3\\
&-i\sigma_3 & 0
\end{pmatrix},\nonumber\\
&\gamma_4=
\begin{pmatrix}
&0 & 1_2\\
&1_2 & 0
\end{pmatrix}
,~~~~~~~\gamma_5=
\begin{pmatrix}
&1_2 & 0\\
&0 & -1_2
\end{pmatrix}, 
\label{so5gammaI1}
\end{align}
where $1_2$ denotes $2\times 2$ unit matrix. 
 From  (\ref{2ndHopfmapexplicit}), we have    
\begin{equation}
x_ax_a=(\phi^{\dagger}\phi)^2=1. 
\label{conditionunitfoursphere}
\end{equation}
Thus, $x_a$ (\ref{2ndHopfmapexplicit}) are coordinates on four-sphere. 

Coordinates on fuzzy four-sphere $S_F^4$ are constructed as \cite{hep-th/9602115}
\begin{equation}
X_a=\Phi^{\dagger}\gamma_a \Phi, 
\end{equation}
where $\Phi=(\Phi_1,\Phi_2,\Phi_3,\Phi_4)^t$ represents a  four-component Schwinger operator 
satisfying $[\Phi_{\alpha},\Phi^{\dagger}_{\beta}]=\delta_{\alpha\beta}$ and 
$[\Phi_{\alpha},\Phi_{\beta}]=0$ $(\alpha,\beta=1,2,3,4)$.  
Square of the radius of fuzzy four-sphere is derived as  
\begin{equation}
X_aX_a=(\Phi^{\dagger}\Phi)(\Phi^{\dagger}\Phi+4). 
\label{squareradiusfuzzyfour} 
\end{equation}
The zero-point energy corresponds to the number of the four-components of the Schwinger operator.  
Let $n$ be the eigenvalues of the number operator $\hat{n}=\Phi^{\dagger}\Phi$. The corresponding eigenstates are fully symmetric representation:   
\begin{equation}
|l_1,l_2,l_3,l_4\rangle= \frac{1}{\sqrt{l_1!~l_2!~l_3!~l_4!}}
{\Phi_1^{\dagger}}^{l_1}{\Phi_2^{\dagger}}^{l_2}{\Phi_3^{\dagger}}^{l_3}{\Phi_4^{\dagger}}^{l_4}|0\rangle, 
\label{fuzzysymmetricrepso5}
\end{equation}
with $l_1+l_2+l_3+l_4=n$ for non-negative integers $l_1$, $l_2$, $l_3$, $l_4$. The degeneracy is  
\begin{equation}
D(n)=\frac{1}{3!}(n+1)(n+2)(n+3). 
\label{so5degenesym}
\end{equation}
Notice, for the fully symmetric representation, square of the radius (\ref{squareradiusfuzzyfour}) is equal to the $SO(5)$ Casimir:  
\begin{equation}
X_aX_a=(\Phi^{\dagger}\Phi)(\Phi^{\dagger}\Phi+4)=2\sum_{a<b}X_{ab}X_{ab}, 
\label{convsumgammageneso5}
\end{equation}
where $X_{ab}$ are the $SO(5)$ generators given by   
\begin{equation}
[X_a,X_b]=4iX_{ab}, 
\end{equation}
or 
\begin{equation}
X_{ab}=\Phi^{\dagger}\gamma_{ab}\Phi
\end{equation}
with 
\begin{equation}
\gamma_{ab}=-i\frac{1}{4}[\gamma_a,\gamma_b]. 
\end{equation}
Here, $\gamma_{ab}$ are explicitly 
\begin{align}
&\gamma_{12}=
\frac{1}{2}
\begin{pmatrix}
&\sigma_3 & 0\\
& 0 & \sigma_3
\end{pmatrix}
,~~~~~~~~~
\gamma_{13} =
\frac{1}{2}
\begin{pmatrix}
&-\sigma_2  & 0 \\
&0  &-\sigma_2 
\end{pmatrix}
,~~~~\gamma_{14}=
\frac{1}{2}
\begin{pmatrix}
&\sigma_1 & 0\\
& 0 & -\sigma_1
\end{pmatrix}
,\nonumber\\
&\gamma_{15} =
\frac{1}{2}
\begin{pmatrix}
& 0 & -\sigma_1\\
& -\sigma_1 & 0 
\end{pmatrix}
,~~~~~\gamma_{23} =
\frac{1}{2}
\begin{pmatrix}
& \sigma_1 & 0\\
& 0 & \sigma_1
\end{pmatrix}
,~~~~~~~~
\gamma_{24} =
\frac{1}{2}
\begin{pmatrix}
& \sigma_2 & 0 \\
& 0 & -\sigma_2
\end{pmatrix}
,\nonumber\\
&\gamma_{25} =
\frac{1}{2}
\begin{pmatrix}
& 0 & -\sigma_2\\
& -\sigma_2 & 0
\end{pmatrix}
,~~~~~
\gamma_{34} =
\frac{1}{2}
\begin{pmatrix}
&\sigma_3  & 0\\
& 0 & -\sigma_3
\end{pmatrix}
,~~~~~~\gamma_{35} =
\frac{1}{2}
\begin{pmatrix}
& 0 & -\sigma_3\\
&  -\sigma_3&0 
\end{pmatrix}
,\nonumber\\
&\gamma_{45} =
\frac{1}{2}
\begin{pmatrix}
& 0 & i1_2\\
& -i 1_2& 0
\end{pmatrix}
.\label{so5geneI1}
\end{align}
Inversely,  the sum of $SO(5)$ generators can be ``converted'' to that of  gamma matrices as long as the fully symmetric representation is adopted. Such conversion is crucial in constructing fuzzy four-superspheres as we shall see.  
  
In total, the fifteen operators, $X_a$ and $X_{ab}$, satisfy a closed algebra: 
\begin{align}
&[{X}_a,{X}_b]=4i{X}_{ab},~~~ [{X}_a,{X}_{bc}]= -i(\delta_{ab}X_c-\delta_{ac}X_b),\nonumber\\
&[X_{ab},{X}_{cd}]=i
(\delta_{ac}{X}_{bd}-\delta_{ad}{X}_{bc}+\delta_{bc}{X}_{ad}-\delta_{bd}{X}_{ac}).
\label{SO(6)algebradetail}
\end{align}
By identifying ${X}_{a6}=\frac{1}{2}{X}_a$ and ${X}_{ab}={X}_{ab}$, one may find that (\ref{SO(6)algebradetail}) is equivalent to $so(6)\simeq su(4)$ algebra,  
\begin{equation}
[{X}_{AB},{X}_{CD}]
=i(\delta_{AC}{X}_{BD}-\delta_{AD}{X}_{BC}+\delta_{BC}{X}_{AD}-\delta_{BD}{X}_{AC}),
\end{equation}
where $A,B=1,2,\dots,6$. Thus, the underlying algebra of fuzzy four-sphere is considered as $su(4)$. 
  The $SU(4)$ structure of the fuzzy four-sphere can also be deduced from the $SU(4)$ invariance of the right-hand side of (\ref{squareradiusfuzzyfour}).
 The states  $|l_1,l_2,l_3,l_4\rangle$ (\ref{fuzzysymmetricrepso5}) ring the four-sphere at latitudes  
\begin{equation}
X_5=n-2k,
\end{equation}
where $k=0,1,2,\cdots,n$,  and is related to $l_1$, $l_2$, $l_3$, $l_4$ as 
\begin{equation}
k=l_3+l_4=n-l_1-l_2
\end{equation}
or 
\begin{equation}
l_1+l_2=n-k,\quad\quad l_3+l_4=k. 
\label{l1l2l3l4krelations}
\end{equation}
From (\ref{l1l2l3l4krelations}), one may find, unlike the fuzzy two-sphere case, at $X_5=n-2k$, there is   degeneracy 
\begin{equation}
D_k(n)=d(n-k)\cdot d(k)=(n-k+1)(k+1), 
\label{degeeachlatitude}
\end{equation}
where $d(k)$ is the number of the states on fuzzy two-sphere with radius $k$ (\ref{su2repre}). (\ref{so5degenesym}) is reproduced as 
\begin{equation} 
D(n)=\sum_{k=0}^n D_k(n)=\sum_{k=0}^n d(n-k)\cdot d(k).
\label{degeefuzzyfourfuzzytwo}
\end{equation}  
With increase of $k$, 
$D_k(n)$ monotonically increases from the north-pole to the equator $k=n/2$, and monotonically decreases from the equator to the south-pole. 
$D_k(n)$ is symmetric under $k\leftrightarrow {n}-k$, which corresponds to the inversion symmetry of  sphere with respect to the equator. 
Since $d(n-k)$ and $d(k)$ represent the degrees of freedom of fuzzy two-spheres with radii $n-k$ and $k$, respectively,   (\ref{degeeachlatitude}) and (\ref{degeefuzzyfourfuzzytwo}) imply the existence of the ``internal'' degrees of freedom of  fuzzy four-sphere: fuzzy four-sphere is constituted of four-sphere and fibre consisting of two fuzzy two-spheres (whose radii are $(n+X_5)/2$ and $(n-X_5)/2$ at the latitude $X_5$).  Schematically, 
\begin{equation}
S_F^{4}(n)|_{X_5=n-2k}\simeq   S_F^2(n-k)\otimes S^2_F(k). 
\end{equation}
In particular, at the north-pole, $i.e.$ $X_5=n$, we have only one fuzzy two-sphere fibre with radius $n$: $S_F^4(n)|_{X_5=n}\simeq S_F^2(n)$.  
With $SO(5)$ generators $X_{ab}$, 
 coordinates of the two ``internal'' fuzzy two-spheres are respectively given by  
\begin{equation}
R_i=\frac{1}{2}\epsilon_{ijk}X_{jk}+X_{i4},\quad\quad\quad R'_i=\frac{1}{2}\epsilon_{ijk}X_{jk}-X_{i4}, 
\end{equation}
and they satisfy 
\begin{align}
[R_i,R_j]=-2i\epsilon_{ijk}R_k,~~~~~[R_i',R'_j]=2i\epsilon_{ijk}R'_k,~~~~~[R_i,R'_j]=0. 
\end{align}
Then, naturally, $|l_1,l_2,l_3,l_4\rangle$ are regarded as the states on the fuzzy manifold  spanned by $X_a$ and $X_{ab}$. 
The three independent quantities of $l_1$, $l_2$, $l_3$, $l_4$,  specify three latitudes of the four-sphere and two ``internal'' fuzzy two-spheres:  
\begin{align}
&X_5=l_1+l_2-l_3-l_4,\nonumber\\
&{R}_3=l_1-l_2,\nonumber\\
&{R}'_3=l_3-l_4.
\label{x5r3byls}
\end{align}
Inversely, 
 $|l_1,l_2,l_3,l_4\rangle$ is uniquely specified by the eigenvalues of $X_5$, $X_{12}$ and $X_{34}$:  
\begin{align}
&l_1=\frac{1}{4}n+\frac{1}{4}X_5+\frac{1}{2}R_3,~~~~~l_2=\frac{1}{4}n+\frac{1}{4}X_5-\frac{1}{2}R_3,\nonumber\\
&l_3=\frac{1}{4}n-\frac{1}{4}X_5+\frac{1}{2}R'_3,~~~~~l_4=\frac{1}{4}n-\frac{1}{4}X_5-\frac{1}{2}R'_3.
\end{align}
 Thus, as emphasized in Refs.\cite{Ho2002,Kimura2002,Kimura2003}, 
the fuzzy four-sphere has such ``extra-fuzzy space'' that  
 does not have counterpart in the original four-sphere\footnote{One could truncate the 
extra fuzzy spaces, however in such a case, non-associative product has to be implemented \cite{hep-th/0105006}.}. 
The existence of th fuzzy fibre $S_F^2$ can naturally be understood in the context of the 2nd Hopf map.  The $SO(5)$ spinor $\phi$ denotes coordinates on $S^7\sim S^4\otimes S^3$, and the  $U(1)$ phase of $\phi$ is factored out to obtain  $\mathbb{C}P^3\simeq S^7/S^1\sim S^4\otimes S^2$ \cite{Bernevig2002effective}: we have $S^2$-fibred $S^4$  as the classical counterpart of $S_F^4$, not just $S^4$. 
Such enhancement mechanism is inherited to the supersymmetric cases. 

\subsection{${N}=1$ fuzzy four-supersphere}\label{subsecn1fuzzyfoursphere}

Here, we utilize $UOSp(1|4)$ algebra to construct fuzzy four-superspheres with $N=1$ SUSY. 

\subsubsection{$UOSp(1|4)$ algebra}\label{subsecuosp14}

The $UOSp(1|4)$ algebra is constituted  of fourteen generators, ten of which are bosonic  $\Gamma_{ab}=-\Gamma_{ba}$ $(a,b=1,2,\cdots,5)$, and the remaining four are fermionic   $\Gamma_{\alpha}$ $(\alpha=1,2,3,4)$,  
\begin{equation}
\dim[uosp(1|4)]=10|4=14. 
\end{equation}
The $UOSp(1|4)$ algebra is given by 
\begin{align}
&[\Gamma_{ab},\Gamma_{cd}]=i(\delta_{ac}\Gamma_{bd}-\delta_{ad}\Gamma_{bc}-\delta_{bc}\Gamma_{ad}+\delta_{bd}\Gamma_{ac}),\nonumber\\
&[\Gamma_{ab},\Gamma_{\alpha}]=({\gamma}_{ab})_{\beta\alpha}\Gamma_{\beta},\nonumber\\
&\{\Gamma_{\alpha},\Gamma_{\beta}\}=\sum_{a<b}(C{\gamma}_{ab})_{\alpha\beta}\Gamma_{ab},\label{3rdalgebra}
\end{align}
where  $C$ is the $SO(5)$ charge conjugation matrix 
\begin{equation}
C=
\begin{pmatrix}
 \epsilon & 0\\
 0 & \epsilon\\
\end{pmatrix} 
\label{so5chargeconjmat}
\end{equation}
with $\epsilon=i\sigma_2$ (see Appendix \ref{appenchargeconjugationso5} for detail properties of ${C}$). 
$\Gamma_{ab}$ act as $SO(5)$ generators and $\Gamma_{\alpha}$ as a $SO(5)$ spinor. 
The $UOSp(1|4)$ quadratic Casimir is given by  
\begin{equation}
\mathcal{C}=\sum_{a<b}\Gamma_{ab}\Gamma_{ab}+C_{\alpha\beta}\Gamma_{\alpha}\Gamma_{\beta}, 
\label{osp14quadra}
\end{equation}
and  Scasimir is  
\begin{equation}
\mathcal{S}=\frac{1}{4\sqrt{2}}(3-4C_{\alpha\beta}\Gamma_{\alpha}\Gamma_{\beta}). 
\label{osp14Scasimir}
\end{equation}
Similar to the $UOSp(1|2)$ case, the Scasimir satisfies 
\begin{equation}
[\Gamma_{ab},\mathcal{S}]=\{\Gamma_{\alpha},\mathcal{S}\}=0, 
\end{equation}
and 
\begin{equation}
\mathcal{S}^2=\mathcal{C}+\frac{9}{8}. 
\end{equation}

The fundamental representation matrices of $uosp(1|4)$ are constructed as follows. 
First,  we introduce   
\begin{equation}
\Gamma_a=
\begin{pmatrix}
\gamma_a & 0 \\
 0 & 0 
\end{pmatrix}  
\label{gammauosp14}
\end{equation}
with  $\gamma_a$ (\ref{so5gammaI1}), to yield $SO(5)$ generators 
\begin{equation}
\Gamma_{ab}=-i\frac{1}{4}[\Gamma_a,\Gamma_b], 
\end{equation}
or 
\begin{equation}
{\Gamma}_{ab}=
\begin{pmatrix}
 {\gamma}_{ab} & 0\\
 0 & 0
\end{pmatrix}, 
\label{gammablarge}
\end{equation}
with $\gamma_{ab}$ (\ref{so5geneI1}). 
The fermionic generators are 
\begin{equation}
\Gamma_{\alpha}=\frac{1}{\sqrt{2}} 
\begin{pmatrix}
0_4 & \tau_{\alpha} \\
-(C\tau_{\alpha})^t & 0 
\end{pmatrix}, 
\end{equation}
where 
\begin{align}
\tau_1=
\begin{pmatrix}
1\\
0\\
0\\
0
\end{pmatrix},~~
\tau_2=
\begin{pmatrix}
0\\
1\\
0\\
0
\end{pmatrix},~~
\tau_3=
\begin{pmatrix}
0\\
0\\
1\\
0
\end{pmatrix},~~
\tau_4=
\begin{pmatrix}
0\\
0\\
0\\
1
\end{pmatrix}. 
\label{taucomponents}
\end{align}
More explicitly, 
\begin{align}
& \Gamma_{\theta_1}= \frac{1}{\sqrt{2}}
\begin{pmatrix}
& 0 & 0 & 0& 0& 1 \\
& 0 & 0 & 0& 0& 0 \\
& 0 & 0 & 0& 0& 0 \\
& 0 & 0 & 0& 0& 0 \\
& 0 & 1 & 0& 0& 0 \\
\end{pmatrix},~~~~~
\Gamma_{\theta_2}=\frac{1}{\sqrt{2}}
\begin{pmatrix}
& 0 & 0 & 0& 0& 0 \\
& 0 & 0 & 0& 0& 1 \\
& 0 & 0 & 0& 0& 0 \\
& 0 & 0 & 0& 0& 0 \\
& -1 & 0 & 0& 0& 0 \\
\end{pmatrix},\nonumber\\
&\Gamma_{\theta_3}=\frac{1}{\sqrt{2}}
\begin{pmatrix}
& 0 & 0 & 0& 0& 0 \\
& 0 & 0 & 0& 0& 0 \\
& 0 & 0 & 0& 0& 1 \\
& 0 & 0 & 0& 0& 0 \\
& 0 & 0 & 0& 1& 0 \\
\end{pmatrix},~~~~~
\Gamma_{\theta_4}=\frac{1}{\sqrt{2}}
\begin{pmatrix}
& 0 & 0 & 0& 0& 0 \\
& 0 & 0 & 0& 0& 0 \\
& 0 & 0 & 0& 0& 0 \\
& 0 & 0 & 0& 0& 1 \\
& 0 & 0 & -1& 0& 0 
\end{pmatrix}.
\label{lalphaosp14}
\end{align}
They satisfy the ``hermiticity'' condition 
\begin{equation}
\Gamma_a^{\ddagger}=\Gamma_a,\quad\quad \Gamma_{ab}^{\ddagger}=\Gamma_{ab},\quad\quad \Gamma_{\alpha}^{\ddagger}=C_{\alpha\beta}\Gamma_{\beta}. 
\end{equation}
One may regard $\Gamma_a$ and $\Gamma_{\alpha}$ as a supersymmetric extension of $SO(5)$ gamma matrices.

\subsubsection{${N}=1$ graded 2nd Hopf map}\label{subsecn=12ndHopf}

Generalizing the procedure in Sec.\ref{subsec:n=11stHopffuzzytwo},  we construct $N=1$ graded version of the 2nd Hopf map.  
We first introduce $UOSp(1|4)$ spinor 
\begin{equation}
\psi
=
(\psi_1, 
\psi_2, 
\psi_3, 
\psi_4, 
\eta)^t, 
\label{ospfundspinor41}
\end{equation}
where  $\psi_1$, $\psi_2$, $\psi_3$, $\psi_4$, are  Grassmann even while $\eta$ is  Grassmann odd. 
$\psi$ is normalized as 
\begin{equation}
\psi^{\ddagger}\psi=1,
\label{noramlizationofn1gradedspinor}
\end{equation}
where 
\begin{equation}
\psi^{\ddagger}=(\psi_1^*,\psi_2^*,\psi_3^*,\psi_4^*,-\eta^*)
\end{equation}
with pseudo-complex conjugation $*$. From (\ref{noramlizationofn1gradedspinor}), we find that $\psi$ denotes  coordinates on $S^{7|2}$.  
With the supersymmetric version of $SO(5)$ gamma matrices $\Gamma_a$ (\ref{gammauosp14})  and $\Gamma_{\alpha}$ (\ref{lalphaosp14}),  we give $N=1$ graded 2nd Hopf map as  
\begin{equation}
\psi\quad \longrightarrow \quad x_a=\psi^{\ddagger}\Gamma_a\psi,\quad \theta_{\alpha}=\psi^{\ddagger}\Gamma_{\alpha}\psi. 
\label{defxathetausp14}
\end{equation}
In detail,  
\begin{align}
&x_1=i\psi_1^*\psi_4+i\psi_2^*\psi_3-i\psi_3^*\psi_2-i\psi_4^*\psi_1,\nonumber\\ 
&x_2=\psi_1^*\psi_4-\psi_2^*\psi_3-\psi_3^*\psi_2+\psi_4^*\psi_1,\nonumber\\
&x_3=i\psi_1^*\psi_3-i\psi_2^*\psi_4-i\psi_3^*\psi_1+i\psi_4^*\psi_2,\nonumber\\
&x_4=\psi_1^*\psi_3+\psi_2^*\psi_4+\psi_3^*\psi_1+\psi_4^*\psi_2,\nonumber\\
&x_5=\psi_1^*\psi_1+\psi_2^*\psi_2-\psi_3^*\psi_3-\psi_4^*\psi_4,\nonumber\\
&\theta_{1}=\frac{1}{\sqrt{2}}(\psi_1^*\eta-\eta^*\psi_2),\nonumber\\
&\theta_2=\frac{1}{\sqrt{2}}(\psi_2^*\eta+\eta^*\psi_1),\nonumber\\
&\theta_3=\frac{1}{\sqrt{2}}(\psi_3^*\eta-\eta^*\psi_4),\nonumber\\
&\theta_4=\frac{1}{\sqrt{2}}(\psi_4^*\eta+\eta^*\psi_3).\label{explicit2ndgradedhopf}
\end{align}
From $(\eta^*)^*=-\eta$, we have $x_a^*=x_a$ and $\theta_{\alpha}^*=C_{\alpha\beta}\theta_{\beta}$. It is straightforward to see 
\begin{equation}
x_a x_a +2C_{\alpha\beta}\theta_{\alpha}\theta_{\beta}=(\psi^{\ddagger}\psi)^2=1. 
\label{cond4dsupersphere}
\end{equation}
If $x_a$ and $\theta_{\alpha}$ were independent,   
(\ref{cond4dsupersphere}) was the definition of four-supersphere with four (pseudo-real) fermionic coordinates, $S^{4|4}$. 
However, $\theta_1$, $\theta_2$, $\theta_3$ and $\theta_4$ are $\it{not}$ independent to each other, since they are constructed from only one Grassmann odd quantity $\eta$  that carries two real (Grassmann odd) degrees of freedom. Indeed, 
\begin{align}
&\theta_1\theta_2=-\frac{1}{2}\eta^*\eta (\psi_1^*\psi_1+\psi_2^*\psi_2),\nonumber\\
&\theta_3\theta_4=-\frac{1}{2}\eta^*\eta (\psi_3^*\psi_3+\psi_4^*\psi_4),  
\label{productsofthetas} 
\end{align}
and then,  
 for instance, $\theta_1\theta_2\theta_3=0$. 
Also we find  
\begin{equation}
C_{\alpha\beta}\theta_{\alpha}\theta_{\beta}=-\eta^*\eta(\psi_1^*\psi_1+\psi_2^*\psi_2+\psi_3^*\psi_3+\psi_4^*\psi_4)=-\eta^*\eta, 
\end{equation}
 and the relation (\ref{cond4dsupersphere}) can be rewritten as 
\begin{equation}
x_ax_a-2\eta^*\eta=1, 
\label{defs42}
\end{equation}
which corresponds to $S^{4|2}$. 
Thus, $x_a$ and $\theta_{\alpha}$ are regarded as coordinates on $S^{4|2}$ rather than $S^{4|4}$.  
As a consequence, (\ref{defxathetausp14}) represents   
\begin{equation}
S^{7|2}~~ 
\overset{S^3}
 \longrightarrow ~~S^{4|2} \subset S^{4|4}. 
 \label{mapn=1gradhopf}
\end{equation}
The cancellation of $S^3$ can be understood by the following arguments.  
The original normalized $SO(5)$ spinor  is embedded in the $UOSp(1|4)$ spinor as   
\begin{equation}
\phi=
\begin{pmatrix}
\phi_1 \\
\phi_2 \\
\phi_3 \\
\phi_4
\end{pmatrix}
=\frac{1}{\sqrt{1+\eta^*\eta}}
\begin{pmatrix}
\psi_1 \\
\psi_2 \\
\psi_3 \\
\psi_4
\end{pmatrix}.  
\end{equation}
From (\ref{noramlizationofn1gradedspinor}), the normalization of $\phi$ follows 
\begin{equation} 
\phi^{\dagger}\phi=1. 
\end{equation}
Then, the map 
\begin{equation}
\phi ~~\rightarrow ~~y_a=\phi^{\dagger}\gamma_a \phi, 
\label{2ndembeddedhopf}
\end{equation}
signifies the 2nd Hopf map (\ref{2ndHopfmapabstract}). $y_a$ are coordinates of $S^4$; the body of $S^{4|2}$.   
  With $y_a$, $\phi$ is expressed as 
\begin{equation}
\phi
=\frac{1}{\sqrt{2(1+y_5)}}
\begin{pmatrix}
 (1+y_5) 
\begin{pmatrix} 
u\\
 v
 \end{pmatrix} 
\\
 (y_4-i y_i \sigma_i  )
 \begin{pmatrix} 
u\\
 v
\end{pmatrix} 
\end{pmatrix}, 
\label{so5phiexpli}
\end{equation}
where $(u,v)^t$ is an arbitrary two-component spinor subject to the normalization  $u^*u+v^*v=1$ representing $S^3$-fibre.   Such $S^3$-fibre is canceled in (\ref{2ndembeddedhopf}) to yield the coordinates on $S^4$. In the graded 2nd Hopf map (\ref{explicit2ndgradedhopf}), the cancellation of $S^3$ can also be shown. 
 Write the Grassmann odd component $\eta$ as 
\begin{equation}
\eta=u\mu+v\nu,
\label{etan12ndhopf}
\end{equation}
with $\mu$ and $\nu$ being real and imaginary Grassmann odd quantities that satisfy 
\begin{equation}
\mu^*=\nu,~~~~~~\nu^*=-\mu.
\end{equation}
By inserting (\ref{so5phiexpli}) and (\ref{etan12ndhopf}) to (\ref{explicit2ndgradedhopf}), one may show 
\begin{align}
&x_a=({1-\mu\nu})~y_a,\nonumber\\
&\begin{pmatrix}
\theta_1\\
\theta_2
\end{pmatrix}=\frac{\sqrt{1+y_5}}{2}\begin{pmatrix}
\mu\\
\nu
\end{pmatrix},\nonumber\\
&\begin{pmatrix}
\theta_3\\
\theta_4
\end{pmatrix}=\frac{1}{2\sqrt{1+y_5}} (y_4+iy_i {\sigma_i}^t)\begin{pmatrix}
\mu\\
\nu
\end{pmatrix},  
\label{expressionxatheta2ndhop}
\end{align}
where $\eta^*\eta=-\mu\nu$ was utilized. 
Notice that $S^3$-fibre denoted by $(u,v)$ vanishes in the expression of $x_a$ and $\theta_{\alpha}$ (\ref{expressionxatheta2ndhop}).   Furthermore, $\theta_{\alpha=3,4}$ are not independent with $\theta_{\alpha=1,2}$, but related as  
\begin{equation}
\begin{pmatrix}
\theta_3\\
\theta_4
\end{pmatrix}=\frac{1}{1+y_5} (y_4+iy_i {\sigma_i}^t)\begin{pmatrix}
\theta_1\\
\theta_2
\end{pmatrix}.
\end{equation}
The $N=1$ graded Hopf fibration, $S^{7|2}\sim S^{4|2}\otimes S^3$, is obvious from the expression 
\begin{align}
\psi&=\frac{1}{\sqrt{1-\eta^*\eta}}
\begin{pmatrix}
\phi_1\\
\phi_2\\
\phi_3\\
\phi_4\\
\eta
\end{pmatrix}=\frac{1}{\sqrt{2(1+y_5)}}
\begin{pmatrix}
\sqrt{1-\mu\nu}~(1+y_5)
\begin{pmatrix}
u\\
v
\end{pmatrix}
\\
\sqrt{1-\mu\nu}~(y_4-iy_i\sigma_i)
\begin{pmatrix}
u\\
v
\end{pmatrix}
\\
\sqrt{2(1+y_5)}~(u\mu+v\nu)
\end{pmatrix},
\end{align}
where $S^3$-fibre, $(u,v)^t$, is canceled in (\ref{explicit2ndgradedhopf}), and  $y_a$ and  $\mu, \nu$, respectively account for bosonic and fermionic coordinates on $S^{4|2}$. 
With $\theta_1$ and $\theta_2$, $\psi$ is rewritten as 
\begin{align}
\psi&=\frac{1}{\sqrt{2(1+y_5)}}
\begin{pmatrix}
\sqrt{1-\frac{4}{1+y_5}\theta_1\theta_2}~(1+y_5)
\begin{pmatrix}
u\\
v
\end{pmatrix}
\\
\sqrt{1-\frac{4}{1+y_5}\theta_1\theta_2}~(y_4-iy_i\sigma_i)
\begin{pmatrix}
u\\
v
\end{pmatrix}
\\
2\sqrt{2}~(u\theta_1+v\theta_2)
\end{pmatrix},
\end{align}
where $y_a$ are related to $x_a$ as 
\begin{equation}
y_a=\biggl(1-\frac{4}{1+x_5}\theta_1\theta_2\biggr)~x_a.
\end{equation}

\subsubsection{${N}=1$ fuzzy four-supersphere }

The target manifold of the graded Hopf map is $S^{4|2}$, and we denote the corresponding  fuzzy four-supersphere  as  $S_F^{4|2}$. 
Coordinates of $S_F^{4|2}$, $X_a$ and $\Theta_{\alpha}$,  are constructed as 
\begin{equation}
X_a=
\Psi^{\dagger}
\Gamma_a 
\Psi,~~~~ 
\Theta_{\alpha}= \Psi^{\dagger} 
\Gamma_{\alpha} \Psi,
\label{bosonicscwingern1four}
\end{equation}
where  $\Psi$ is a five-component graded Schwinger operator 
\begin{equation}
\Psi=
(\Psi_1, \Psi_2, 
\Psi_3, 
\Psi_4, 
\tilde{\Psi})^t,  
\label{uosp14spinorschwinger}
\end{equation}
with $\Psi_{\alpha}$  $(\alpha=1,2,3,4)$ being bosonic operators and $\tilde{\Psi}$ a fermionic operator: 
\begin{equation}
[\Psi_{\alpha},\Psi_{\beta}^{\dagger}] =\delta_{\alpha\beta},~~~~\{\tilde{\Psi},\tilde{\Psi}^{\dagger}\}=1, ~~~~[\Psi_{\alpha}^{\dagger},\tilde{\Psi}]=[\Psi_{\alpha},\tilde{\Psi}]=\{\tilde{\Psi},\tilde{\Psi}\}=0.
\end{equation}
Square of the radius of  fuzzy four-supersphere is derived as  
\begin{equation}
 X_aX_a+2C_{\alpha\beta}\Theta_{\alpha}\Theta_{\beta}=(\Psi^{\dagger}\Psi)(\Psi^{\dagger}\Psi+3).  \label{n1fuzzy4dradius}
\end{equation}
With the Schwinger construction, the Casimir (\ref{osp14quadra}) 
is represented as 
\begin{equation}
\mathcal{C}=\sum_{a<b}X_{ab}X_{ab}+C_{\alpha\beta}\Theta_{\alpha}\Theta_{\beta}=\frac{1}{2}(\Psi^{\dagger}\Psi)(\Psi^{\dagger}\Psi+3),  
\label{casimirosp14fulre}
\end{equation}
where 
\begin{equation}
{X}_{ab}= \Psi^{\dagger}  
\Gamma_{ab}  \Psi. 
\end{equation}
We used 
\begin{align}
&X_a X_a=2\sum_{a<b}X_{ab}X_{ab}=\hat{n}_B(\hat{n}_B+4),\nonumber\\
&C_{\alpha\beta}\Theta_{\alpha}\Theta_{\beta}=-\frac{1}{2}\hat{n}_B+\hat{n}_B\hat{n}_F+2\hat{n}_F, 
\end{align}
with $\hat{n}_B=\sum_{\alpha=1,2,3,4}\Psi_{\alpha}^{\dagger}\Psi_{\alpha}$ and $\hat{n}_F=\tilde{\Psi}^{\dagger}\tilde{\Psi}$.  
The Casimir (\ref{casimirosp14fulre}) is equivalent to (\ref{n1fuzzy4dradius}) except 
the proportional factor. 
The graded fully symmetric representation is expressed as  
\begin{subequations}
\begin{align}
&|l_1,l_2,l_3,l_4\rangle =\frac{1}{\sqrt{l_1!
~l_2!~l_3!~l_4!}}
 {\Psi_1^{\dagger}}^{l_1} {\Psi_2^{\dagger}}^{l_2} {\Psi_3^{\dagger}}^{l_3} {\Psi_4^{\dagger}}^{l_4}|0\rangle,\label{uosp14fullysymbosonstates}\\ 
&|m_1,m_2,m_3,m_4)=\frac{1}{\sqrt{m_1!~m_2!~m_3!~m_4!}} 
{\Psi_1^{\dagger}}^{m_1} {\Psi_2^{\dagger}}^{m_2} {\Psi_3^{\dagger}}^{m_3} {\Psi_4^{\dagger}}^{m_4}\tilde{\Psi}^{\dagger}|0\rangle,
\label{uosp14fullysymfermistates}
\end{align}\label{uosp14fullysymstates}
\end{subequations}
where $l_1+l_2+l_3+l_4=m_1+m_2+m_3+m_4+1=n$. 
The Scasimir (\ref{osp14Scasimir}) is expressed as   
\begin{equation}
\mathcal{S}=\frac{1}{2\sqrt{2}}(1-2\hat{n}_F)(2\hat{n}+3),   
\end{equation}
with $\hat{n}=\Psi^{\dagger}\Psi=\hat{n}_B+\hat{n}_F$, 
and the bosonic (\ref{uosp14fullysymbosonstates}) and fermionic (\ref{uosp14fullysymfermistates}) states are classified by the sign of Scasimir eigenvalues,   
\begin{equation}
\mathcal{S}=\pm \frac{1}{2\sqrt{2}} (2n+3). 
\end{equation}
The dimensions of bosonic and fermionic states are respectively given by 
\begin{subequations}
\begin{align}
&D_B=D(n)\equiv  \frac{1}{3!}(n+1)(n+2)(n+3),\label{osp14bosonicdof}\\
&D_F=D(n-1)= \frac{1}{3!}n(n+1)(n+2), \label{osp14fermidof} 
\end{align}
\end{subequations}
and the total dimension is
\begin{equation}
D_T=D_B+D_F=\frac{1}{6}(n+1)(n+2)(2n+3). 
\label{totaldimn=1fuzzyfour}
\end{equation}
 Similar to the case of $N=1$ fuzzy two-supersphere, the bosonic degrees of freedom (\ref{osp14bosonicdof}) are accounted for by fuzzy four-sphere with radius $n$ and the fermionic degrees of freedom (\ref{osp14fermidof}) are by fuzzy four-sphere with radius $n-1$.   
Thus, the ${N}=1$ fuzzy four-supersphere is a ``superposition'' of two fuzzy four-spheres with radii $n$ and $n-1$. 
Schematically, 
\begin{equation}
S_F^{4|2}(n)~\simeq~ S_F^{4}(n)\oplus S_F^{4}(n-1).  
\end{equation}

$X_5$ eigenvalues for the states (\ref{uosp14fullysymstates}) are  
\begin{equation}
X_5=n-k,
\end{equation}
with $k=0,1,2,\cdots,2n$. 
The degeneracies for even $k=2l$ and for odd $k=2l+1$ are   respectively  given by 
\begin{align}
&D_{k=2l}(n)=d(n-l)\cdot d(l)=(n-l+1)(l+1),\nonumber\\
&D_{k=2l+1}(n)=d(n-l-1)\cdot d(l)=(n-l)(l+1),  
\end{align}
which give rise to  
\begin{equation}
\sum_{l=0}^{n}D_{k=2l}(n)=D_B,~~~~~~\sum_{l=0}^{n-1}D_{k=2l+1}(n)=D_F. 
\end{equation}
Therefore, at latitude $X_5=n-2l$, we have fuzzy fibre consisting of two fuzzy two-spheres with radii $n-l$ and $l$, while at latitude  
$X_5=n-2l-1$ two fuzzy two-spheres with radii $n-l-1$ and $l$.  In other words, as fuzzy fibre at  $X_5=n-2l$, we have two fuzzy two-spheres with radii $(n+X_5)/2$ and $(n-X_5)/2$, while at  $X_5=n-2l-1$, two fuzzy two-spheres with radii $(n+X_5)/2-1/2$ and $(n-X_5)/2-1/2$.

\subsubsection{Algebraic structure}\label{subsubalgebrasu41sphere}

The ${N}=1$ fuzzy four-supersphere (\ref{n1fuzzy4dradius}) is invariant under the $SU(4|1)$ rotation of the Schwinger operator $\Psi$.   
This implies hidden $SU(4|1)$ structure of ${N}=1$ fuzzy four-supersphere.  
Here, we demonstrate the $SU(4|1)$ structure of fuzzy four-supersphere based on algebraic approach. 
Notice that the fuzzy four-supersphere coordinates $X_a$, $\Theta_{\alpha}$ do not satisfy a closed algebra by themselves, 
\begin{equation}
[X_a,X_b]=4iX_{ab},~~~~[X_a,\Theta_{\alpha}]=(\gamma_a)_{\beta\alpha}\varTheta_{\beta},~~~~
\{\Theta_{\alpha},\Theta_{\beta}\}=\sum_{a<b} (C\gamma_{ab})_{\alpha\beta}X_{ab}.  
\label{unclosedrelsu41first}
\end{equation}
The ``new'' operators that appear on the right-hand sides of (\ref{unclosedrelsu41first}) are 
\begin{equation}
X_{ab}=\Psi^{\dagger}\Gamma_{ab}\Psi,~~~~\varTheta_{\alpha}=\Psi^{\dagger}D_{\alpha}\Psi, 
\end{equation}
with $\Gamma_{ab}$ (\ref{gammablarge}) and $D_{\alpha}$\footnote{ $D_{\alpha}$ have the properties
\begin{equation}
D_{\alpha}^{\ddagger}=-C_{\alpha\beta}D_{\beta}, ~~~~~D_{\alpha}=-C_{\alpha\beta}\Gamma_{\beta}^{\dagger}.
\end{equation}
$D_{\alpha}$ can be constructed by 
\begin{equation}
D_{\alpha}=\frac{2}{5}\sum_{a<b}(\gamma_{ab})_{\beta\alpha}\{\Gamma_{ab},\Gamma_{\beta}\},   
\end{equation} 
similarly to the $su(2|1)$ case (see Appendix \ref{appendatyposp(2|2)}).}
\begin{equation}
D_{\alpha}=\frac{1}{\sqrt{2}} 
\begin{pmatrix}
0_4 & \tau_{\alpha} \\
(C\tau_{\alpha})^t & 0
\end{pmatrix}. 
\label{defdalphasu41}
\end{equation}
 $X_{ab}$ and $\varTheta_{\alpha}$  respectively act as $SO(5)$ generators and spinor.  
Commutation relations including them are 
\begin{align}
&[X_a,\Theta_{\alpha}]=(\gamma_a)_{\beta\alpha} \varTheta_{\beta},~~~~~~~~~~~~~~[X_a,\varTheta_{\alpha}]=(\gamma_a)_{\beta\alpha}\Theta_{\beta},\nonumber\\
&[X_{ab},\Theta_{\alpha}] =  (\gamma_{ab})_{\beta\alpha} \Theta_{\beta},~~~~~~~~~~~~[X_{ab},\varTheta_{\alpha}]= (\gamma_{ab})_{\beta\alpha} \varTheta_{\beta},\nonumber\\ 
&\{\Theta_{\alpha},\Theta_{\beta}\}=\sum_{a<b}(C\gamma_{ab})_{\alpha\beta}X_{ab},~~~~\{\varTheta_{\alpha},\varTheta_{\beta}\}=-\sum_{a<b}(C\gamma_{ab})_{\alpha\beta}X_{ab},\nonumber\\
&\{\Theta_{\alpha},\varTheta_{\beta}\}=\frac{1}{4}(C\gamma_a)_{\alpha\beta}X_a+\frac{1}{4}C_{\alpha\beta}Z.  \label{lalphalalpharelations}
\end{align}
The last equation further yield a new operator 
\begin{equation}
Z=\Psi^{\dagger}H\Psi, 
\end{equation}
with\footnote{$H$ is constructed as
\begin{equation}
H=\frac{6}{5}\biggl(C_{\alpha\beta}\Gamma_{\alpha}\Gamma_{\beta}+\frac{4}{3}\biggr).
\end{equation}
}   
\begin{equation}
H=
\begin{pmatrix}
1_4  & 0 \\
 0 & 4 
\end{pmatrix}. 
\label{gamma5times5}
\end{equation}
The commutation relations concerned with $Z$ are given by 
\begin{equation}
[Z,X_a]=[Z,X_{ab}]=0,~~~~[Z,\Theta_{\alpha}] =-3\varTheta_{\alpha},~~~~[Z,\varTheta_{\alpha}]=-3\Theta_{\alpha}. 
\label{concernwithz}
\end{equation}
(\ref{concernwithz}) does not yield further new operators.  After all, for the closure of the algebra of the fuzzy coordinates $X_a$ and $\Theta_{\alpha}$, we have to introduce new fuzzy coordinates $X_{ab}$, $\varTheta_{\alpha}$ and $Z$,\footnote{With use of $\varTheta_{\alpha}$ and $Z$,  the $UOSp(1|4)$ invariant quantity is given by  
\begin{align}
C_{\alpha\beta}\varTheta_{\alpha}\varTheta_{\beta}+\frac{1}{6}Z^2=\frac{1}{6}(\Psi^{\dagger}\Psi)(\Psi^{\dagger}\Psi+3).    
\end{align}
} and such twenty four operators amount to the $SU(4|1)$ algebra (see Appendix \ref{append:su41}).  
The basic concept of the non-commutative geometry is ``algebraic construction of geometry''. 
Thus, the algebraic structure underlying the fuzzy four-supersphere is considered as $SU(4|1)$ rather than $UOSp(1|4)$. We revisit the $SU(4|1)$ structure in Sec.\ref{Sectsupercoherent}.

\subsection{${N}=2$ fuzzy four-supersphere}
 
 We proceed to the construction of $N=2$ version of fuzzy four-sphere, $S_F^{4|8}$ based on the $UOSp(2|4)$ algebra. 
 
\subsubsection{$UOSp(2|4)$ algebra}

 The dimension of the $UOSp(2|4)$ algebra is 
\begin{equation}
\dim[uosp(2|4)]=11|8=19. 
\end{equation}
We denote the eleven bosonic  generators as $\Gamma_{ab}=-\Gamma_{ba}$ $(a,b=1,2,3,4,5)$ and $\Gamma$, and the eight fermionic generators as $\Gamma_{\alpha}$ and $\Gamma_{\alpha}'$ $(\alpha=\theta_1,\theta_2,\theta_3,\theta_4)$. 
 The $UOSp(2|4)$ algebra is given by    
 \begin{align}
&[\Gamma_{ab},\Gamma_{cd}]= i
(\delta_{ac}{\Gamma}_{bd}-\delta_{ad}{\Gamma}_{bc}+\delta_{bc}{\Gamma}_{ad}-\delta_{bd}{\Gamma}_{ac})\nonumber,\\
 &[\Gamma_{ab},\Gamma_{\alpha}]= (\gamma_{ab})_{\beta\alpha}\Gamma_{\beta}, ~~~~~[\Gamma_{ab},\Gamma'_{\alpha}]= (\gamma_{ab})_{\beta\alpha}\Gamma'_{\beta}, \nonumber\\
&\{ \Gamma_{\alpha},\Gamma_{\beta}\}=\{ \Gamma'_{\alpha},\Gamma'_{\beta}\}=\sum_{a<b}(C\gamma_{ab})_{\alpha\beta}\Gamma_{ab},\nonumber\\
 &\{\Gamma_{\alpha},\Gamma'_{\beta}\}= \frac{1}{2}C_{\alpha\beta}\Gamma,~~~~~~~[\Gamma_{ab},\Gamma]=0,\nonumber\\
 &[\Gamma_{\alpha},\Gamma]=-\Gamma'_{\alpha},~~~~~~~~~~~~~[\Gamma'_{\alpha},\Gamma]=\Gamma_{\alpha}.   
 \label{osp24algebranew}
 \end{align}
The $UOSp(2|4)$ quadratic Casimir is 
 \begin{equation}
\mathcal{C}=\sum_{a<b}\Gamma_{ab}\Gamma_{ab}+C_{\alpha\beta}\Gamma_{\alpha}\Gamma_{\beta}+C_{\alpha\beta}\Gamma'_{\alpha}\Gamma'_{\beta}+\frac{1}{2}\Gamma^2. 
\label{osp42casimir}
 \end{equation}
The fundamental representation of $UOSp(2|4)$ generators is expressed by the following $6\times 6$ matrices  
 \begin{align}
 &\Gamma_{ab}=\begin{pmatrix}
 \gamma_{ab} & 0 \\
 0 & 0_2
 \end{pmatrix},~~~~~~~~~~~~~~~~~~\Gamma=\begin{pmatrix}
 0 _4 & 0 & 0 \\
 0 & 0 & 1\\
0 & -1 & 0  
 \end{pmatrix},\nonumber\\
&\Gamma_{\alpha}=\frac{1}{\sqrt{2}}\begin{pmatrix}
 0_4 & \tau_{\alpha} & 0 \\
 -(C\tau_{\alpha})^t & 0 & 0 \\
 0 & 0 & 0 
 \end{pmatrix},~~\Gamma'_{\alpha}=\frac{1}{\sqrt{2}}\begin{pmatrix}
 0_4 &  0 &  \tau_{\alpha}  \\
  0 & 0 & 0 \\
 -(C\tau_{\alpha})^t & 0 & 0  
 \end{pmatrix},  
 \label{osp24fermionicgenemat}
 \end{align}
 with $\gamma_{ab}$ (\ref{so5geneI1}) and $\tau_{\alpha}$ (\ref{taucomponents}). The corresponding gamma matrices are also 
 \begin{equation}
 \Gamma_a=
 \begin{pmatrix}
 \gamma_{a} & 0 \\
 0 & 0_2
 \end{pmatrix}, 
 \label{so5gammaosp24}
 \end{equation}
 with $\gamma_a$ (\ref{so5gammaI1}). 
 
\subsubsection{${N}=2$ fuzzy four-supersphere}
With a Schwinger operator $\Psi=(\Psi_1,\Psi_2,\Psi_3,\Psi_4,\tilde{\Psi}_1,\tilde{\Psi}_2)^t$, we introduce  
${N}=2$ fuzzy four-supersphere coordinates  
\begin{equation}
X_{a}=\Psi^{\dagger}\Gamma_a\Psi,~~~~\Theta_{\alpha}=\Psi^{\dagger}\Gamma_{\alpha}\Psi,~~~~\Theta'_{\alpha}=\Psi^{\dagger}\Gamma'_{\alpha}\Psi,~~~~{G}=\Psi^{\dagger}\Gamma\Psi.
\end{equation}
As emphasized in Sec.\ref{subsec2ndHopfandfuzzy}, in Schwinger construction, the $SO(5)$ Casimir can be replaced with the inner product of $SO(5)$ gamma matrices,   
 \begin{equation}
 \sum_{a<b}X_{ab}X_{ab}=\frac{1}{2}X_aX_a,  
 \end{equation}
and from (\ref{osp42casimir}),  square of the radius of $N=2$ fuzzy four-supersphere is obtained as  
\begin{equation}
X_aX_a+2C_{\alpha\beta} \Theta_{\alpha} \Theta_{\beta}+2C_{\alpha\beta} \Theta'_{\alpha} \Theta_{\beta}'+G^2=(\Psi^{\dagger}\Psi)(\Psi^{\dagger}\Psi+2).     
\end{equation}
 For $\Psi^{\dagger}\Psi=n$,  
the graded fully symmetric representation is constructed as  
 \begin{subequations}
 \begin{align}
&\!\!\!\!\!\!\!\!\!|l_1,l_2,l_3,l_4\rangle =\frac{1}{\sqrt{l_1!~ l_2!~l_3!~l_4!}}
 {\Psi_1^{\dagger}}^{l_1}{\Psi_2^{\dagger}}^{l_2}{\Psi_3^{\dagger}}^{l_3}{\Psi_4^{\dagger}}^{l_4}|0\rangle,\label{gradedstateboson1osp24}\\
&\!\!\!\!\!\!\!\!\!|m_1,m_2,m_3,m_4 )\! =\!\frac{1}{\sqrt{m_1!~m_2!~m_3!~m_4!}}
 {\Psi_1^{\dagger}}^{m_1}{\Psi_2^{\dagger}}^{m_2}{\Psi_3^{\dagger}}^{m_3}{\Psi_4^{\dagger}}^{m_4}\tilde{\Psi}_1^{\dagger}|0\rangle,\label{gradedstatefermion1osp24}\\
&\!\!\!\!\!\!\!\!\!|m'_1,m'_2,m'_3,m'_4 )\! =\!
\frac{1}{\sqrt{m'_1!~m'_2!~m'_3!~m'_4!}} 
{\Psi_1^{\dagger}}^{m'_1}{\Psi_2^{\dagger}}^{m'_2}{\Psi_3^{\dagger}}^{m'_3}{\Psi_4^{\dagger}}^{m'_4}\tilde{\Psi}_2^{\dagger}|0\rangle,\label{gradedstatefermion2osp24}\\
&\!\!\!\!\!\!\!\!\!|n_1,n_2,n_3,n_4 \rangle =\frac{1}{\sqrt{n_1!~n_2!~n_3!~n_4!}}
 {\Psi_1^{\dagger}}^{n_1}{\Psi_2^{\dagger}}^{n_2}{\Psi_3^{\dagger}}^{n_3}{\Psi_4^{\dagger}}^{n_4}\tilde{\Psi}_1^{\dagger}\tilde{\Psi}_2^{\dagger}|0\rangle, \label{gradedstateboson2osp24}
 \end{align} \label{osp24gradedsymmetricrepre}
 \end{subequations}
 where $l_1+l_2+l_3+l_4=m_1+m_2+m_3+m_4+1=m'_1+m'_2+m'_3+m'_4+1=n_1+n_2+n_3+n_4+2=n$. 
 The first two are $UOSp(1|4)$ representation of the index $n$ (\ref{uosp14fullysymstates}) while the other two are that of  $n-1$.  
In passing from $|l_1,l_2,l_3,l_4\rangle$ to $|n_1,n_2,n_3,n_4\rangle$ via either $|m_1,m_2,m_3,m_4)$ or  $|m'_1,m'_2,m'_3,m'_4)$, we perform supersymmetric transformations twice, and hence we have ${N}=2$ SUSY.  
 Dimensions of bosonic and fermionic states are respectively  
 \begin{align}
 &D_B=D(n)+D(n-2)=\frac{1}{3}(n+1)(n^2+2n+3),\nonumber\\
 &D_F=2D(n-1)=\frac{1}{3}(n+2)(n+1)n. 
 \end{align}
 The total dimension is 
 \begin{align} 
 D_T=D_B+D_F =\frac{1}{3}(2n^2+4n+3)(n+1). 
 \label{totaldimn=2fuzzyfour}
 \end{align}
As in the case of fuzzy two-supersphere, ${N}=2$ fuzzy four-supersphere is a ``superposition'' of two $N=1$ fuzzy four-superspheres.  
Schematically, 
\begin{align}
S_F^{4|4}(n)&\simeq S_F^{4|2}(n)\oplus S_F^{4|2}(n-1)\nonumber\\
&\simeq S_F^{4}(n)\oplus  S_F^{4}(n-1)\oplus  S_F^{4}(n-1)\oplus S_F^4(n-2).  
\end{align}
The last expression corresponds to the degrees of freedom of (\ref{osp24gradedsymmetricrepre}). 
The states (\ref{osp24gradedsymmetricrepre}) are eigenstates of $X_5$ with eigenvalues  
\begin{equation}
X_5=n-k, 
\end{equation}
where $k=0,1,2,\cdots,2n$. 
The degeneracy at $X_5=n-2l$ $(l=0,1,2,\cdots,n)$ is accounted for by the bosonic states (\ref{gradedstateboson1osp24}) and (\ref{gradedstateboson2osp24}):  
\begin{equation}
D_B^{k=2l}=D_l(n)+D_{l-1}(n-2)=2l(n-l)+n+1
\end{equation}
with $D_l(n)$ (\ref{degeeachlatitude}), 
while that at $X_5=n-2l-1$ $(l=0,1,2,\cdots,n-1)$ is accounted for by the fermionic states (\ref{gradedstatefermion1osp24}) and (\ref{gradedstatefermion2osp24}): 
\begin{equation}
D_F^{k=2l+1}=2D_l(n-1)=2(l+1)(n-l). 
\end{equation}

\subsubsection{${N}=2$ graded 2nd Hopf map}\label{subsec:n=2graded2nd}
 
 The derivation of the corresponding  Hopf map is straightforward. 
 With a normalized $UOSp(2|4)$ spinor $\psi$  
 \begin{equation}
 \psi= (\psi_1,\psi_2,\psi_3,\psi_4,\eta_1,\eta_2)^t, 
 \end{equation}
subject to $\psi^{\ddagger}\psi=1$,  $N=2$ graded 2nd Hopf map is given by   
 \begin{equation}
 x_a=\psi^{\ddagger}\Gamma_a \psi,~~~~\theta_{\alpha}=\psi^{\ddagger} \Gamma_{\alpha}\psi,~~~~\theta'_{\alpha}=\psi^{\ddagger} \Gamma'_{\alpha}\psi,~~~~{g}=\psi^{\ddagger} \Gamma \psi.
 \label{n2graded2ndhopf}
 \end{equation}
The normalization of $\psi$ indicates that $\psi$ is coordinates on $S^{7|4}$. (\ref{n2graded2ndhopf}) satisfy   
 \begin{equation}
 x_ax_a+2C_{\alpha\beta}\theta_{\alpha}\theta_{\beta}+2C_{\alpha\beta}\theta'_{\alpha}\theta'_{\beta}+g^2 
=(\psi^{\ddagger}\psi)^2=1. 
\label{relations44}
 \end{equation}
Then, we have eight (pseudo-)Majorana fermionic coordinates, $\theta_{\alpha}$ and $\theta'_{\alpha}$. However they are not independent, since they contain only four real Grassmann odd degrees of freedom coming from  $\eta_1$ and $\eta_2$.   
With the renormalization, 
\begin{align}
&x_a\rightarrow \sqrt{1-g^2}~x_a=(1-\frac{1}{2}g^2)x_a,~~~~\theta\rightarrow \sqrt{1-g^2}~\theta_{\alpha}=\theta_{\alpha},\nonumber\\
&\theta'_{\alpha}\rightarrow \sqrt{1-g^2}~\theta'_{\alpha}=\theta'_{\alpha},~~~~~~~~~~~~~~~~~g\rightarrow \sqrt{1-g^2}~{g}={g}, 
\label{renormalizationosp24}
\end{align}
  (\ref{relations44}) is restated as 
\begin{equation}
 x_ax_a+2C_{\alpha\beta}\theta_{\alpha}\theta_{\beta}+2C_{\alpha\beta}\theta'_{\alpha}\theta'_{\beta}=1.   
 \label{condrenors44}
 \end{equation}
 This would represent $S^{4|8}$ provided $\theta_{\alpha}$ and $\theta_{\alpha}'$ were independent. 
 The original $SO(5)$ normalized spinor $\phi$ (\ref{so5phiexpli}) is embedded as 
 \begin{equation}
 \phi=
 \begin{pmatrix}
 \phi_1\\
 \phi_2\\
 \phi_3\\
 \phi_4
 \end{pmatrix}=\frac{1}{\sqrt{1+\eta_1^*\eta_1+\eta_2^*\eta_2}}
 \begin{pmatrix}
 \psi_1\\
 \psi_2\\
 \psi_3\\
 \psi_4
 \end{pmatrix}.   
 \end{equation}
One may demonstrate the cancellation of $S^3$-fibre in the map (\ref{n2graded2ndhopf}) by following similar arguments in Sec.\ref{subsecn=12ndHopf}.  
Write the two Grassmann odd components as 
\begin{align}
&\eta_1=u\mu_1+v\nu_1,\nonumber\\
&\eta_2=u\mu_2+v\nu_2,
\end{align}
where $u$ and $v$ denote the coordinates on $S^3$ $(u^*u+v^*v=1)$, and $\mu_{1,2}$ and $\nu_{1,2}$ are respectively real and imaginary components of $\eta_{1,2}$.  
From (\ref{n2graded2ndhopf}), we have 
\begin{align}
x_a&=({1-\mu_1\nu_1-\mu_2\nu_2})~y_a,\nonumber\\
\begin{pmatrix}
\theta_1\\
\theta_2
\end{pmatrix}&=\frac{1}{2}
\sqrt{(1+y_5)(1-\mu_1\nu_1-\mu_2\nu_2)}\begin{pmatrix}
\mu_1\\
\nu_1
\end{pmatrix},~~~~~\begin{pmatrix}
\theta_3\\
\theta_4
\end{pmatrix}=
\frac{1}{2}
{\sqrt{\frac{1-\mu_1\nu_1-\mu_2\nu_2}{1+y_5}}}(y_4+iy_i {\sigma_i}^t)\begin{pmatrix}
\mu_1\\
\nu_1
\end{pmatrix},\nonumber\\
\begin{pmatrix}
\theta'_1\\
\theta'_2
\end{pmatrix}&=\frac{1}{2}
\sqrt{(1+y_5)(1-\mu_1\nu_1-\mu_2\nu_2)}\begin{pmatrix}
\mu_2\\
\nu_2
\end{pmatrix},~~~~~\begin{pmatrix}
\theta'_3\\
\theta'_4
\end{pmatrix}=
\frac{1}{2}
{\sqrt{\frac{1-\mu_1\nu_1-\mu_2\nu_2}{1+y_5}}}(y_4+iy_i{\sigma_i}^t)\begin{pmatrix}
\mu_2\\
\nu_2
\end{pmatrix}, 
\label{explicitrepxathetan2hopf}
\end{align}
where $\eta_1^*\eta_1=-\mu_1\nu_1$ and $\eta_2^*\eta_2=-\mu_2\nu_2$ were utilized. 
Notice that $u$ and $v$ do not appear in (\ref{explicitrepxathetan2hopf}).  Besides, $\theta_{3,4}$ and $\theta'_{3,4}$ are respectively related to $\theta_{1,2}$ and $\theta'_{1,2}$ as 
\begin{equation}
\begin{pmatrix}
\theta_3\\
\theta_4
\end{pmatrix}=\frac{1}{1+y_5}(y_4+iy_i {\sigma_i}^t)
\begin{pmatrix}
\theta_1 \\
\theta_2
\end{pmatrix},~~~~~~\begin{pmatrix}
\theta'_3\\
\theta'_4
\end{pmatrix}=\frac{1}{1+y_5}(y_4+iy_i {\sigma_i}^t)
\begin{pmatrix}
\theta'_1 \\
\theta'_2
\end{pmatrix}.
\end{equation}
Thus, with the representation  
\begin{align}
\psi&={\sqrt{1+\eta^*_1\eta_1+\eta_2^*\eta_2}}
\begin{pmatrix}
\phi_1\\
\phi_2\\
\phi_3\\
\phi_4\\
\sqrt{1-\eta_2^*\eta_2}~\eta_1\\
\sqrt{1-\eta_1^*\eta_1}~\eta_2
\end{pmatrix}=\sqrt{\frac{1-\mu_1\nu_1-\mu_2\nu_2}{{2(1+y_5)}}}
\begin{pmatrix}
(1+y_5)
\begin{pmatrix}
u\\
v
\end{pmatrix}\\
(y_4-iy_i\sigma_i)
\begin{pmatrix}
u\\
v
\end{pmatrix}\\
\sqrt{2(1+y_5)(1+\mu_2\nu_2)}~(u\mu_1+v\nu_1)\\
\sqrt{2(1+y_5)(1+\mu_1\nu_1)}~(u\mu_2+v\nu_2)\\
\end{pmatrix},
\label{n2hopfosp24exspino}
\end{align}
the $S^3$-fibre denoted by $(u,v)$ is canceled in $x_a$, $\theta_{\alpha}$ and $\theta'_{\alpha}$ (\ref{n2graded2ndhopf}).  
From (\ref{explicitrepxathetan2hopf}), we have  
\begin{equation}
\theta_{1}\theta_{2}=-\frac{1+y_5}{4(1+\mu_2\nu_2)}\mu_1\nu_1,~~~~\theta'_{1}\theta'_{2}=-\frac{1+y_5}{4(1+\mu_1\nu_1)}\mu_2\nu_2,
\end{equation}
and hence 
\begin{equation}
\theta_{1}\theta_{2}+\theta'_{1}\theta'_{2}=-\frac{1+y_5}{4(1+\mu_1\nu_1+\mu_2\nu_2)}(\mu_1\nu_1+\mu_2\nu_2),
\end{equation}
or inversely,
\begin{equation}
\mu_1\nu_1+\mu_2\nu_2=\frac{4}{1+y_5-4(\theta_1\theta_2+\theta_1'\theta_2')}(\theta_1\theta_2+\theta_1'\theta_2').
\end{equation}
Then,  
\begin{align}
&\begin{pmatrix}
\mu_1\\
\nu_1
\end{pmatrix}
=\frac{2}{\sqrt{1+y_5-4(\theta_1\theta_2+\theta'_1\theta'_2)}}
\begin{pmatrix}
\theta_1\\
\theta_2
\end{pmatrix}, \nonumber\\
&\begin{pmatrix}
\mu_2\\
\nu_2
\end{pmatrix}
=\frac{2}{\sqrt{1+y_5-4(\theta_1\theta_2+\theta'_1\theta'_2)}}
\begin{pmatrix}
\theta'_1\\
\theta'_2
\end{pmatrix}. 
\end{align}
Therefore, with the coordinates on $S^{4|4}$, $y_a$. $\theta_{1,2}$ and $\theta'_{1,2}$, (\ref{n2hopfosp24exspino}) is rewritten as 
\begin{equation}
\psi
=\frac{1}{\sqrt{2(1+y_5-4(\theta_1\theta_2+\theta_1'\theta_2'))}}\begin{pmatrix}
{\sqrt{{1-\frac{8}{1+y_5}(\theta_1\theta_2+\theta_1'\theta_2')}}}~
(1+y_5)
\begin{pmatrix}
u\\
v
\end{pmatrix}\\
{\sqrt{{1-\frac{8}{1+y_5}(\theta_1\theta_2+\theta_1'\theta_2')}}~
(y_4-iy_i\sigma_i)}
\begin{pmatrix}
u\\
v
\end{pmatrix}\\
2\sqrt{2}~(u\theta_1+v\theta_2)\\
2\sqrt{2}~(u\theta'_1+v\theta'_2)\\
\end{pmatrix},
\end{equation}
where $y_a$ are related to $x_a$ as 
\begin{align}
y_a&=\biggl(1-\frac{4}{1+y_5-4(\theta_1\theta_2+\theta_1'\theta_2')}(\theta_1\theta_2+\theta_1'\theta_2')\biggr)~x_a\nonumber\\
&=\biggl(1-\frac{4}{1+x_5} (\theta_1\theta_2+\theta_1'\theta_2')
 -\frac{16}{(1+x_5)^3}(1+2x_5)(\theta_1\theta_2+\theta_1'\theta_2')^2 \biggr)~x_a. 
\end{align}

Meanwhile, $g(=-\eta_1^*\eta_2+\eta_1^*\eta_2)$ is given by  
\begin{equation}
g=2\mu_1\mu_2 uv^*-(\mu_1\nu_2+\nu_1\mu_2)(u^*u-v^*v)-2\nu_1\nu_2 u^*v,  
\end{equation}
which depends on the $S^3$-fibre, $(u,v)$.   
 $S^3$-fibre is canceled in $g^2$:  
\begin{equation}
g^2=2\mu_1\nu_1\mu_2\nu_2=\frac{32}{(1+y_5)^2}\theta_1\theta_2\theta_1'\theta'_2.
\end{equation}
Thus, though the $S^3$ cancellation is not ``complete''  in (\ref{n2graded2ndhopf}) (because of $g$), with the renormalization (\ref{renormalizationosp24}) in which only $g^2$ is concerned, $S^3$ is completely projected out to yield coordinates on $S^{4|4}$. 
Consequently, the map (\ref{n2graded2ndhopf}) with (\ref{renormalizationosp24}) represents 
 \begin{equation}
 S^{7|4}~~\overset{S^3}\longrightarrow~~ S^{4|4} \subset S^{4|8}. 
 \end{equation}
The base manifold is $S^{4|4}$, and then the corresponding fuzzy manifold is $S_F^{4|4}$. 

\section{More supersymmetries}\label{SectMoreSUSY}

One may incorporate more supersymmetries based on  $UOSp(N|4)$ algebras with $N\ge 3$.  The dimension of the $UOSp(N|4)$ algebra is 
\begin{equation}
\dim [uosp(N|4)]= 10+\frac{1}{2}N(N-1)|4N =10+\frac{1}{2}N(N+7).
\end{equation}
We denote bosonic generators as $\Gamma_{ab}=-\Gamma_{ba}$ $(a,b=1,2,3,4,5)$, $\tilde{\Gamma}_{lm}=-\tilde{\Gamma}_{ml}$ $(l,m=1,2,\cdots,N)$ and fermionic generators as $\Gamma_{l\alpha}$ $(\alpha=1,2,3,4)$.  They satisfy 
\begin{align}
&[\Gamma_{ab},\Gamma_{cd}]= i
(\delta_{ac}{\Gamma}_{bd}-\delta_{ad}{\Gamma}_{bc}+\delta_{bc}{\Gamma}_{ad}-\delta_{bd}{\Gamma}_{ac}),\nonumber\\
&[\Gamma_{ab},\Gamma_{l\alpha}]=(\gamma_{ab})_{\beta\alpha} \Gamma_{l\beta},\nonumber\\
&[\Gamma_{ab},\tilde{\Gamma}_{lm}] = 0,  \nonumber\\
&\{\Gamma_{l\alpha},\Gamma_{m\beta}\}= \sum_{a<b} (C\gamma_{ab})_{\alpha\beta} \Gamma_{ab}\delta_{lm}+\frac{1}{4}C_{\alpha\beta}\tilde{\Gamma}_{lm}, \nonumber\\
&[\Gamma_{l\alpha},\tilde{\Gamma}_{mn}]=   (\gamma_{mn})_{pl}\Gamma_{p\alpha},\nonumber\\
&[\tilde{\Gamma}_{lm},\tilde{\Gamma}_{np}]= -
\delta_{ln}{\tilde{\Gamma}}_{mp}+\delta_{lp}\tilde{\Gamma}_{mn}-\delta_{mp}\tilde{\Gamma}_{ln}+\delta_{mn}\tilde{\Gamma}_{lp}, 
\label{algebraosp4k}
\end{align}
where $C$ is the $SO(5)$ charge conjugation matrix (\ref{so5chargeconjmat}) and 
$\gamma_{lm}=-\gamma_{ml}$ $(l<m)$ are $SO(N)$ generators given by 
\begin{equation}
(\gamma_{lm})_{np}=\delta_{ln}\delta_{mp}-\delta_{lp}\delta_{mn}.
\label{explicitspksmall}
\end{equation}
The $UOSp(N|4)$ quadratic Casimir is 
\begin{equation}
\mathcal{C}=\sum_{a<b}\Gamma_{ab}\Gamma_{ab}+C_{\alpha\beta}\sum_{l=1}^N \Gamma_{l\alpha}\Gamma_{l\beta}+\frac{1}{2}\sum_{l<m=1}^N \tilde{\Gamma}_{lm}\tilde{\Gamma}_{lm}.
\end{equation}

The fundamental representation matrices of $uosp(N|4)$ are given by 
\begin{equation}
\Gamma_{ab}=\begin{pmatrix}
\gamma_{ab} & 0 \\
0 & 0_N 
\end{pmatrix},~~~~\Gamma_{l\alpha}=\begin{pmatrix}
0_{3+l} & \tau_{\alpha} & 0 \\
 -(C\tau_{\alpha})^t & 0 & 0 \\
 0 & 0 & 0_{N-l} 
\end{pmatrix},~~~~\tilde{\Gamma}_{lm}=\begin{pmatrix} 
0_4 & 0 \\
0 & \gamma_{lm}
\end{pmatrix},  
\label{fundrepgeneuspn4}
\end{equation}
where $0_k$ signify $k\times k$ zero-matrices, and 
$\tau_{\alpha}$ are given by (\ref{taucomponents}). 
Notice that $\tilde{\Gamma}_{lm}$ are taken to be anti-hermitian, $\tilde{\Gamma}_{lm}^{\dagger}=-\tilde{\Gamma}_{lm}$. 
We apply the Schwinger construction to (\ref{fundrepgeneuspn4}) and define      
\begin{equation}
X_a=\Psi^{\dagger}\Gamma_a\Psi,~~~~X_{ab}=\Psi^{\dagger}\Gamma_{ab}\Psi,~~~~\Theta_{\alpha}^{(l)}=\Psi^{\dagger}\Gamma_{l\alpha}\Psi,~~~~{Y}_{lm}=\Psi^{\dagger}\tilde{\Gamma}_{lm}\Psi, \label{Schwingerrepgeneral}
\end{equation}
where   
$\Psi=(\Psi_1,\Psi_2,\Psi_3,\Psi_4,\tilde{\Psi}_1,\tilde{\Psi}_2,\cdots,\tilde{\Psi}_N)^t$ in which $\Psi_{\alpha}$ $(\alpha=1,2,3,4)$ are bosonic while $\tilde{\Psi}_{l}$ $(l=1,2,\cdots,N)$ are fermionic.   
Square of the radius of $N$-SUSY fuzzy four-supersphere is derived as  
\begin{equation}
X_a X_a +2\sum_{l=1}^N C_{\alpha\beta}\Theta_{\alpha}^{(l)}\Theta_{\beta}^{(l)}+\sum_{l<m=1}^N Y_{lm}Y_{lm}=\hat{n}(\hat{n}+4-N),  
\label{square42kfuzzy}
\end{equation}
with $\hat{n}=\Psi^{\dagger}\Psi$.  Here, we utilized 
\begin{align}
&\sum_{a=1}^5 X_aX_a=2\sum_{a<b=1}^5 X_{ab}X_{ab}=\hat{n}_B(\hat{n}_B+4),\nonumber\\
&\sum_{l=1}^N C_{\alpha\beta}\Theta_{\alpha}^{(l)}\Theta_{\beta}^{(l)}=-\frac{N}{2}\hat{n}_B+\hat{n}_B\hat{n}_F+2\hat{n}_F,\nonumber\\
&\sum_{l<m=1}^N Y_{lm}Y_{lm}=\hat{n}_F(\hat{n}_F-N),
\end{align}
with $\hat{n}_B=\sum_{\alpha=1}^4\Psi_{\alpha}^{\dagger}\Psi_{\alpha}$ and $\hat{n}_F=\sum_{\sigma=1}^N \tilde{\Psi}_{\sigma}^{\dagger}\tilde{\Psi}_{\sigma}$. 
 $X_a$, $\Theta_{\alpha}^{(l)}$, $Y_{lm}$ do not satisfy a closed algebra by themselves. As a similar manner to Sec.\ref{subsubalgebrasu41sphere}, one may readily show that the minimally extended algebra that includes $X_a$, $\Theta_{\alpha}^{(l)}$ and $Y_{lm}$ is  $su(4|N)$.

For $UOSp(N|4)$ with  $\Psi^{\dagger}\Psi=n$, the graded fully symmetric representation is constructed as 
\begin{align}
&\!\!\!\!\!\!|l_1,l_2,l_3,l_4\rangle =\frac{1}{\sqrt{l_1!~l_2!~l_3!~l_4!}} 
{\Psi_1^{\dagger}}^{l_1} {\Psi_2^{\dagger}}^{l_2} {\Psi_3^{\dagger}}^{l_3} {\Psi_4^{\dagger}}^{l_4} |0\rangle, \nonumber\\
&\!\!\!\!\!\!|m_1,m_2,m_3,m_4)_{i_1} =\frac{1}{\sqrt{m_1!~m_2!~m_3!~m_4!}} 
{\Psi_1^{\dagger}}^{m_1} {\Psi_2^{\dagger}}^{m_2} {\Psi_3^{\dagger}}^{m_3} {\Psi_4^{\dagger}}^{m_4} \tilde{\Psi}_{i_1}^{\dagger}|0\rangle \nonumber\\
&\!\!\!\!\!\!|n_1,n_2,n_3,n_4\rangle_{i_1<i_2} =\frac{1}{\sqrt{n_1!~n_2!~n_3!~n_4!}} 
{\Psi_1^{\dagger}}^{n_1} {\Psi_2^{\dagger}}^{n_2} {\Psi_3^{\dagger}}^{n_3} {\Psi_4^{\dagger}}^{n_4} \tilde{\Psi}_{i_1}^{\dagger}  \tilde{\Psi}_{i_2}^{\dagger} |0\rangle \nonumber\\
&\vdots\nonumber\\
&\!\!\!\!\!\!|q_1,q_2,q_3,q_4\rangle_{i_1<i_2< \cdots< i_{N-1}}=
\frac{1}{\sqrt{q_1!q_2!q_3!q_4!}} 
{\Psi_1^{\dagger}}^{q_1} {\Psi_2^{\dagger}}^{q_2} {\Psi_3^{\dagger}}^{q_3} {\Psi_4^{\dagger}}^{q_4} ~\tilde{\Psi}_{i_1}^{\dagger}  \tilde{\Psi}_{i_2
}^{\dagger}\tilde{\Psi}^{\dagger}_{i_3} \cdots \tilde{\Psi}_{i_{N-1}}^{\dagger}  |0\rangle,\nonumber\\
&\!\!\!\!\!\!|r_1,r_2,r_3,r_4)=
\frac{1}{\sqrt{r_1!r_2!r_3!r_4!}} 
{\Psi_1^{\dagger}}^{r_1} {\Psi_2^{\dagger}}^{r_2} {\Psi_3^{\dagger}}^{r_3} {\Psi_4^{\dagger}}^{r_4} ~\tilde{\Psi}_{1}^{\dagger}  \tilde{\Psi}_{2
}^{\dagger}\tilde{\Psi}_3 \cdots \tilde{\Psi}_{N-1}^{\dagger}  \tilde{\Psi}_N^{\dagger}|0\rangle, \label{gradedfulrepgeneral}
\end{align}
where $l_1+l_2+l_3+l_4=m_1+m_2+m_3+m_4+1=n_1+n_2+n_3+n_4+2=\cdots=q_1+q_2+q_3+q_4+N-1=r_1+r_2+r_3+r_4+N=n$.  
Therefore, with $D(n)$ (\ref{so5degenesym}),  
the dimension of (\ref{gradedfulrepgeneral}) is derived as 
\begin{equation}
D_T=\sum_{l=0}^N {}_NC_l\cdot  D(n-l)=\frac{1}{3}  (2n+4-N)\biggl((2n+4-N)^2-4+3N\biggr) 2^{N-4},  
\label{totaldimingeneral}
\end{equation}
for $n\ge N-3$. (One may readily confirm that (\ref{totaldimingeneral}) reproduces the previous results (\ref{totaldimn=1fuzzyfour}), (\ref{totaldimn=2fuzzyfour}) for $N=1,2$.) 
For odd $N$, $N=2l+1$, the degeneracies of  bosonic and fermionic states are respectively given by 
$D_B=\sum_{k=0}^{l}{}_{2l+1}C_{2k} \cdot D(n-2k)$ and 
$D_F=\sum_{k=0}^{l}{}_{2l+1}C_{2k+1} \cdot D(n-2k-1)$. 
Meanwhile, for even $N$, $N=2l$, the degeneracies are respectively 
$D_B=\sum_{k=0}^{l}{}_{2l}C_{2k} \cdot D(n-2k)$ and 
$D_F=\sum_{k=0}^{l-1}{}_{2l}C_{2k+1} \cdot D(n-2k-1)$. 
Schematically,  
 $S_F^{4|2N}(n)$ is expressed as a superposition of fuzzy four-superspheres with lower supersymmetries, $S_F^{4|2N-2l}$, with  different radii, $n$, $n-1$, $n-2$, $\cdots$, $n-l$: 
\begin{align}
S_F^{4|2N}(n)&~\simeq~ \sum_{m=0}^l ~ {}_{l}C_m\cdot S_F^{4|2N-2l}(n-m)\nonumber\\
&~\simeq~ S_F^{4|2N-2l}(n)\oplus l\cdot S_F^{4|2N-2l}(n-1) \oplus \frac{l(l-1)}{2!} \cdot S_F^{4|2N-2l}(n-2) 
\oplus \cdots \oplus S_F^{4|2N-2l}(n-l).
\end{align}
Explicitly, 
\begin{align}
S_F^{4|2N}(n)&~\simeq~ S^{4|2N-2}_F(n) \oplus S_F^{4|2N-2}(n-1)\nonumber\\
&~\simeq ~S_F^{4|2N-4}(n)\oplus  2 S_F^{4|2N-4}(n-1)\oplus  S_F^{4|2N-4}(n-2),\nonumber\\
&~\simeq ~S_F^{4|2N-6}(n)\oplus  3 S_F^{4|2N-6}(n-1)\oplus  3S_F^{4|2N-6}(n-2)\oplus S_F^{4|2N-6}(n-3),\nonumber\\
&~\simeq~ \cdots. 
\end{align}

Replacing the Schwinger operator with a normalized $UOSp(4|N)$ spinor, $i.e.$, $\Psi\rightarrow \psi$ and $\Psi^{\dagger} \rightarrow \psi^{\ddagger}$ ($\psi^{\ddagger}\psi=1$)    
 in (\ref{Schwingerrepgeneral}), we introduce $x_a$, $\theta_{\alpha}^{(l)}$ and $y_{lm}$ 
that satisfy 
\begin{equation}
x_a x_a +2\sum_{l=1}^N C_{\alpha\beta}\theta_{\alpha}^{(l)}\theta_{\beta}^{(l)}+\sum_{l<m=1}^N y_{lm}y_{lm}=(\psi^{\ddagger}\psi)^2=1. 
\label{classicalrelationgeneral}
\end{equation}
The original $SO(5)$ normalized spinor is embedded as 
\begin{equation}
\phi=
\begin{pmatrix}
\phi_1\\
\phi_2\\
\phi_3\\
\phi_4
\end{pmatrix}
=\frac{1}{\sqrt{1+\eta_1^*\eta_1+\eta_2^*\eta_2+\cdots+\eta_N^*\eta_N}}
\begin{pmatrix}
\psi_1\\
\psi_2\\
\psi_3\\
\psi_4
\end{pmatrix}. 
\end{equation}
The normalized $UOSp(4|N)$ spinor $\psi$ has the dimension $(7|2N)$.  One may readily demonstrate the cancellation of the $S^3$-fibre in the graded Hopf map by following the similar arguments presented in the previous sections (especially Sec.\ref{subsec:n=2graded2nd}), and hence   
the present graded Hopf map signifies\footnote{
The normalization corresponding to (\ref{renormalizationosp24}) is $(x_a,\theta_{\alpha}^{(l)})\rightarrow \sqrt{1-\sum_{l<m=1}^N y_{lm}y_{lm}} ~(x_a,\theta_{\alpha}^{(l)})$. After this normalization, the coordinates satisfy $ x_a x_a +2\sum_{l=1}^N C_{\alpha\beta}\theta_{\alpha}^{(l)}\theta_{\beta}^{(l)}=1.$}    
\begin{equation}
S^{7|2N} ~\overset{S^3}\longrightarrow ~S^{4|2N}. 
\end{equation}

Compare  (\ref{classicalrelationgeneral})  with (\ref{square42kfuzzy}).
Due to the existence of fermionic degrees of freedom, the zero-point energy in (\ref{square42kfuzzy}) decreases with increase of the number of supersymmetry.  
For $N=4$, the square of the radius of fuzzy supersphere (\ref{square42kfuzzy}) ``saturates'' the classical bound (\ref{classicalrelationgeneral}). 
In this sense,  ${N}=4$ is the ``maximum'', otherwise the square of the radius takes negative value for sufficiently small $n$ that satisfies $n < N-4$.   We have already discussed $N=0,1,2$ cases. In the following subsections, we argue the remaining cases, $N=3$ and $4$.

\subsection{${N}=3$ graded 2nd Hopf map and fuzzy four-supersphere}

The dimension of the $UOSp(3|4)$ algebra is 
\begin{equation}
\dim[uosp(3|4)]= 13|12=25.  
\end{equation}
From (\ref{explicitspksmall}), we derive the $SO(3)$ generators
$\gamma_{ij}$ 
 $(i,j=1,2,3)$ as  
\begin{equation}
\gamma_{12}=
\begin{pmatrix}
0 & 1 & 0 \\
-1 & 0 & 0 \\
0 & 0 & 0 
\end{pmatrix},~\gamma_{23}=
\begin{pmatrix}
0 & 0 & 0 \\
0 & 0 & 1 \\
0 & -1 & 0 
\end{pmatrix},~\gamma_{31}=
\begin{pmatrix}
0 & 0 & -1 \\
0 & 0 & 0 \\
1 & 0 & 0 
\end{pmatrix}. 
\end{equation}
With the identification $\tilde{\Gamma}_i=-\frac{1}{2}\epsilon_{ijk}\tilde{\Gamma}_{jk}$, $\tilde{\Gamma}_i$ satisfy the $SU(2)$ algebra 
\begin{equation}
[\tilde{\Gamma}_i,\tilde{\Gamma}_j]=\epsilon_{ijk}\tilde{\Gamma}_k. 
\end{equation}
Then, with   
\begin{equation}
Y_i=\Psi^{\dagger}\tilde{\Gamma}_i\Psi, 
\end{equation}
square of the radius of ${N}=3$ fuzzy four-supersphere is obtained as 
\begin{equation}
X_a X_a +2\sum_{i=1}^3 C_{\alpha\beta}\Theta_{\alpha}^{(i)}\Theta_{\beta}^{(i)}+\sum_{i=1}^3 Y_{i}Y_{i}=(\Psi^{\dagger}\Psi)(\Psi^{\dagger}\Psi+1). 
\end{equation}
The corresponding classical relation is 
\begin{equation}
x_a x_a +2\sum_{i=1}^3 C_{\alpha\beta}\theta_{\alpha}^{(i)}\theta_{\beta}^{(i)}+\sum_{i=1}^3 y_{i} y_{i}=(\psi^{\ddagger}\psi)^2=1. 
\end{equation}
For $\Psi^{\dagger}\Psi=n$,  
 the dimensions of the bosonic and fermionic states in (\ref{gradedfulrepgeneral}) with $N=3$ are respectively given by  
\begin{align}
&D_B=D(n)+3D(n-2)=
\frac{1}{3}(2n^2+n+3)(n+1),\nonumber\\
&D_F=3D(n-1)+D(n-3)=
\frac{1}{3}(2n^2+3n+4)n,
\end{align}
and the total dimension is 
\begin{equation} 
D_T=D_B+D_F=
\frac{1}{3}(2n+1)(2n^2+2n+3). 
\end{equation}

\subsection{${N}=4$ graded 2nd Hopf map and fuzzy four-supersphere}

The dimension of the $UOSp(4|4)$ algebra is 
\begin{equation}
\dim [uosp(4|4)]= 16|16=32.  
\end{equation}
From (\ref{explicitspksmall}), the $SO(4)$ generators are obtained as   
\begin{align}
&\gamma_{12}=
\begin{pmatrix}
0 & 1 & 0 & 0 \\
-1 & 0 & 0 & 0 \\
0 & 0 & 0 & 0 \\
0 & 0 & 0 & 0 
\end{pmatrix},~~\gamma_{13}=
\begin{pmatrix}
0 & 0 & 1 & 0 \\
0 & 0 & 0 & 0  \\
-1 & 0 & 0 & 0 \\
0 & 0 & 0 & 0 
\end{pmatrix},~~\gamma_{14}=
\begin{pmatrix}
0 & 0 & 0 & 1 \\
0 & 0 & 0 & 0 \\
0 & 0 & 0 & 0 \\
-1 & 0 & 0 & 0 
\end{pmatrix},\nonumber\\
&\gamma_{23}=
\begin{pmatrix}
0 & 0 & 0 & 0 \\
0 & 0 & 1 & 0 \\
0 & -1 & 0 & 0 \\
0 & 0 & 0 & 0 
\end{pmatrix},~~\gamma_{24}=
\begin{pmatrix}
0 & 0 & 0 & 0 \\
0 & 0 & 0 & 1  \\
0 & 0 & 0 & 0 \\
0 & -1 & 0 & 0 
\end{pmatrix},~~\gamma_{34}=
\begin{pmatrix}
0 & 0 & 0 & 0 \\
0 & 0 & 0 & 0 \\
0 & 0 & 0 & 1 \\
0 & 0 & -1 & 0 
\end{pmatrix}.  
\end{align}
Since $so(4)\simeq su(2)\oplus su(2)$, two independent sets of $SU(2)$ generators can be constructed with the $SO(4)$ generators $\tilde{\Gamma}_{ij}$ $(i,j,k=1,2,3)$: 
\begin{align}
\tilde{\Gamma}_i=-\frac{1}{4}\epsilon_{ijk}\tilde{\Gamma}_{jk}+\frac{1}{2}\tilde{\Gamma}_{i4},~~~~~~~\tilde{\Gamma}_i^{'}=-\frac{1}{4}\epsilon_{ijk}\tilde{\Gamma}_{jk}-\frac{1}{2}\tilde{\Gamma}_{i4},  
\end{align}
which satisfy 
\begin{align}
[\tilde{\Gamma}_i,\tilde{\Gamma}_j]=\epsilon_{ijk}\tilde{\Gamma}_k,~~~~~[\tilde{\Gamma}_i^{'},\tilde{\Gamma}_j^{'}]=\epsilon_{ijk}\tilde{\Gamma}_k^{'},~~~~[\tilde{\Gamma}_i,\tilde{\Gamma}_j^{'}]=0. 
\end{align}
Then, square of the radius of $N=4$ fuzzy four-supersphere is written as 
\begin{equation}
X_aX_a+C_{\alpha\beta}\sum_{l=1}^4 \Theta_{\alpha}^{(l)} \Theta_{\beta}^{(l)}+\sum_{i=1}^3Y_i Y_i+\sum_{i=1}^3 {Y}'_i {Y}'_i=(\Psi^{\dagger}\Psi)^2,  
\label{squareofradiusofs48}
\end{equation}
and the corresponding classical relation is 
\begin{equation}
x_ax_a+C_{\alpha\beta}\sum_{l=1}^4 \theta_{\alpha}^{(l)} \theta_{\beta}^{(l)}+\sum_{i=1}^3 y_i y_i+\sum_{i=1}^3 y_i' y_i'=(\psi^{\ddagger}\psi)^2=1. 
\end{equation}
The cancellation of the zero-point energy in (\ref{squareofradiusofs48}) suggests  equal numbers of bosonic and the fermionic states. Indeed,   
\begin{align}
&D_B=D(n)+6D(n-2)+D(n-4)=\frac{4}{3}n(n^2+2) ,\nonumber\\
&D_F=4D(n-1)+4D(n-3)=\frac{4}{3}n(n^2+2).   
\end{align}
 The total dimension is 
\begin{equation}
D_T=D_B+D_F= \frac{8}{3}n(n^2+2). 
\end{equation}

\section{Symmetry enhancement as quantum fluctuations}\label{Sectsupercoherent}

As discussed in Sec.\ref{subsubalgebrasu41sphere}, the algebraic structure of $N=1$ fuzzy four-supersphere is given by $su(4|1)$. 
In this section, we provide a physical interpretation of the $SU(4|1)$ structure by 
 evaluating quantum fluctuations of fuzzy two- and four-superspheres exemplified by correlation functions. The method is taken from Balachandran et al.\cite{hep-th/0203259,hep-th/0204170}.  We only discuss ${N}=1$ fuzzy two- and four-superspheres, but generalizations to more SUSY cases are straightforward.

\subsection{$N=1$ fuzzy two-supersphere}\label{subsectwopointtwosupersphere}

We first define the supercoherent state on ${N}=1$ fuzzy two-supersphere. 
With the coordinates  $X_i$ (\ref{SchRepF2sphere}) and $x_i$ (\ref{1stgradedHopf}), the super-coherent state,     
$|\omega\rangle$, is defined so as to satisfy  
\begin{equation}
(x_i X_i +\epsilon_{\alpha\beta}\theta_{\alpha}\Theta_{\beta})|\omega\rangle =n|\omega\rangle. 
\end{equation}
$|\omega\rangle$ is derived as  
\begin{equation}
|\omega\rangle=\frac{1}{\sqrt{n!}} (\Psi^{\dagger}\psi)^n|0\rangle=\frac{1}{\sqrt{n!}}(\psi_1\Psi_1^{\dagger}+\psi_2\Psi_2^{\dagger}-\eta \tilde{\Psi}^{\dagger})^n|0\rangle, 
\label{explicitomega12}
\end{equation}
where $\Psi$ is the graded Schwinger operator and $\psi$ the normalized spinor related to $X_i$, $\Theta_{\alpha}$ and $x_i$, $\theta_{\alpha}$ by  (\ref{SchRepF2sphere}) and (\ref{1stgradedHopf}) respectively\footnote{ $\Psi$ and $\psi$ respectively satisfy 
\begin{align}
&L_i\Psi\cdot X_i +\epsilon_{\alpha\beta} L_{\alpha}\Psi\cdot \Theta_{\beta}=\frac{1}{2}\Psi(\Psi^{\dagger}\Psi+1),\nonumber\\
&L_i\psi\cdot x_i +\epsilon_{\alpha\beta}L_{\alpha}\psi\cdot \theta_{\beta}=\frac{1}{2}\psi. 
\end{align}
}.
$|\omega\rangle$ is $n$th order polynomials expanded by the graded fully symmetric representation (\ref{UOSp12irredu}).  
$|\omega\rangle$ is normalized as 
\begin{equation}
\langle\!\langle \omega|\omega\rangle=1, 
\end{equation}
with dual state $\langle\!\langle\omega|$ given by 
\begin{equation}
\langle\!\langle \omega|=\frac{1}{\sqrt{n!}}\langle 0| (\psi^{\ddagger}\Psi)^n.
\end{equation}

The expectation values of $X_i$ and $\Theta_{\alpha}$ are calculated as 
\begin{equation}
\langle\!\langle \omega|X_i|\omega\rangle =nx_i,~~\langle\!\langle \omega|\Theta_{\alpha}|\omega\rangle =n\theta_{\alpha}. 
\end{equation}
Meanwhile, the correlation functions are   
\begin{align}
&\langle\!\langle \omega|X_i X_j|\omega \rangle=n^2 x_i x_j+ 4n \psi^{\ddagger}L_i P_- L_j\psi,\nonumber\\ 
&\langle\!\langle \omega|X_i \Theta_{\alpha}|\omega \rangle=n^2 x_i \theta_{\alpha}+4n \psi^{\ddagger}L_i P_- L_{\alpha}\psi,\nonumber\\ 
&\langle\!\langle \omega|\Theta_{\alpha} \Theta_{\beta}|\omega \rangle=n^2 \theta_{\alpha}\theta_{\beta}+4n\psi^{\ddagger}L_{\alpha} P_- L_{\beta}\psi,
\label{correlationfuncosp12}
\end{align}
where $P_{-}$ denotes a projection operator \footnote{ 
With $P_+=\psi\psi^{\ddagger}$, $P_-$ (\ref{defp-osp14}) satisfies the following relations, 
\begin{align}
&P_+\psi=\psi,~~~P_-\psi=0,\nonumber\\
&P_++P_-=1,~~~{P_{\pm}}^2=P_{\pm},~~~P_+P_-=P_-P_+=0. 
\end{align}
} 
\begin{equation}
P_{-}=1-\psi\psi^{\ddagger}. 
\label{defp-osp14}
\end{equation}
In the classical limit $n\rightarrow \infty$, the first terms of the order $n^2$ are dominant and the fuzzy two-supersphere is reduced to the ordinary commutative supersphere. 
The second terms of the order $n$ exhibit quantum fluctuations particular to fuzzy geometry. 
The second terms are evaluated as   
\begin{align}
&4\psi^{\ddagger}L_i P_- L_j\psi=-x_ix_j+i\epsilon_{ijk}x_k+\delta_{ij},\label{twopointfuzzytwosphere}\nonumber\\ 
&4\psi^{\ddagger}L_i P_- L_{\alpha}\psi=- x_i \theta_{\alpha}+\frac{1}{2} (\sigma_i)_{\beta\alpha} (\theta_{\beta}+\vartheta_{\beta} ),\nonumber\\ 
&4\psi^{\ddagger}L_{\alpha} P_- L_{\beta}\psi=-\theta_{\alpha}\theta_{\beta}+\frac{1}{2}(\epsilon\sigma_i)_{\alpha\beta}x_i+\frac{3}{2}\epsilon_{\alpha\beta}z-2\epsilon_{\alpha\beta}. 
\end{align} 
Here, $\vartheta_{\alpha}$ and $z$ are defined by 
\begin{equation}
\vartheta_{\alpha}=2\psi^{\ddagger}D_{\alpha}\psi,~~~~~z=\psi^{\ddagger} H\psi,  
\end{equation}
where $D_{\alpha}$ and $H$ are 
\begin{equation}
D_{\alpha}=\frac{1}{2}\begin{pmatrix}
0 & -\tau_{\alpha} \\
-(\epsilon\tau_{\alpha})^t & 0 
\end{pmatrix},~~~~H=\begin{pmatrix}
1 & 0 & 0 \\
0 & 1 & 0 \\
0 & 0 & 2
\end{pmatrix},  
\end{equation}
with $\tau_{1}=(1,0)^t$, $\tau_2=(0,1)^t$ 
and $\epsilon=i\sigma_2$.    
They are considered as ``new emerging coordinates'' by quantum fluctuation.    
The corresponding fuzzy coordinates of  $x_i$, $\theta_{\alpha}$, $\vartheta_{\alpha}$ and $z$ are $X_i$, $\Theta_{\alpha}$, $\varTheta_{\alpha}$ and $Z$ defined in (\ref{theschwingerconsatyp}) (see Appendix \ref{appendatyposp(2|2)}), and they amount to the $SU(2|1)$  algebra.  Thus, the hidden $SU(2|1)$ structure appears as quantum fluctuation of fuzzy two-supersphere.

\subsection{$N=1$ fuzzy four-supersphere}

With similar manner to Sec.\ref{subsectwopointtwosupersphere}, 
the supercoherent state on $N=1$ fuzzy four-supersphere, $|\omega\rangle$, is introduced as 
\begin{equation}
(x_a X_a +2C_{\alpha\beta}\theta_{\alpha}\Theta_{\beta})|\omega\rangle =n|\omega\rangle, 
\end{equation}
where  $X_a,\Theta_{\alpha}$ (\ref{bosonicscwingern1four}) and $x_a,\theta_{\alpha}$ (\ref{defxathetausp14}) are 
 coordinates on $S^{4|2}_F$ and $S^{4|2}$, respectively. 
Explicitly, $|\omega\rangle$ is given by 
\begin{equation}
|\omega\rangle=\frac{1}{\sqrt{n!}} (\Psi^{\dagger}\psi)^n|0\rangle=\frac{1}{\sqrt{n!}}(\psi_1\Psi_1^{\dagger}+\psi_2\Psi_2^{\dagger}+\psi_3\Psi_3^{\dagger}+\psi_4\Psi_4^{\dagger}-\eta \tilde{\Psi}^{\dagger})^n|0\rangle, 
\end{equation}
where $\Psi$ and $\psi$ are respectively the graded Schwinger operator (\ref{uosp14spinorschwinger}) and normalized spinor (\ref{ospfundspinor41})\footnote{$\Psi$ and $\psi$ satisfy 
\begin{align}
&\Gamma_a\Psi\cdot X_a +2C_{\alpha\beta} \Gamma_{\alpha}\Psi\cdot \Theta_{\beta}=\Psi(\Psi^{\dagger}\Psi+3),\nonumber\\
&\Gamma_a\psi\cdot x_a +2C_{\alpha\beta}\Gamma_{\alpha}\psi\cdot \theta_{\beta}=\psi. 
\end{align}
}. 
$|\omega\rangle$ can be expanded by the graded fully symmetric representation (\ref{uosp14fullysymstates}).   
The dual state   
$\langle\!\langle\omega|$ satisfying 
\begin{equation}
\langle\!\langle \omega|\omega\rangle=1, 
\end{equation}
is given by 
\begin{equation}
\langle\!\langle \omega|=\frac{1}{\sqrt{n!}}\langle 0| (\psi^{\ddagger}\Psi)^n.
\end{equation}

The expectation values of $X_a$ and $\Theta_{\alpha}$ are 
\begin{equation}
\langle\!\langle \omega|X_a|\omega\rangle =nx_a,~~~~~~\langle\!\langle \omega|\Theta_{\alpha}|\omega\rangle =n\theta_{\alpha}. 
\end{equation}
The correlation functions are 
\begin{align}
&\langle\!\langle \omega|X_a X_b|\omega \rangle=n^2 x_a x_b+ n \psi^{\ddagger}\Gamma_a P_- \Gamma_b\psi,\nonumber\\ 
&\langle\!\langle \omega|X_a \Theta_{\alpha}|\omega \rangle=n^2 x_a \theta_{\alpha}+n \psi^{\ddagger}\Gamma_a P_- L_{\alpha}\psi,\nonumber\\ 
&\langle\!\langle \omega|\Theta_{\alpha} \Theta_{\beta}|\omega \rangle=n^2 \theta_{\alpha}\theta_{\beta}+n\psi^{\ddagger}\Gamma_{\alpha} P_- \Gamma_{\beta}\psi,
\label{correlationsfuncsxxosp14}
\end{align}
where 
\begin{equation}
P_{-}=1-\psi\psi^{\ddagger}. 
\end{equation}
The second terms of the right-hand side of (\ref{correlationsfuncsxxosp14}) are calculated as  
\begin{align}
&\psi^{\ddagger}\Gamma_aP_-\Gamma_b\psi=-x_a x_b +2i x_{ab} +\delta_{ab}z+\frac{4}{3}\delta_{ab},\nonumber\\ 
&\psi^{\ddagger}\Gamma_aP_-\Gamma_{\alpha}\psi=-x_a\theta_{\alpha}+\frac{1}{2}(\gamma_a)_{\beta\alpha}(\theta_{\beta}+\vartheta_{\beta}),\nonumber\\ 
&\psi^{\ddagger}\Gamma_{\alpha}P_-\Gamma_{\beta}\psi=-\theta_{\alpha}\theta_{\beta}-\frac{1}{2}\sum_{a<b}(C\gamma_{ab})_{\alpha\beta}x_{ab}-\frac{1}{8}(C\gamma_a)_{\alpha\beta}x_a+\frac{5}{24}C_{\alpha\beta}z-\frac{1}{3}C_{\alpha\beta}. 
\end{align}
Thus, for fuzzy four-supersphere, we have new coordinates, $x_{ab}$, $\vartheta_{\alpha}$, $z$,  defined by 
\begin{equation}
x_{ab}=\psi^{\ddagger}\Gamma_{ab}\psi,~~~~\vartheta_{\alpha}=\psi^{\ddagger}D_{\alpha}\psi, ~~~~z=\psi^{\ddagger}H \psi. 
\end{equation}
Here, $\Gamma_{ab}$, $D_{\alpha}$ and $H$ are respectively (\ref{gammablarge}), (\ref{defdalphasu41}) and (\ref{gamma5times5}).  
The fuzzy coordinates corresponding to twenty four coordinates, $x_a$, $\theta_{\alpha}$, $x_{ab}$, $\vartheta_{\alpha}$, $z$, are, $X_a$,  $\Theta_{\alpha}$, $X_{ab}$, $\varTheta_{\alpha}$, $Z$, defined in Sec.\ref{subsubalgebrasu41sphere}, which satisfy the $SU(4|1)$ algebra (see also Appendix \ref{append:su41}).   
Thus, we confirmed, similar to the fuzzy two-supersphere case, the enhanced $SU(4|1)$ structure is brought by quantum fluctuation of fuzzy four-supersphere.

\section{Summary}\label{SectSummary}

 We performed a systematic study of fuzzy superspheres and graded Hopf maps based on  $UOSp(N|2)$ and  $UOSp(N|4)$, respectively.  
 For the positive definiteness of square of the radius, the construction of fuzzy two-superspheres is restricted to ${N}=1,2$, and  fuzzy four-superspheres to ${N}=1,2,3, 4$ (see Tables \ref{fuzzytwosuperspheres} and \ref{fuzzyfoursuperspheres}).  The graded Hopf maps were introduced as the classical counterpart of the fuzzy superspheres. 
We derived an explicit realization of the 1st and 2nd graded Hopf maps: 
\begin{equation}
 S^{3|2N}~\overset{S^1}\longrightarrow ~S^{2|2N},~~~~~~~~~~~S^{7|2N}~\overset{S^3}\longrightarrow ~ S^{4|2N}.
\end{equation}
The particular feature of the present construction is based on the super Lie algebraic structures.     
With use of the graded Schwinger operators, super Lie group symmetries are naturally incorporated and the graded fully  symmetric representation is readily derived. 
Adoption of the graded fully symmetric representation brings a 
 particular feature to fuzzy superspheres: fuzzy superspheres are represented as a ``superposition'' of fuzzy superspheres with lower supersymmetries.    
The algebras of the fuzzy two- and four-superspheres are enhanced from the original algebras,  $uosp(N|2)$ and $uosp(N|4)$, to the larger algebras, $su(2|N)$ and  $su(4|N)$, respectively.  
We also argued such enhancement in view of quantum fluctuation of fuzzy spheres by evaluating correlation functions.  

\begin{table}
\renewcommand{\arraystretch}{1.2}
\centering \vspace{1mm}
\begin{tabular}{|c|c|c|c|}
\hline
 Fuzzy manifold & $S_F^2$ & $S^{2|2}_F$ & $S_{F}^{2|4}$ \\
 \hline  Number of supersymmetry  &  ${N}=0$  & ${N}=1$ &  ${N}=2$
\\
 \hline
Original symmetry &  $SO(3)$ &  $UOSp(1|2)$ & $UOSp(2|2)$
\\ \hline
Enhanced symmetry &  $SU(2)$ &  $SU(2|1)$ & $SU(2|2)$
\\ \hline
 Square of the radius & ${n(n+2)}$  &   ${n(n+1)}$ & $n^2$
\\
\hline
\end{tabular} 
\caption{Fuzzy two-superspheres and symmetries. }\label{fuzzytwosuperspheres}
\end{table}
\begin{table}
\renewcommand{\arraystretch}{1.2}
\centering 
\vspace{1mm}
\begin{tabular}{|c|c|c|c|c|c|}
\hline
 Fuzzy manifold & $S_F^4$ & $S^{4|2}_F$ & $S_{F}^{4|4}$ & $S^{4|6}_F$ & $S^{4|8}_F$ \\
 \hline  Number of supersymmetry  &  ${N}=0$  & ${N}=1$ &  ${N}=2$ & ${N}=3$  & ${N}=4$
\\
 \hline
Original symmetry &  $SO(5)$ &  $UOSp(1|4)$ & $UOSp(2|4)$ & $UOSp(3|4)$ & $UOSp(4|4)$
\\ \hline
Enhanced symmetry &  $SU(4)$ &  $SU(4|1)$ & $SU(4|2)$ & $SU(4|3)$ & $SU(4|4)$
\\\hline
 Square of the radius & ${n(n+4)}$  &   ${n(n+3)}$ &   ${n(n+2)}$  &   ${n(n+1)}$ & $n^2$
\\
\hline
\end{tabular}\caption{Fuzzy four-superspheres and symmetries. }\label{fuzzyfoursuperspheres}
\end{table}

Since the present work is a natural generalization of precedent low dimensional fuzzy superspheres, one could pursue similar applications performed in low dimensions, such as  realization in string theory, construction of supersymmetric gauge theories on fuzzy superspheres.  
Applications to topologically non-trivial many-body models would be interesting, too. 
The Hopf maps have applications in many branches of physics \cite{Nakaharabook} and also in quantum computation  \cite{Mosseri2001}. It may be intriguing to see the roles of the graded Hopf maps in the context of superqubits \cite{2010Duff}.   

In this work, we focused on the construction of fuzzy supersphere whose bosonic dimension is two or four. This is because of the restriction of isomorphism between unitary-symplectic and orthogonal groups,   
$USp(2)\simeq SO(3)$,  $USp(4)\simeq SO(5)$. 
At the present, we do not know how to generalize the present construction to even higher dimensions.  
Another remaining mathematical issue we have not fully discussed is the bundle structure of the graded Hopf maps.  
At least, these may deserve further investigations.   

\section*{Acknowledgment}

The author would like to thank Professor Harald Grosse for helpful email correspondences and Professor Christian Fronsdal for useful discussions at Miami 2010 conference. 
This research was partially supported by a Grant-in-Aid for Scientific Research  from the Ministry of Education, Science, Sports and Culture of Japan (Grant No.23740212). 

\section*{Appendix}

\appendix

\section{$SU(M|N)$ algebra and fuzzy complex projective superspace }

We summarize formulae about $SU(M|N)$ algebra. 
The dimension is 
\begin{equation}
\dim[su(M|N)]=M^2+N^2-1|2MN=M^2+2MN+N^2-1. 
\end{equation}
The maximal bosonic subalgebra of $su(M|N)$ is $su(M)\oplus su(N)\oplus u(1)$, and 
its fundamental representation matrices are given by  
\begin{align}
S_A=
\begin{pmatrix}
s_A & 0 \\
0 & 0 
\end{pmatrix},
~~~~T_{P}=
\begin{pmatrix}
0 & 0 \\
0 & t_{P}
\end{pmatrix},~~~~H=\frac{1}{N}
\begin{pmatrix}
N \cdot 1_M & 0 \\
0 & M\cdot 1_N
\end{pmatrix}, 
\label{sumngeneratorsmatrices}
\end{align}
with $A=1,2,\cdots,M^2-1$ and $P=1,2,\cdots,N^2-1$.  $s_A$ and $t_P$ in (\ref{sumngeneratorsmatrices}) satisfy  
\begin{equation}
[s_A,s_B]=if_{ABC}s_C,~~~~~~~[t_{P},t_{Q}]=if'_{PQR}t_{R}, 
\end{equation}
with $SU(M)$ and $SU(N)$ structure constants, $f_{ABC}$ and $f'_{PQR}$. 
The fermionic generators are 
\begin{equation}
Q_{\alpha\sigma}=\begin{pmatrix}
0_{M+\sigma-1} & \tau_{\alpha} & 0 \\
0 & 0 & 0 \\
0 & 0 & 0_{N-\sigma} \\
\end{pmatrix},~~~~~~~\tilde{Q}_{\sigma\alpha}=
\begin{pmatrix}
0_{M+\sigma-1} & 0 & 0  \\ 
\tau_{\alpha}^t & 0 & 0 \\
0  & 0 & 0_{N-\sigma} \\
\end{pmatrix}, \label{Qmatrices}
\end{equation}
where $\alpha$ stands for the $SU(M)$ spinor index 
$(\alpha=1,2,\cdots,M)$, and $\sigma$ does the $SU(N)$ index $(\sigma=1,2,\cdots,N)$, and 
\begin{equation}
\tau_{\alpha}=({0,\cdots, 0,} \overset{\alpha}{\check{1}}, {0, \cdots, 0})^t.  
\end{equation}
Therefore, the only non-zero components of $Q_{\alpha\sigma}$ and $\tilde{Q}_{\sigma\alpha}$ (\ref{Qmatrices}) are, $(\alpha, M+\sigma)$ and $(M+\sigma,\alpha)$, respectively:   
\begin{equation}
(Q_{\alpha\sigma})_{\beta\tau}=\delta_{\alpha\beta}\delta_{M+\sigma,\tau},~~~~~~(\tilde{Q}_{\sigma\alpha})_{\beta\tau}=\delta_{\beta,M+\sigma}\delta_{\tau \alpha}. 
\end{equation}
Then, 
\begin{equation}
(Q_{\alpha\sigma})^t=\tilde{Q}_{\sigma\alpha}.
\end{equation}
The $SU(M|N)$ algebra is given by  
\begin{align}
&[S_A,S_B]=if_{ABC}S_C,~~~~~~~[S_A,Q_{\alpha\sigma}]=(s_A)_{\beta\alpha}Q_{\beta\sigma},~~~~~~~~~~~[S_A,\tilde{Q}_{\sigma\alpha}]=-(s_A)_{\alpha\beta}\tilde{Q}_{\sigma\beta},\nonumber\\
&[S_A,T_{P}]=0,~~~~~~~~~~~~~~~~~\{Q_{\alpha\sigma},Q_{\beta\tau}\}=\{\tilde{Q}_{\sigma\alpha},\tilde{Q}_{\tau\beta}\}=0,\nonumber\\
&\{Q_{\alpha\sigma},\tilde{Q}_{\tau\beta}\}=2\delta_{\sigma\tau} (s_A)_{\beta\alpha}S_A +2\delta_{\alpha\beta}(t_{P})_{\sigma\tau}T_{P} +\frac{1}{M}\delta_{\sigma\tau}\delta_{\alpha\beta}H,\nonumber\\
&[Q_{\alpha\sigma},T_{P}]=(t_{P})_{\sigma\tau}Q_{\alpha\tau},~~~~~[\tilde{Q}_{\sigma\alpha},T_{P}]=-(t_{P})_{\tau\sigma}\tilde{Q}_{\tau\alpha},\nonumber\\
&[T_{P},T_{Q}]=if'_{PQR}T_{R}, ~~~~~~~~[S_A,\Gamma]=[T_{P},H]=0,\nonumber\\
&[Q_{\alpha\sigma},H]=\frac{M-N}{N}Q_{\alpha\sigma},~~~~[\tilde{Q}_{\sigma\alpha},H]=-\frac{M-N}{N}\tilde{Q}_{\sigma\alpha},  
\label{SUMNcommutationrelations}
\end{align}
and the Casimir is  
\begin{equation}
\mathcal{C}=2\sum_{A=1}^{M^2-1}{S}_A{S}_A-\sum_{\alpha=1}^M\sum_{\sigma=1}^{N}({Q}_{\alpha\sigma}{\tilde{Q}}_{\sigma\alpha}-{\tilde{Q}}_{\sigma\alpha}{Q}_{\alpha\sigma})-2\sum_{P=1}^{N^2-1}{T}_P{T}_P- \frac{N}{M(M-N)}H^2. 
\label{casimirsumn}
\end{equation}
 We apply the Schwinger construction to  
$X=S_A, Q_{\alpha\sigma},\tilde{Q}_{\sigma\alpha},T_P$, $H$:  
\begin{align}
\hat{X}=\Psi^{\dagger}X\Psi, 
\label{cpfuzzyconst}
\end{align}
where 
\begin{equation}
\Psi=
(
\Psi_1,
\Psi_2,
\cdots,
\Psi_M,
\tilde{\Psi}_1, 
\tilde{\Psi}_2, 
\cdots,
\tilde{\Psi}_N)^t,   
\end{equation}
satisfying 
\begin{equation}
[\Psi_{\alpha},\Psi_{\beta}^{\dagger}]=\delta_{\alpha\beta},~~~~\{\tilde{\Psi}_{\sigma},\tilde{\Psi}^{\dagger}_{\tau}\}=\delta_{\sigma\tau},~~~~[\Psi_{\alpha},\tilde{\Psi}_{\sigma}]=0. 
\end{equation}
Inserting (\ref{cpfuzzyconst}) to (\ref{casimirsumn}), the Casimir is expressed as 
\begin{equation}
\mathcal{C}=\frac{M-N-1}{M-N}\hat{n}(\hat{n}+M-N),  
\label{cppqcasimir}
\end{equation}
with $\hat{n}=\Psi^{\dagger}\Psi$. Here, we used 
\begin{align}
&\sum_{A=1}^{M^2-1}{\hat{S}}_A{\hat{S}}_A=\frac{M-1}{2M}\hat{n}_B(\hat{n}_B+M),\nonumber\\
&\sum_{P=1}^{N^2-1}{\hat{T}}_{P}{\hat{T}}_{P}=-\frac{N+1}{2N}\hat{n}_F(\hat{n}_F-N),\nonumber\\
&\sum_{\alpha=1}^M\sum_{\sigma=1}^{N}(\hat{Q}_{\alpha\sigma}\hat{\tilde{Q}}_{\sigma\alpha}-\hat{\tilde{Q}}_{\sigma\alpha}\hat{Q}_{\alpha\sigma})=N \hat{n}_B-2\hat{n}_B\hat{n}_F-M \hat{n}_F,\nonumber\\
&\hat{H}^2=\frac{1}{N^2}(N\hat{n}_B+M\hat{n}_F)^2, 
\end{align}
with  $\hat{n}_B=\sum_{\alpha=1}^M \Psi^{\dagger}_{\alpha}\Psi_{\alpha}$, $\hat{n}_F=\sum_{\sigma=1}^N \tilde{\Psi}^{\dagger}_{\sigma}\tilde{\Psi}_{\sigma}$.   
The Casimir eigenvalues are regarded as the square of the radius of fuzzy complex projective superspaces, $\mathbb{C}P^{M-1|N}_F$. Notice that the coefficient of the right-hand side of (\ref{cppqcasimir}) vanishes for $M=N+1$, and is not well defined for $M=N$. 
In such cases, taking away of the common vanishing or divergent coefficient, 
we may regard the square of the radius of $\mathbb{C}P^{M-1|N}_F$ as  
\begin{equation}
n(n+M-N). 
\end{equation}

The classical counterpart of (\ref{cpfuzzyconst}) reads as  
\begin{equation}
x=\psi^{\dagger}X\psi, 
\label{sqpqhopfmap}
\end{equation}
where $\psi$ is a normalized $SU(M|N)$ spinor $\psi=(\psi_1,\psi_2,\cdots,\psi_M,\tilde{\psi}_1,\tilde{\psi}_2,\cdots,\tilde{\psi}_N)^t$ with $\psi^{\dagger}\psi=1$  regarded as coordinates on $S^{2M-1|2N}$.  $U(1)$ phase of $\psi$ is canceled in (\ref{sqpqhopfmap}), and thus (\ref{sqpqhopfmap}) signifies a generalized graded 1st Hopf map,  
\begin{equation}
S^{2M-1|2N} \overset{S^1}\longrightarrow \mathbb{C}P^{M-1|N}. 
\end{equation}
See  Ref.\cite{arXiv:hep-th/0311159} for more details about  $\mathbb{C}P^{M-1|N}_F$.

\subsection{$SU(2|N)$ algebra and $\mathbb{C}P_F^{1|N}$}

The dimension of the $SU(2|N)$ algebra is 
\begin{equation}
\\dim [su(2|N)]=N^2+3|4N=N^2+4N+3. 
\end{equation}
The bosonic generators (\ref{sumngeneratorsmatrices}) are given by 
\begin{equation}
L_i=\frac{1}{2}
\begin{pmatrix}
\sigma_i & 0 \\
0 & 0 
\end{pmatrix}, ~~~T_P=\begin{pmatrix}
0 & 0 \\
0 & t_P
\end{pmatrix},~~~~H=\frac{1}{N}\begin{pmatrix}
N\cdot 1_2 & 0 \\
0 & 2\cdot 1_N
\end{pmatrix},   
\label{su2n33mat}
\end{equation}
which respectively correspond to  $SU(2)$, $SU(N)$ and $U(1)$ generators. To clarify relations to the subalgebra 
$uosp(N|2)$, we separate the $SU(N)$ generators into symmetric and antisymmetric matrices: 
\begin{equation}
{T_S}^t=T_S,~~~~~~~{T_I}^t=-T_I,
\end{equation}
with $S=1,2,\cdots,N(N+1)/2-1$ and $I=1,2,\cdots,N(N-1)/2$. Note $T_I$ are pure imaginary antisymmetric matrices that satisfy the $SO(N)$ algebra by themselves. 
Instead of $Q_{\alpha\sigma}$ and $\tilde{Q}_{\sigma\alpha}$ (\ref{Qmatrices}), we introduce  
the following fermionic generators  
\begin{align}
&L_{\alpha\sigma}=\frac{1}{2}
\begin{pmatrix}
0_{1+\sigma} & \tau_{\alpha} & 0  \\
-(\epsilon\tau_{\alpha})^t & 0 & 0 \\
0 & 0 & 0_{N-\sigma}  
\end{pmatrix},~~~D_{\alpha\sigma}=\frac{1}{2}
\begin{pmatrix}
0_{1+\sigma} & -\tau_{\alpha} & 0  \\
-(\epsilon\tau_{\alpha})^t & 0 & 0 \\
0 & 0 & 0_{N-\sigma}  
\end{pmatrix}, 
\label{lalphasigmadalphasigmasu2n}
\end{align}
with $\epsilon=i\sigma_2$. $L_{\alpha\sigma}$ and $D_{\alpha\sigma}$ are related to $Q_{\alpha\sigma}$ and $\tilde{Q}_{\sigma\alpha}$ as 
\begin{equation}
Q_{\alpha\sigma}=L_{\alpha\sigma}-D_{\alpha\sigma},~~~~~\tilde{Q}_{\sigma\alpha}=-\epsilon_{\alpha\beta}(L_{\beta\sigma}+D_{\beta\sigma}),  
\end{equation}
or 
\begin{equation}
L_{\alpha\sigma}=\frac{1}{2}(Q_{\alpha\sigma}+\epsilon_{\alpha\beta}\tilde{Q}_{\sigma\beta}),~~~~~D_{\alpha\sigma}=-\frac{1}{2}(Q_{\alpha\sigma}-\epsilon_{\alpha\beta}\tilde{Q}_{\sigma\beta}). 
\end{equation}
Therefore, 
\begin{equation}
Q_{\alpha\sigma}\tilde{Q}_{\sigma\alpha}-\tilde{Q}_{\sigma\alpha}Q_{\alpha\sigma}=-2\epsilon_{\alpha\beta}(L_{\alpha\sigma}L_{\beta\sigma}-D_{\alpha\sigma}D_{\beta\sigma}). 
\end{equation}
$L_{\alpha\sigma}$ and $D_{\alpha\sigma}$ act as $UOSp(N|2)$ spinor as we shall see below.  
With $L_{\alpha\sigma}$ and $D_{\alpha\sigma}$, the $SU(2|N)$ algebra is rewritten as 
\begin{align}
&[L_i,L_j]=i\epsilon_{ijk}L_k,\nonumber\\
&[L_i,L_{\alpha\sigma}]=\frac{1}{2}(\sigma_i)_{\beta\alpha}L_{\beta\sigma},~~~~~[L_i,D_{\alpha\sigma}]=\frac{1}{2}(\sigma_i)_{\beta\alpha}D_{\beta\sigma},\nonumber\\
&\{L_{\alpha\sigma},L_{\beta\tau}\}=-\{D_{\alpha\sigma},D_{\beta\tau}\}=
\frac{1}{2}\delta_{\sigma\tau}(\epsilon\sigma_i)_{\alpha\beta}L_i-\epsilon_{\alpha\beta} (t_{I})_{\sigma\tau}T_{I},\nonumber\\
&\{L_{\alpha\sigma},D_{\beta\tau}\}=-\epsilon_{\alpha\beta}(t_{S})_{\sigma\tau} T_{S}-\frac{1}{4}\epsilon_{\alpha\beta}\delta_{\sigma\tau}H,\nonumber\\
&[L_{\alpha\sigma},T_{S}]=-(t_{S})_{\sigma\tau}D_{\alpha\tau},~~~~~[L_{\alpha\sigma},T_{I}]=(t_{I})_{\sigma\tau}L_{\alpha\tau},\nonumber\\
&[D_{\alpha\sigma},T_{S}]=-(t_{S})_{\sigma\tau}L_{\alpha\tau},~~~~~[D_{\alpha\sigma},T_{I}]=(t_{I})_{\sigma\tau}D_{\alpha\tau},\nonumber\\
&[L_{\alpha\sigma},H]=-\frac{2-N}{N}D_{\alpha\sigma},~~~~~[D_{\alpha\sigma},H]=-\frac{2-N}{N}L_{\alpha\sigma},\nonumber\\
&[T_P,T_Q]=if'_{PQR}T_R, 
\label{algebrasu2n}
\end{align}
where $t_S$ and $t_I$ are respectively $N\times N$ symmetric and antisymmetric matrices of $SU(N)$ generators $t_P$ (\ref{sumngeneratorsmatrices}). 
From (\ref{algebrasu2n}), one may see that $L_i$, $L_{\alpha\sigma}$, $T_I$ satisfy a closed subalgebra, the  $uosp(N|2)$. 

We introduce the fuzzy coordinates of $\mathbb{C}P_F^{1|N}$ as   
\begin{align}
&X_i=2\Psi^{\dagger}L_i\Psi,~~~~~~\Theta_{\alpha}^{(\sigma)}= 2\Psi^{\dagger}L_{\alpha\sigma}\Psi,~~~~~\varTheta_{\alpha}^{(\sigma)}=2\Psi^{\dagger}D_{\alpha\sigma}\Psi,\nonumber\\
&Y_{P}=2\Psi^{\dagger}T_{P}\Psi,~~~~~Z=\Psi^{\dagger}H\Psi, 
\end{align}
with the Schwinger operator $\Psi=(\Psi_1,\Psi_2,\tilde{\Psi}_1,\cdots,\tilde{\Psi}_N)^t$.  
Square of the radius of $\mathbb{C}P_F^{1|N}$ is derived as   
\begin{align}
&\sum_{i=1}^3 X_iX_i+\sum_{\alpha,\beta=1}^{2}\sum_{\sigma=1}^N \epsilon_{\alpha\beta}(\Theta_{\alpha}^{(\sigma)}\Theta_{\beta}^{(\sigma)}-\varTheta_{\alpha}^{(\sigma)}\varTheta_{\beta}^{(\sigma)})-\sum_{P=1}^{N^2-1}Y_{P}Y_{P}-\frac{N}{2-N}Z^2\nonumber\\
&=\frac{2(1-N)}{2-N}\hat{n}(\hat{n}+2-N), 
\end{align}
where $\hat{n}=\Psi^{\dagger}\Psi$. Here, we used 
\begin{align}
&X_iX_i=\hat{n}_B(\hat{n}_B+2),\nonumber\\
&\epsilon_{\alpha\beta}(\Theta_{\alpha}^{(\sigma)}\Theta_{\beta}^{(\sigma)}-\varTheta_{\alpha}^{(\sigma)}\varTheta_{\beta}^{(\sigma)})=-2N\hat{n}_B+4\hat{n}_B\hat{n}_F+4\hat{n}_F,\nonumber\\
&Y_{P}Y_{P}=-\frac{2(N+1)}{N}\hat{n}_F(\hat{n}_F-N),\nonumber\\
&Z^2=(\hat{n}_B+\frac{2}{N}\hat{n}_F)^2, 
\end{align}
with $\hat{n}_B=\Psi^{\dagger}_1\Psi_1+\Psi^{\dagger}_2\Psi_2$ and $\hat{n}_F=\sum_{\sigma=1}^N\tilde{\Psi}^{\dagger}_{\sigma}\tilde{\Psi}_{\sigma}$. 
For $\mathbb{C}P^{1|N}_F$, square of the radius is proportional to  
\begin{equation}
 {n}({n}+2-N).   
 \label{propsquarecp1n}
\end{equation}
Notice that, for $n< N-2$, (\ref{propsquarecp1n}) becomes negative.   This situation is similar to $S_F^{2|2N}$ (see the discussions below (\ref{classicaln2twosphere})). 

\subsubsection{$su(2|1)$}\label{appendatyposp(2|2)} 

The dimension of the $SU(2|1)$ algebra is 
\begin{equation}
\dim [su(2|1)]=4|4=8. 
\end{equation}
From (\ref{algebrasu2n}), the $SU(2|1)$ algebra reads as   
\begin{align}
&[L_i,L_j]=i\epsilon_{ijk}L_k,~~~~[L_i,L_{\alpha}]=\frac{1}{2}(\sigma_i)_{\beta\alpha}L_{\beta},~~~~[L_i,D_{\alpha}]=\frac{1}{2}(\sigma_i)_{\beta\alpha}D_{\beta},\nonumber\\
&\{L_{\alpha},L_{\beta}\}=-\{D_{\alpha},D_{\beta}\}=\frac{1}{2}\delta_{\sigma\tau}(\epsilon\sigma_i)_{\alpha\beta}L_i,~~~~~\{L_{\alpha},D_{\beta}\}=-\frac{1}{4}\epsilon_{\alpha\beta}H,\nonumber\\
&[L_{\alpha},H]=-D_{\alpha},~~~~~~~[D_{\alpha},H]=-L_{\alpha}. 
\label{su21algebraex}
\end{align}
The $SU(2|1)$ algebra (\ref{su21algebraex}) is isomorphic to the $UOSp(2|2)$ algebra (\ref{osp22algebranew}) with the identification $(L_{\alpha},iD_{\alpha})=L_{\alpha\sigma}$ and $H=2i\Gamma$.  
The maximal bosonic subalgebra of $su(2|1)$ is $su(2)\oplus u(1)$. 
The $uosp(1|2)$ is realized as the subalgebra by $L_i$ and $L_{\alpha}$ in (\ref{su21algebraex}). 
The $SU(2|1)$ irreducible representation is specified by ``superspin'' indices  $j$ (integers or half-integers) and $g$ (complex value). For details, see Refs.\cite{Scheunert1977,Marcu1980}. 
$SU(2|1)$ has two Casimirs, quadratic and cubic. The quadratic Casimir is given by      
\begin{equation}
\mathcal{C}=L_iL_i +\epsilon_{\alpha\beta}L_{\alpha}L_{\beta}-\epsilon_{\alpha\beta}D_{\alpha}D_{\beta}-\frac{1}{4}H^2, \label{quadraticsu21cas}
\end{equation}
and its eigenvalues are
\begin{equation}
\mathcal{C}=j^2-g^2. 
\label{eigensu21quadaty}
\end{equation}

\begin{itemize}
\item Atypical representation
\end{itemize}

When $g=\pm j$, the irreducible representation is called atypical representation. Since there is automorphism, $D_{\alpha}\rightarrow -D_{\alpha}$ and $H\rightarrow -H$ in (\ref{su21algebraex}),  we discuss only the case $g=+j$.  
The dimension of the atypical representation is $4j+1$,  
which is  already irreducible for the subgroup $UOSp(1|2)$. 
In the present case, the quadratic Casimir eigenvalues (\ref{eigensu21quadaty}) vanish  identically. 
(Also, the cubic Casimir eigenvalues vanish since the eigenvalues are proportional to $g(j^2-g^2)$ \cite{Scheunert1977}.) 
Thus, the two Casimirs do not specify atypical representation.
 The fundamental representation of $su(2|1)$ is the simplest atypical representation given by the following $3\times 3$ matrices\footnote{$D_{\alpha}$ and $H$ in (\ref{simplestospmat}) are constructed as 
\begin{equation}
D_{\alpha}=-\frac{1}{2(j+\frac{1}{4})}(\sigma_i)_{\beta\alpha}\{L_i,L_{\beta}\},~~~~~~~
H=\frac{1}{j+\frac{1}{4}}\biggl(\epsilon_{\alpha\beta}L_{\alpha}L_{\beta}+2j\biggl(j+\frac{1}{2}\biggr)\biggr),\label{gamma}
\end{equation}
with $j=1/2$. }:  
\begin{equation}
L_i=\frac{1}{2}\begin{pmatrix}
\sigma_i & 0 \\
0 & 0 
\end{pmatrix},~~L_{\alpha}=\frac{1}{2}\begin{pmatrix}
0_2 & \tau_{\alpha} \\
-(\epsilon\tau_{\alpha})^t & 0 
\end{pmatrix}, ~~D_{\alpha}=\frac{1}{2}\begin{pmatrix}
0_2 & -\tau_{\alpha} \\
-(\epsilon\tau_{\alpha})^t & 0 
\end{pmatrix},~~H=
\begin{pmatrix}
1 & 0 & 0 \\
0 & 1 & 0 \\
0 & 0 & 2 
\end{pmatrix},  \label{simplestospmat}
\end{equation}
where $\epsilon=i\sigma_2$, $\tau_1=(1,0)^t$ and $\tau_2=(0,1)^t$.  For the matrices (\ref{simplestospmat}), one may readily check that $\mathcal{C}$ (\ref{quadraticsu21cas}) vanishes. 
In the Schwinger construction 
\begin{align}
X_i=2\Psi^{\dagger}L_i \Psi,~~~~\Theta_{\alpha}=2\Psi^{\dagger}L_{\alpha} \Psi,~~~~\varTheta_{\alpha}=2\Psi^{\dagger}D_{\alpha} \Psi,~~~~Z=\Psi^{\dagger}H \Psi,  
\label{theschwingerconsatyp}
\end{align}
 $UOSp(1|2)$ invariant quantities are given by    
\begin{equation}
X_i X_i+\epsilon_{\alpha\beta}\Theta_{\alpha}\Theta_{\beta}=\epsilon_{\alpha\beta}\varTheta_{\alpha}\varTheta_{\beta}+Z^2=\hat{n}(\hat{n}+1),   
\end{equation}
where $\hat{n}=\Psi^{\dagger}\Psi$. 
Then, the $SU(2|1)$ Casimir vanishes identically: 
\begin{equation}
X_i X_i+\epsilon_{\alpha\beta}\Theta_{\alpha}\Theta_{\beta}-\epsilon_{\alpha\beta}\varTheta_{\alpha}\varTheta_{\beta}-Z^2=0, 
\end{equation}
which implies that the Schwinger construction corresponds to  atypical representation with $g=j=n/2$.

\begin{itemize}
\item Typical representation
\end{itemize}

The typical representation refers to $g\neq \pm j$.  
The simplest matrices of the typical representation are the following $4\times 4$ matrices\footnote{The typical representation  matrices (\ref{typicalrepre4times4})  are superficially different from those in Ref.\cite{Scheunert1977}:  
\begin{align}
L_i=\frac{1}{2}
\begin{pmatrix}
\sigma_i & 0  \\
0 & 0_2
\end{pmatrix},~~~L_{\alpha}=\frac{1}{2}
\begin{pmatrix}
0_2 & \tau_{\alpha} & 0 \\
-(\epsilon\tau_{\alpha})^t & 0 & 0  \\
0 & 0 & 0 
\end{pmatrix},~~~D'_{\alpha}=\frac{1}{2}
\begin{pmatrix}
0_2& 0  & -\tau_{\alpha}  \\
0 & 0 & 0  \\
-(\epsilon\tau_{\alpha})^t & 0 & 0 
\end{pmatrix},~~~\Gamma=\frac{1}{2}\begin{pmatrix}
0_2 & 0 \\
0 & \sigma_1
\end{pmatrix}.
\end{align}
In that case, the corresponding quadratic Casimir is given by  
\begin{equation}
\mathcal{C}={L}_i{L}_i+\epsilon_{\alpha\beta}{L}_{\alpha}{L}_{\beta}-\epsilon_{\alpha\beta}{D}'_{\alpha}{D}'_{\beta}-{\Gamma}^2.
\end{equation}
} 
\begin{align}
L_i=\frac{1}{2}
\begin{pmatrix}
\sigma_i & 0  \\
0 & 0_2
\end{pmatrix},~~~L_{\alpha}=\frac{1}{2}
\begin{pmatrix}
0_2 & \tau_{\alpha} & 0 \\
-(\epsilon\tau_{\alpha})^t & 0 & 0  \\
0 & 0 & 0 
\end{pmatrix},~~~L'_{\alpha}=\frac{1}{2}
\begin{pmatrix}
0_2& 0  & \tau_{\alpha}  \\
0 & 0 & 0  \\
-(\epsilon\tau_{\alpha})^t & 0 & 0 
\end{pmatrix},~~~\Gamma=\frac{1}{2}\begin{pmatrix}
0_2 & 0 \\
0 & \epsilon
\end{pmatrix}. 
\label{typicalrepre4times4}
\end{align}
In the Schwinger construction with $\Psi=(\Psi_1,\Psi_2,\tilde{\Psi}_1,\tilde{\Psi}_2)^t$, 
\begin{equation}
\hat{L}_i=\Psi^{\dagger}L_i\Psi,~~~~\hat{L}_{\alpha}=\Psi^{\dagger}L_{\alpha}\Psi,~~~~
\hat{L}'_{\alpha}=\Psi^{\dagger}L'_{\alpha}\Psi,~~~~\hat{\Gamma}=\Psi^{\dagger}\Gamma\Psi,  
\end{equation}
the quadratic Casimir is derived as  
\begin{equation}
\mathcal{C}=\hat{L}_i\hat{L}_i+\epsilon_{\alpha\beta}\hat{L}_{\alpha}\hat{L}_{\beta}+\epsilon_{\alpha\beta}\hat{L}'_{\alpha}\hat{L}'_{\beta}+\hat{\Gamma}^2
=\frac{1}{4}(\Psi^{\dagger}\Psi)^2. 
\label{casimiroftyposp22}
\end{equation}
 
The eigenvalue is $n^2/4$ and the corresponding eigenstates are given by (\ref{fullysymmosp22rep}). The $SU(2|1)$ typical representation for $(j,g)$ consists of $|j,j_3,g \rangle$,  $|j-1/2,j_3,g+1/2\rangle$, $|j-1/2,j_3,g-1/2\rangle$ and $|j-1,j_3,g\rangle$  with the Casimir eigenvalue (\ref{eigensu21quadaty}) \cite{Scheunert1977}. 
One may readily see that the Schwinger construction corresponds to $(j,g)=(n/2,0)$.

\subsubsection{$su(2|2)$}

The dimension of the $SU(2|2)$ algebra is 
\begin{equation}
\dim [su(2|2)]=7|8=15.
\end{equation}
The fundamental representation matrices of $su(2|2)$ are 
\begin{align}
&L_i=\frac{1}{2}\begin{pmatrix}
\sigma_i & 0 \\
0 & 0_2
\end{pmatrix},~~L_{\alpha}=\frac{1}{2}\begin{pmatrix}
0_2 & \tau_{\alpha}& 0 \\
-(\epsilon\tau_{\alpha})^t & 0 & 0 \\
0 & 0 & 0  
\end{pmatrix},~~D_{\alpha}=-\frac{1}{2}\begin{pmatrix}
0_2 & \tau_{\alpha} & 0 \\
(\epsilon\tau_{\alpha})^t & 0  & 0 \\
0 & 0 & 0 
\end{pmatrix},\nonumber\\
&L'_{\alpha}=\frac{1}{2}\begin{pmatrix}
0_2 & 0 & \tau_{\alpha}\\ 
0 & 0 & 0  \\
-(\epsilon\tau_{\alpha})^t & 0 & 0 
\end{pmatrix},~~D'_{\alpha}=-\frac{1}{2}\begin{pmatrix}
0_2 & 0 & \tau_{\alpha}\\
0 & 0 & 0 \\ 
(\epsilon\tau_{\alpha})^t & 0 & 0 
\end{pmatrix},~~T_i=\frac{1}{2}
\begin{pmatrix}
0_2 & 0 \\
0 & \sigma_i
\end{pmatrix},\nonumber\\
&H=1_4. 
\label{4times4su22}
\end{align}
With these,  
the $SU(2|2)$ algebra is expressed as\footnote{Since $H=1_4$ commutes with all of the other fourteen generators, $H$ generates the center of the $SU(2|2)$ algebra. 
 The $pSU(2|2)$ algebra is defined by quenching $H$, and then   
\begin{equation}
\dim~ [psu(2|2)]=6|8.
\end{equation}
There do not exist $4\times 4$ matrices that satisfy the $pSU(2|2)$ algebra. The minimum dimension matrices of $psu(2|2)$  are  $14\times 14$ matrices, $i.e.$ the adjoint representation. With $4\times 4$ fundamental representation matrices (\ref{4times4su22}), one may nevertheless discuss 
 $psu(2|2)$ by identifying matrices modulo $H$.} 
\begin{align}
&[L_i,L_j]=i\epsilon_{ijk}L_k,~~[L_i,L_{\alpha\sigma}]=\frac{1}{2}(\sigma_i)_{\beta\alpha}L_{\beta\sigma},~~[L_i,D_{\alpha\sigma}]=\frac{1}{2}(\sigma_i)_{\beta\alpha}D_{\beta\sigma},\nonumber\\
&\{L_{\alpha\sigma},L_{\beta\tau}\}=-\{D_{\alpha\sigma},D_{\beta\tau}\}=\frac{1}{2}\delta_{\sigma\tau}(\epsilon\sigma_i)_{\alpha\beta}L_i+i\frac{1}{2}\epsilon_{\sigma\tau}\epsilon_{\alpha\beta}T_2,\nonumber\\
&\{L_{\alpha\sigma},D_{\beta\tau}\}=-\frac{1}{2}
(\sigma_1)_{\sigma\tau}\epsilon_{\alpha\beta}T_1-\frac{1}{2}(\sigma_3)_{\sigma\tau}\epsilon_{\alpha\beta}T_3-\frac{1}{4}\delta_{\sigma\tau}\epsilon_{\alpha\beta}H,\nonumber\\
&[L_{\alpha\sigma},T_1]=-\frac{1}{2}(\sigma_1)_{\tau\sigma}D_{\alpha\tau},~~[D_{\alpha\sigma},T_1]=-\frac{1}{2}(\sigma_1)_{\tau\sigma}L_{\alpha\tau},\nonumber\\
&[L_{\alpha\sigma},T_2]=-\frac{1}{2}(\sigma_2)_{\tau\sigma}L_{\alpha\tau},~~[D_{\alpha\sigma},T_2]=-\frac{1}{2}(\sigma_2)_{\tau\sigma}D_{\alpha\tau},\nonumber\\
&[L_{\alpha\sigma},T_3]=-\frac{1}{2}(\sigma_3)_{\tau\sigma}D_{\alpha\tau},~~[D_{\alpha\sigma},T_3]=-\frac{1}{2}(\sigma_3)_{\tau\sigma}L_{\alpha\tau},\nonumber\\
&[L_i,T_j]=[L_{\alpha\sigma},H]=[D_{\alpha\sigma},H]=0,
\label{algebasu22}
\end{align}
where $L_{\alpha\sigma}=(L_{\alpha},L'_{\alpha})$ and $D_{\alpha\sigma}=(D_{\alpha},D'_{\alpha})$.  
 The $UOSp(2|2)$ algebra (\ref{osp22algebranew}) is a subalgebra of  (\ref{algebasu22}) realized by $L_i$, $L_{\alpha\sigma}$ and $\Gamma=iT_2$.

\subsection{$SU(4|N)$ algebra and $\mathbb{C}P_F^{3|N}$}

The dimension of the $SU(4|N)$ algebra is given by 
\begin{equation}
\dim[su(4|N)]=15+N^2|8N=N^2+8N+15.  
\end{equation}
For instance, 
 \begin{align}
 &\dim[su(4|1)]= 16|8=24,~~~~~~~~\dim[su(4|2)]=19|16=35,\nonumber\\
 &\dim [su(4|3)]= 24 |24=48,~~~~~~~\dim [su(4|4)]= 31 |32=63. 
 \end{align}

To clarify relations to $uosp(N|4)$, we adopt the following ``decomposition''. 
We separate the $SU(4)$ generators, $S_A$ $(A=1,2,\cdots,15)$, into $SO(5)$ vector and antisymmetric rank 2 tensor:    
\begin{equation}
\frac{1}{2\sqrt{2}}\Gamma_a=\frac{1}{\sqrt{2}}
\begin{pmatrix}
\gamma_a & 0 \\
0 & 0_N
\end{pmatrix},~~~~~\frac{1}{\sqrt{2}}\Gamma_{ab}=\frac{1}{\sqrt{2}}
\begin{pmatrix}
\gamma_{ab} & 0 \\
0 & 0_N
\end{pmatrix}, 
\end{equation}
with $\gamma_a$ (\ref{so5gammaI1}) and $\gamma_{ab}$ (\ref{so5geneI1}). Notice that $\gamma_a$ and $\gamma_{ab}$ have different properties under transpose    
\begin{equation}
(C\gamma_a)^t=-C\gamma_a,~~~~~~~~~(C\gamma_{ab})^t=C\gamma_{ab},  
\label{propertiesofchargematrix}
\end{equation}
where $C$ stands for the $SO(5)$ charge conjugation matrix (\ref{so5chargeconjmat}). 
We also separate the $SU(N)$ generators $T_P$ $(P=1,2,\cdots,N^2-1)$ into symmetric and antisymmetric matrices: 
\begin{equation}
{T_S}^t=T_S,~~~~~~{T_I}^t=-T_I,
\end{equation}
with $S=1,2,\cdots,N(N+1)/2-1$ and $I=1,2,\cdots,N(N-1)/2$. $T_I$ satisfy the $SO(N)$ algebra by themselves. 
 Also, the $U(1)$ generator is given by 
\begin{equation}
H=\frac{1}{N}\begin{pmatrix}
N\cdot 1_4 & 0 \\
0 & 4\cdot 1_N
\end{pmatrix}. 
\end{equation}
We introduce the $SU(4|N)$ fermionic generators as  
\begin{align}
&\Gamma_{\alpha\sigma}=\frac{1}{\sqrt{2}}
\begin{pmatrix}
0_{3+\sigma} & \tau_{\alpha} & 0  \\
-(C\tau_{\alpha})^t & 0 & 0 \\
0 & 0 & 0_{N-\sigma}  
\end{pmatrix},~~~~~D_{\alpha\sigma}=\frac{1}{\sqrt{2}}
\begin{pmatrix}
0_{3+\sigma} & \tau_{\alpha} & 0  \\
(C\tau_{\alpha})^t & 0 & 0 \\
0 & 0 & 0_{N-\sigma}  
\end{pmatrix}, 
\end{align}
which are related to (\ref{Qmatrices}) as 
\begin{align}
Q_{\alpha\sigma}=\frac{1}{\sqrt{2}}(\Gamma_{\alpha\sigma}+D_{\alpha\sigma}),~~~~~~~\tilde{Q}_{\sigma\alpha}=-\frac{1}{\sqrt{2}}C_{\alpha\beta}(\Gamma_{\beta\sigma}-D_{\beta\sigma}),  
\end{align}
or 
\begin{align}
\Gamma_{\alpha\sigma}=\frac{1}{\sqrt{2}} (Q_{\alpha\sigma}+C_{\alpha\beta}\tilde{Q}_{\sigma\beta}),~~~~~~~~D_{\alpha\sigma}=\frac{1}{\sqrt{2}} (Q_{\alpha\sigma}-C_{\alpha\beta}\tilde{Q}_{\sigma\beta}). 
\end{align}
Therefore, 
\begin{equation}
Q_{\alpha\sigma}\tilde{Q}_{\sigma\alpha}-\tilde{Q}_{\sigma\alpha}Q_{\alpha\sigma}=-C_{\alpha\beta}(\Gamma_{\alpha\sigma}\Gamma_{\beta\sigma}-D_{\alpha\sigma}D_{\beta\sigma}). 
\end{equation}
The $SU(4|N)$ commutation relations (\ref{SUMNcommutationrelations}) concerned with $\Gamma_{\alpha\sigma}$ and $D_{\alpha\sigma}$ read as 
\begin{align}
&[\Gamma_{a},\Gamma_{\alpha\sigma}]=(\gamma_a)_{\beta\alpha}D_{\beta\sigma},~~~~~~~~~~
[\Gamma_{ab},\Gamma_{\alpha\sigma}]=(\gamma_{ab})_{\beta\alpha}\Gamma_{\beta\sigma},\nonumber\\
&[\Gamma_{a},D_{\alpha\sigma}]=(\gamma_a)_{\beta\alpha}\Gamma_{\beta\sigma},~~~~~~~~~~
[\Gamma_{ab},D_{\alpha\sigma}]=(\gamma_{ab})_{\beta\alpha}D_{\beta\sigma}, \nonumber\\
&\{\Gamma_{\alpha\sigma},\Gamma_{\beta\tau}\}=-\{D_{\alpha\sigma},D_{\beta\tau}\}=\delta_{\sigma\tau}(C\gamma_{ab})_{\alpha\beta}\Gamma_{ab}-2C_{\alpha\beta}(t_{I})_{\sigma\tau}T_{I},\nonumber\\
&\{\Gamma_{\alpha\sigma},D_{\beta\tau}\}=\frac{1}{4}\delta_{\sigma\tau} (C\gamma_{a})_{\alpha\beta}\Gamma_a+2C_{\alpha\beta}(t_{S})_{\sigma\tau}T_{S}+\frac{1}{4}C_{\alpha\beta}\delta_{\sigma\tau}H,\nonumber\\
&[\Gamma_{\alpha\sigma},T_{S}]=(t_{S})_{\sigma\tau}D_{\alpha\tau},~~~~~~~~~~[\Gamma_{\alpha\sigma},T_{I}]=(t_{I})_{\sigma\tau}\Gamma_{\alpha\tau},\nonumber\\
&[D_{\alpha\sigma},T_{S}]=(t_{S})_{\sigma\tau}\Gamma_{\alpha\tau},~~~~~~~~~~[D_{\alpha\sigma},T_{I}]=(t_{I})_{\sigma\tau}D_{\alpha\tau},\nonumber\\
&[\Gamma_{\alpha\sigma},H]=\frac{4-N}{N}D_{\alpha\sigma},~~~~~~~~~~[D_{\alpha\sigma},H]=\frac{4-N}{N}\Gamma_{\alpha\sigma}, 
\label{SU(4N)commutationrel}
\end{align}
where $t_S$ and $t_I$ are respectively $N\times N$ symmetric and antisymmetric matrices of $SU(N)$ generators $t_P$ (\ref{sumngeneratorsmatrices}). 
Thus, one may find that $\Gamma_{ab}$, $\Gamma_{\alpha\sigma}$, $T_I$ satisfy a closed algebra, the $uosp(N|4)$. 

Square of the radius of  $\mathbb{C}P^{3|N}_F$ is derived as  
\begin{align}
&\sum_{A=1}^{15}\hat{S}_A\hat{S}_A-\frac{1}{2}\sum_{\alpha=1}^4\sum_{\sigma=1}^k (\hat{Q}_{\alpha\sigma}\hat{\tilde{Q}}_{\sigma\alpha}-\hat{\tilde{Q}}_{\sigma\alpha}\hat{Q}_{\alpha\sigma})-\sum_{P=1}^{N^2-1}\hat{T}_P \hat{T}_P-\frac{N}{8(4-N)}\hat{Z}^2\nonumber\\ 
=&\frac{1}{8}\sum_{a=1}^5 X_aX_a+\frac{1}{2}\sum_{a<b=1}^5 X_{ab}X_{ab}+\frac{1}{2}\sum_{\alpha,\beta=1}^4\sum_{\sigma=1}^{N}C_{\alpha\beta}(\Theta_{\alpha}^{(\sigma)}\Theta_{\beta}^{(\sigma)}-\varTheta_{\alpha}^{(\sigma)}\varTheta_{\beta}^{(\sigma)})-\frac{1}{4}\sum_{P=1}^{N^2-1}Y_{P}Y_{P}-\frac{N}{8(4-N)}Z^2\nonumber\\
=&\frac{3-N}{2(4-N)}\hat{n}(\hat{n}+4-N), 
\end{align}
where $\hat{n}=\Psi^{\dagger}\Psi$, 
\begin{align}
&X_a=\Psi^{\dagger}\Gamma_a\Psi,~~~~~~~X_{ab}=\Psi^{\dagger}\Gamma_{ab}\Psi,~~~~~\Theta_{\alpha}^{(\sigma)}= \Psi^{\dagger}\Gamma_{\alpha\sigma}\Psi,\nonumber\\
&\varTheta_\alpha^{(\sigma)}=\Psi^{\dagger}D_{\alpha\sigma}\Psi,~~~~Y_{P}=2\Psi^{\dagger}T_{P}\Psi,~~~~Z=\Psi^{\dagger}H\Psi.    
\end{align}
We utilized  
\begin{align}
&\sum_{A=1}^{15}\hat{S}_A\hat{S}_A=\frac{1}{8}\sum_{a} X_aX_a+\frac{1}{2}\sum_{a<b} X_{ab}X_{ab}=\frac{3}{8}\hat{n}_B(\hat{n}_B+4),\nonumber\\
&\sum_{\alpha=1}^4\sum_{\sigma=1}^{N}(\hat{Q}_{\alpha\sigma}\hat{\tilde{Q}}_{\sigma\alpha}-\hat{\tilde{Q}}_{\sigma\alpha}\hat{Q}_{\alpha\sigma})=- C_{\alpha\beta}(\Theta_{\alpha}^{(\sigma)}\Theta_{\beta}^{(\sigma)}-\varTheta_{\alpha}^{(\sigma)}\varTheta_{\beta}^{(\sigma)})=N\hat{n}_B-2\hat{n}_B\hat{n}_F-4\hat{n}_F,\nonumber\\
&\sum_{P=1}^{N^2-1}\hat{T}_P\hat{T}_P=\frac{1}{4}\sum_{P=1}^{N^2-1}Y_{P}Y_{P}=-\frac{N+1}{2N}\hat{n}_F(\hat{n}_F-N),\nonumber\\
&\hat{Z}^2=(\hat{n}_B+\frac{4}{N}\hat{n}_F)^2. 
\end{align}
Thus, for $\mathbb{C}P^{3|N}_F$, square of the radius is proportional to 
\footnote{For instance, for $N=0$, (\ref{squareradiuscp3n}) is reduced to square of the radius of fuzzy  $\mathbb{C}P^3$. 
Since $\mathbb{C}P^3$ is locally $S^4\otimes S^2$, one may think of the need of two quantities to determine the radius of each  $S^4$ and $S^2$. However, in the construction of fuzzy $\mathbb{C}P^3$, we utilized the fully symmetric representation specified by the sole number $n$. Consequently, the ``internal'' fuzzy $S^2$ and $S^4$ of fuzzy $\mathbb{C}P^3$ are represented by matrices of same size specified by $n$, and hence their radii are similarly by $n$ 
(see also Sec.\ref{subsec2ndHopfandfuzzy}). 
} 
\begin{equation}
{n}({n}+4-N).  
\label{squareradiuscp3n}
\end{equation}
 
\subsubsection{$su(4|1)$}\label{append:su41}

The  $24(=16|8)$ matrices of  $su(4|1)$ are given by  
\begin{align}
&\Gamma_a=\begin{pmatrix}
\gamma_{a} & 0 \\
0 & 0 
\end{pmatrix},~~~\Gamma_{ab}=\begin{pmatrix} 
\gamma_{ab} & 0 \\
0 & 0 
\end{pmatrix},~~~H=\begin{pmatrix}
1_4 & 0 \\
0 & 4
\end{pmatrix},\nonumber\\
&\Gamma_{\alpha}=\frac{1}{\sqrt{2}}\begin{pmatrix}
0_4 & \tau_{\alpha}\\
-(C\tau_{\alpha})^t & 0 
\end{pmatrix},~~D_{\alpha}=\frac{1}{\sqrt{2}}
\begin{pmatrix}
0_4 & \tau_{\alpha}\\
(C\tau_{\alpha})^t & 0 
\end{pmatrix}, 
\label{su41matricegene}
\end{align}
which satisfy  
 \begin{align}
 &[\Gamma_a,\Gamma_b]=4i\Gamma_{ab},~~~~~~~~~~[\Gamma_a,\Gamma_{bc}]=-i(\delta_{ab}\Gamma_c-\delta_{ac}\Gamma_b),~~~~[\Gamma_{ab},\Gamma_{cd}]=i(\delta_{ac}\Gamma_{bd}-\delta_{ad}\Gamma_{bc}+\delta_{bc}\Gamma_{ad}-\delta_{bd}\Gamma_{ac}),\nonumber\\
 &[\Gamma_a,\Gamma_{\alpha}]=(\gamma_a)_{\beta\alpha}D_{\beta},~~~~[\Gamma_a,D_{\alpha}]=(\gamma_a)_{\beta\alpha}\Gamma_{\beta},\nonumber\\
&[\Gamma_{ab},\Gamma_{\alpha}]=(\gamma_{ab})_{\beta\alpha}\Gamma_{\beta},~~~[\Gamma_{ab},D_{\alpha}]=(\gamma_{ab})_{\beta\alpha}D_{\beta},\nonumber\\
 &\{\Gamma_{\alpha},\Gamma_{\beta}\}=\sum_{a<b}(C\gamma_{ab})_{\alpha\beta}\Gamma_{ab},~~~
\{D_{\alpha},D_{\beta}\}=-\sum_{a<b}(C\gamma_{ab})_{\alpha\beta}\Gamma_{ab},~~~
\{\Gamma_{\alpha},D_{\beta}\}=\frac{1}{4}(C\gamma_a)_{\alpha\beta}\Gamma_a+\frac{1}{4}C_{\alpha\beta}H,\nonumber\\
&[\Gamma_{\alpha},H] =3D_{\alpha},~~~~~~~~~~~~[D_{\alpha},H]=3\Gamma_{\alpha}. 
 \end{align}
$\Gamma_{ab}$ and $\Gamma_{\alpha}$ satisfy a closed algebra, the $uosp(1|4)$. 

\subsubsection{$su(4|2)$}\label{append:su42}

The $SU(4|2)$ algebra contains $35(=19|16)$ generators. The bosonic and fermionic generators are given by  
\begin{align}
&\Gamma_a,~~\Gamma_{ab},~~~T_i,~~~H,\nonumber\\
&\Gamma_{\alpha},~~D_{\alpha},~~~\Gamma'_{\alpha},~~D'_{\alpha}, 
\label{su42geneunderosp42}
\end{align}
where $\Gamma_a$, $\Gamma_{ab}$, $\Gamma_{\alpha}$ and $\Gamma'_{\alpha}$
 are defined by (\ref{osp24fermionicgenemat}) and (\ref{so5gammaosp24}),  $T_i$ and $H$ are $U(2)$ generators 
\begin{equation}
T_i=\frac{1}{2}\begin{pmatrix}
0_4 & 0 \\
0 & \sigma_i
\end{pmatrix},~~~~H=\begin{pmatrix}
1_4 & 0 \\
0 & 2\cdot 1_2
\end{pmatrix},  
\end{equation}
and $D_{\alpha}$ and  $D_{\alpha}'$ are 
\begin{equation}
D_{\alpha}=\frac{1}{\sqrt{2}}\begin{pmatrix}
 0_4 & \tau_{\alpha} & 0 \\
 (C\tau_{\alpha})^t & 0 & 0 \\
 0 & 0 & 0 
 \end{pmatrix},~~~~D'_{\alpha}=\frac{1}{\sqrt{2}}\begin{pmatrix}
 0_4 &  0 &  \tau_{\alpha}  \\
  0 & 0 & 0 \\
 (C\tau_{\alpha})^t & 0 & 0  
 \end{pmatrix}.   
 \label{su22dfermionicgenemat}
 \end{equation}
The $SU(4|2)$ generators (\ref{su42geneunderosp42}) satisfy 
\begin{align}
&\{\Gamma_{\alpha\sigma},\Gamma_{\beta\tau}\}=-\{D_{\alpha\sigma},D_{\beta\tau}\}=\delta_{\sigma\tau}\sum_{a<b}(C\gamma_{ab})_{\alpha\beta}\Gamma_{ab}+i\frac{1}{2}\epsilon_{\sigma\tau}C_{\alpha\beta}T_2,\nonumber\\
&\{\Gamma_{\alpha\sigma},D_{\beta\tau}\}=\frac{1}{4}\delta_{\sigma\tau}(C\gamma_a)_{\alpha\beta}\Gamma_a+
(\sigma_1)_{\sigma\tau}C_{\alpha\beta}T_1+(\sigma_3)_{\sigma\tau}C_{\alpha\beta}T_3+\frac{1}{4}\delta_{\sigma\tau}C_{\alpha\beta}H,\nonumber\\
&[\Gamma_{\alpha\sigma},T_1]=\frac{1}{2}(\sigma_1)_{\tau\sigma}D_{\alpha\tau},~~~~~~~[D_{\alpha\sigma},T_1]=\frac{1}{2}(\sigma_1)_{\tau\sigma}\Gamma_{\alpha\tau},\nonumber\\
&[\Gamma_{\alpha\sigma},T_2]=-\frac{1}{2}(\sigma_2)_{\tau\sigma}\Gamma_{\alpha\tau},~~~~~[D_{\alpha\sigma},T_2]=-\frac{1}{2}(\sigma_2)_{\tau\sigma}D_{\alpha\tau},\nonumber\\
&[\Gamma_{\alpha\sigma},T_3]=\frac{1}{2}(\sigma_3)_{\tau\sigma}D_{\alpha\tau},~~~~~~~[D_{\alpha\sigma},T_3]=\frac{1}{2}(\sigma_3)_{\tau\sigma}\Gamma_{\alpha\tau},\nonumber\\
&[\Gamma_{\alpha\sigma},H]=D_{\alpha\sigma},~~~~~~~~~~~~~~~~~~[D_{\alpha\sigma},H]=\Gamma_{\alpha\sigma},\nonumber\\
&[T_i,T_j]=i\epsilon_{ijk}T_k,~~~~~~~~~~~~~~~~~[\Gamma_a,H]=[\Gamma_{ab},H]=[T_i,H]
=0,\nonumber\\
&[\Gamma_a,T_i]=[\Gamma_{ab},T_i]=0,\label{algebasu42}
\end{align}
where $\Gamma_{\alpha\sigma}=(\Gamma_{\alpha},\Gamma'_{\alpha})$ and $D_{\alpha\sigma}=(D_{\alpha},D'_{\alpha})$. 
 The $uosp(2|4)$  (\ref{osp24algebranew}) is realized as a subalgebra of $su(4|2)$  (\ref{algebasu42}) with $\Gamma_{ab},\Gamma_{\alpha\sigma}$ and $\Gamma=2iT_2$.

\section{Charge conjugation matrices of $SO(5)$ and $UOSp(1|4)$}\label{appenchargeconjugationso5}

The complex representation of  $so(5)$, $\gamma_a$ (\ref{so5gammaI1}) and $\gamma_{ab}$ (\ref{so5geneI1}), is given by 
\begin{equation}
\tilde{\gamma}_a=\gamma^*_a=\gamma_a^t,~~~~\tilde{\gamma}_{ab}=-i\frac{1}{4}[\tilde{\gamma}_a,\tilde{\gamma}_b]=-\gamma_{ab}^*=-\gamma_{ab}^t. 
\end{equation}
The $SO(5)$ charge conjugation matrix (\ref{so5chargeconjmat})  acts as 
\begin{equation}
C^t\gamma_aC=\tilde{\gamma}_a,~~~~C^t\gamma_{ab}C=\tilde{{\gamma}}_{ab}. 
\end{equation}
 $C\gamma_{ab}$ and $\gamma_{ab} C$ are symmetric matrices, while 
$C\gamma_{a}$ and $\gamma_{a} C$ are anti-symmetric matrices. 
$C$  has the following properties
\begin{equation}
C^{\dagger}=C^t=C^{-1}=-C,~~~~C^2=-1, 
\end{equation}
and is related to the $USp(4)$ invariant matrix (see Sec.\ref{SecUOSp(MN)}) 
\begin{equation}
J=\begin{pmatrix}
0 & 1_2 \\
-1_2 & 0 
\end{pmatrix}, 
\end{equation}
by unitary transformation, $J=V^{\dagger} C V$, with 
\begin{equation}
V=
\begin{pmatrix}
1 & 0  & 0 & 0 \\
0 & 0 & 1 & 0 \\
0 & 1 & 0 & 0 \\
0 & 0 & 0 & 1
\end{pmatrix}.
\label{unitarymatrels}
\end{equation}
The unitary matrix (\ref{unitarymatrels}) also relates the $so(5)$ matrices (\ref{so5geneI1}) to the bases of $usp(4)$ matrix (\ref{sp4generators}).   
 
 The complex representation of $uosp(1|4)$, $\Gamma_a$, $\Gamma_{ab}$ and $\Gamma_{\alpha}$ defined in Sec.\ref{subsecuosp14}, is given by 
\begin{equation}
\tilde{\Gamma}_a=\Gamma^t_a=\Gamma_a^*,~~~~\tilde{\Gamma}_{ab}=-\Gamma_{ab}^*=-\Gamma_{ab}^t,~~~~\tilde{\Gamma}_{\alpha}=C_{\alpha\beta}\Gamma_{\beta}.  
\end{equation}
The complex representation is related to the original representation as 
\begin{equation}
\mathcal{R}^t \Gamma_a \mathcal{R}=\tilde{\Gamma}_a,\quad\quad\mathcal{R}^t \Gamma_{ab}\mathcal{R}=\tilde{\Gamma}_{ab},\quad\quad \mathcal{R}^t \Gamma_{\alpha} \mathcal{R}=\tilde{\Gamma}_{\alpha},
\end{equation}
with the charge conjugation matrix 
\begin{equation}
\mathcal{R}=
\begin{pmatrix}
 C & 0 \\
 0 & 1 
\end{pmatrix}.
\end{equation}

\section{Relations for matrix products}

\subsection{$su(2|1)$ matrices}

With $3\times 3$ unit matrix $1_3$, 
the $su(2|1)$ fundamental representation matrices (\ref{simplestospmat}) span the space of $3\times 3$ matrices, and hence their products can be given by their linear combination.  
For $uosp(1|2)$ matrices, $L_i$ and $L_{\alpha}$, their products are represented as 
\begin{align}
&L_iL_j=\frac{1}{4}\delta_{ij}+i\frac{1}{2}\epsilon_{ijk}L_k,\nonumber\\
&L_iL_{\alpha}=\frac{1}{4}(\sigma_i)_{\beta\alpha}(L_{\beta}-D_{\beta}),\nonumber\\
&L_{\alpha}L_{\beta}=\frac{1}{4}(\epsilon\sigma_i)_{\alpha\beta}L_i-\frac{1}{2}\epsilon_{\alpha\beta}(1-\frac{3}{4}H).  
\end{align}
For the other $su(2|1)$ fundamental representation matrices, 
\begin{align}
&L_iD_{\alpha}=-\frac{1}{4}(\sigma_i)_{\beta\alpha}(L_{\beta}-D_{\beta}),\nonumber\\
&L_i H= L_i,\nonumber\\
&L_{\alpha}D_{\beta}= \frac{1}{4}(\epsilon\sigma_i)_{\alpha\beta}L_i-\frac{1}{8}\epsilon_{\alpha\beta}H,\nonumber\\
&L_{\alpha}H=-\frac{1}{2}D_{\alpha}+\frac{3}{2}L_{\alpha},\nonumber\\
&D_{\alpha}D_{\beta}= -\frac{1}{4}(\epsilon\sigma_i)_{\alpha\beta}L_i+\frac{1}{2}\epsilon_{\alpha\beta}(1-\frac{3}{4}H),\nonumber\\
&D_{\alpha}H= -\frac{1}{2}L_{\alpha}+\frac{3}{2}D_{\alpha},\nonumber\\
&H^2=3H-2.   
\end{align}

\subsection{$su(4|1)$ matrices}\label{appensubsecsu41}

Similar to the $su(2|1)$ case, with $5\times 5$ unit matrix $1_5$, 
the $su(4|1)$ fundamental representation matrices (\ref{su41matricegene}) span the space of  $5\times 5$ matrices. Then, their products can be expressed by their linear combination: for the products of $\Gamma_a$ and $\Gamma_{\alpha}$,  
\begin{align}
&\Gamma_a\Gamma_b=2i\Gamma_{ab}-\frac{1}{3}\delta_{ab}(H-{4}),\nonumber\\
&\Gamma_a \Gamma_{\alpha}= \frac{1}{2}(\gamma_a)_{\beta\alpha}(\Gamma_{\beta}+D_{\beta}),\nonumber\\
&\Gamma_{\alpha}\Gamma_{\beta}=-\frac{1}{2}\sum_{a<b}(C\gamma_{ab})_{\alpha\beta}\Gamma_{ab}-\frac{1}{8}(C\gamma_a)_{\alpha\beta}\Gamma_a-\frac{1}{3}C_{\alpha\beta}(1-\frac{5}{8}H),
\end{align}
and for the other $su(4|1)$ matrices, 
\begin{align}
&\Gamma_a\Gamma_{bc}=\frac{1}{2}(\epsilon_{abcde}\Gamma_{de}+i\Gamma_b\delta_{ac}-i\Gamma_c\delta_{ab}),\nonumber\\
&\Gamma_a D_{\alpha}= \frac{1}{2}(\gamma_a)_{\beta\alpha}(\Gamma_{\beta}+D_{\beta}), \nonumber\\
&\Gamma_a\Gamma=\Gamma_a,\nonumber\\ 
&\Gamma_{ab}\Gamma_{cd}=2i(\delta_{ab}\Gamma_{cd}-\delta_{ac}\Gamma_{bd}+\delta_{ad}\Gamma_{bc}-\delta_{bc}\Gamma_{da}+\delta_{bd}\Gamma_{ca}-\delta_{cd}\Gamma_{ba})-\epsilon_{abcde}\Gamma_e \nonumber\\
&~~~~~~~~~~~+\frac{1}{3}(\delta_{ab}\delta_{cd}-\delta_{ac}\delta_{bd}+\delta_{ad}\delta_{bc})(4-H),\nonumber\\
&\Gamma_{ab} \Gamma_{\alpha}=\frac{1}{2}(\gamma_{ab})_{\beta\alpha}(\Gamma_{\beta}+D_{\beta}),\nonumber\\
&\Gamma_{ab} D_{\alpha}=\frac{1}{2}(\gamma_{ab})_{\beta\alpha}(\Gamma_{\beta}+D_{\beta}),\nonumber\\
&\Gamma_{ab}H=\Gamma_{ab},\nonumber\\ 
&\Gamma_{\alpha} D_{\beta}=-\frac{1}{2}\sum_{a<b}(C\gamma_{ab})_{\alpha\beta}\Gamma_{ab}+\frac{1}{8}(C\gamma_a)_{\alpha\beta}\Gamma_a +\frac{1}{8}C_{\alpha\beta}H,\nonumber\\
&\Gamma_{\alpha}H=-\frac{3}{2}D_{\alpha}-\frac{5}{2}\Gamma_{\alpha},\nonumber\\
&D_{\alpha} D_{\beta}=-\frac{1}{2}\sum_{a<b}(C\gamma_{ab})_{\alpha\beta}\Gamma_{ab}+\frac{1}{8}(C\gamma_a)_{\alpha\beta}\Gamma_a +\frac{1}{3}C_{\alpha\beta} (1-\frac{5}{8}H),\nonumber\\
&D_{\alpha}H= -\frac{5}{2}D_{\alpha}-\frac{3}{2}\Gamma_{\alpha},\nonumber\\
&H^2= 5H-4. 
\end{align}



\end{document}